\documentclass[%
reprint,
nofootinbib,
mathtools,
amsmath,
amssymb,
prd,
longbibliography,
superscriptaddress]{revtex4-2}

\usepackage{graphicx}
\usepackage{dcolumn}
\usepackage{bm}
\usepackage{mathrsfs}
\usepackage{mathtools}
\usepackage{commath}
\usepackage{color}
\usepackage[dvipsnames]{xcolor}
\usepackage{float}
\usepackage[pdftex,breaklinks,colorlinks,
linkcolor=Blue,
citecolor=teal,
anchorcolor=red,
urlcolor=cyan]{hyperref}
\usepackage{booktabs}
\usepackage{comment}
\usepackage{lipsum}
\usepackage{graphics}
\usepackage{bbm}
\usepackage[dvipsnames]{xcolor}
\usepackage{microtype}
\usepackage{IEEEtrantools}
\usepackage{amsthm}
\usepackage{enumitem}
\usepackage{orcidlink}

\newcommand{\paramD}{\partial_{\vec{\mathcal{V}}}}

\begin{document}

\title{\boldmath Transition-to-plunge self-force waveforms with a spinning primary}

\author{Loïc Honet}
\email{loic.honet@ulb.be}
\affiliation{Universit\'e Libre de Bruxelles, BLU-ULB Brussels Laboratory of the Universe, C.P. 231, B-1050 Bruxelles, Belgium}

\author{Lorenzo K\"uchler}%
\email{l.m.kuchler@soton.ac.uk}
\affiliation{School of Mathematical Sciences and STAG Research Centre, University of Southampton, Southampton SO17 1BJ, United Kingdom}

\author{Adam Pound}%
\email{A.Pound@soton.ac.uk}
\affiliation{School of Mathematical Sciences and STAG Research Centre, University of Southampton, Southampton SO17 1BJ, United Kingdom}

\author{Geoffrey Comp\`ere}%
\email{geoffrey.compere@ulb.be}
\affiliation{Universit\'e Libre de Bruxelles, BLU-ULB Brussels Laboratory of the Universe, C.P. 231, B-1050 Bruxelles, Belgium}

\date{\today}

\begin{abstract}
With the upcoming third-generation gravitational-wave detectors comes the need to build complete, faithful, and fast waveform models for asymmetric-mass-ratio compact binaries. Most efforts within the self-force community have focused on modeling these binaries' inspiral regime, but for ground-based detectors the systems' final merger can represent the dominant part of the signal. Recent work by three of us has extended the multiscale self-force framework through the transition-to-plunge and merger-ringdown regimes for nonspinning binaries. In this paper, we generalize the next-to-next-to-leading-order transition-to-plunge waveform model to include the spin of the primary black hole. We also improve the construction of composite inspiral-transition waveform models by performing a change of variables on the binary's mechanical phase space during the transition to plunge. We provide detailed discussions of our numerical implementation and comparisons with numerical relativity simulations.
\end{abstract}

\maketitle

\tableofcontents

\section{Introduction and outline}\label{sec:level1}

Since the first detection of a gravitational-wave (GW) signal 10 years ago (GW150914)~\cite{LIGOScientific:2016aoc}, the LIGO-Virgo-KAGRA (LVK) collaboration has now seen more than a hundred binary coalescence events among their first three observing runs~\cite{Abbott_2023}. With the upcoming release of the fourth version of the Gravitational-Wave Transient Catalog (GWTC-4) and the fifth observing run planned for 2027~\cite{GraceDB}, many more GW events will soon be reported or discovered by ground-based detectors. Together with the improvement of detectors' sensitivity, this spurs GW modelers to provide fast and faithful waveform models for parameter estimation studies and tests of general relativity~\cite{toubiana2024indistinguishabilitycriterionestimatingpresence,PhysRevD.57.4566,PhysRevD.78.124020,PhysRevD.82.024014,PhysRevD.95.104004,PhysRevResearch.2.023151}.

In particular, one specific event from the third observing run, GW191219\_163120, has been estimated to come from a binary with mass ratio $\sim$1:27. Such a high mass ratio lies beyond what current models are able to cover~\cite{Abbott_2023,Wardell:2021fyy} and points to one of the LVK observational science short-term R\&D objectives: providing fast and accurate waveform models for asymmetric-mass-ratio binaries~\cite{LVKWhitePaper}.

On the other side, future space-based detectors such as LISA will detect GW signals in the millihertz spectrum~\cite{LISA:2024hlh}, allowing us to observe signals emitted by extreme-mass-ratio inspirals (EMRIs). The joint use of space-based detectors with future third-generation (3G) ground-based detectors such as the Einstein Telescope~\cite{ET:2019dnz} will enable the observation of sources such as intermediate mass ratio coalescences (IMRACs) across multiple bandwidths~\cite{LISA:2024hlh}. Those intermediate systems with mass ratios typically ranging from $\sim$1:$10^{2}$ to $\sim$1:$10^{4}$ currently lack accurate waveform models and constitute a real GW modeling challenge for 3G detectors.

The waveform modeling technique that naturally leverages the existence of two disparate masses is the gravitational self-force (GSF or SF) program, where the Einstein field equations (EFEs) and the orbital motion of the secondary black hole are expanded in the binary's small mass ratio. Recent milestones in the self-force community have been, for example, the construction of a first-post-adiabatic (1PA)/second-order self-force (2GSF) waveform model for spinning binaries with a slowly spinning primary black hole and rapidly spinning, precessing secondary~\cite{PaperII} and the development of a fast, data-analysis-ready adiabatic (0PA) model for eccentric equatorial binaries with a rapidly spinning primary in the \texttt{FEW} python package~\cite{Chapman-Bird:2025xtd}, leveraging the SF multiscale expansion framework~\cite{Hinderer:2008dm,Miller:2020bft,Pound:2021qin,Mathews:2025nyb} and hardware acceleration~\cite{Katz:2021yft} for rapidly generating waveforms.

These pieces of work focus primarily on the inspiral phase of the binary, which is expected to be sufficient for modeling EMRI signals. In contrast, IMRACs usually have a merger that occurs in the frequency band of ground-based detectors~\cite{LISA:2024hlh}. Moreover, recent results in second-order self-force~\cite{Warburton:2021kwk,Wardell:2021fyy} show that self-force models can be remarkably precise even at more comparable mass ratios $\sim$1:10~\cite{PaperII,Albertini:2022rfe}, and for such systems the merger can always represent a significant fraction of the signal-to-noise ratio (SNR). Those considerations make it important to extend the multiscale self-force framework beyond the innermost stable circular orbit (ISCO), where the inspiral motion of the binary breaks down.

A recent work by three of us~\cite{Kuchler:2024esj} extensively derived the self-force framework for the transition-to-plunge (or simply ``transition'') motion of nonspinning, quasicircular binaries. This work extended the results of Refs.~\cite{Compere:2021iwh,PhysRevLett.128.029901,Compere:2021zfj} by including a treatment of the Einstein field equations and waveform generation on top of the orbital dynamics, while also reformulating the transition in the phase-space approach~\cite{Pound:2021qin,Mathews:2025nyb,Lewis:2025ydo} that underlies the multiscale expansion's accuracy~\cite{Wardell:2021fyy} and efficiency. 

In GSF theory, the merger-ringdown part of the waveform is generated by the secondary' final, approximately geodesic plunge into the primary after it transitions across the ISCO~\cite{Ori:2000zn,Apte:2019txp}. Again, three of us recently showed how to formulate this regime in the phase-space approach of the multiscale expansion~\cite{Kuchler:2025hwx} (building on Refs.~\cite{Hadar:2009ip,Folacci:2018cic}). This created a unified framework for inspiral, transition, and plunge that can be carried (in principle) to any order in the small mass ratio. In Ref.~\cite{Roy:2025kra}, two of us employed that framework to generate subleading-order merger-ringdown waveforms in a test case of modified gravity. 

These developments have paved the way for building the first inspiral-merger-ringdown (IMR) model for quasicircular, nonspinning binaries beyond leading order in GSF theory~\cite{KuchlerCapra27,KuchlerLISA,KuchlerAEI,KuchlerGR}. (See Refs.~\cite{Rifat:2019ltp,Islam:2022laz,Rink:2024swg} for earlier such waveforms at leading order, following an iterative method initiated in Ref.~\cite{Sundararajan:2007jg} rather than a multiscale approach.)

With a first complete beyond-leading-order GSF IMR waveform model for nonspinning binaries at hand, we now aim to to include additional physical parameters. The work presented in this paper represents one of the intermediate steps towards including the effects of the primary black hole spin in the IMR waveform model presented in Ref.~\cite{KuchlerCapra27,KuchlerLISA,KuchlerAEI,KuchlerGR}: we derive and implement second-post-leading transition-to-plunge (2PLT) waveforms using the phase-space formalism for non-eccentric equatorial motion of a Schwarzschild secondary black hole around a primary Kerr black hole. Moreover, we build a composite waveform model that smoothly interpolates between an adiabatic (0PA) model in the early inspiral and a 2PLT transition model when reaching the ISCO using a matched asymptotic expansions procedure. We address and solve issues already raised in Ref.~\cite{Kuchler:2024esj} about the accuracy of such composite models due to early-time transition residuals in the dynamics.

The plan of the paper is as follows. In Sec.~\ref{sec:forcedEOM} we set up the conventions used throughout the paper and rewrite the equations of motion of a massive particle on a quasicircular orbit around a Kerr primary in a generic multiscale framework on the binary's mechanical phase space. In Sec.~\ref{sec:PAexpansion}, we explicitly expand such equations of motion during the inspiral phase of the binary up to next-to-leading order (1PA) in the self-force expansion. In Sec.~\ref{sec:PLTexpansion}, we perform the expansion of the same equations of motion but during the transition phase of the binary, up to 2PLT order in SF. Next, in Sec.~\ref{sec:matching}, we match the late-time asymptotic expansion of the 0PA inspiral with the early-time asymptotic expansion of the 2PLT transition motion and use it to build a composite 0PA-2PLT model. In Sec.~\ref{sec:reexpansion}, we perform a change of transition phase-space coordinates and explain how it affects the asymptotic matching with the inspiral motion and, crucially, how it improves the accuracy of the composite model. In Sec.~\ref{sec:implementation} we detail explicitly the different components of our composite model, as well as other models we use as benchmarks. Finally, in Sec.~\ref{sec:results} we provide an analysis of the impact of our change of variables on phase space and compare our waveform models with numerical relativity (NR) simulations.

\section{Forced motion, circular geodesics, and field equations}\label{sec:forcedEOM}

In this section, we set up our conventions and formulate the secondary's motion and the Einstein field equations in the phase-space approach.

\subsection{Setup and conventions}

We consider a nonspinning compact object of mass $m_p$ (where $p$ stands for particle since we treat it as such) orbiting a Kerr black hole of mass $\mathring M$ and dimensionless spin $\mathring\chi$ on quasicircular, equatorial orbits, where we use a ring to denote constant, initial values of evolving parameters. We will refer to the two objects as the ``secondary'' and the ``primary'', respectively. We define the small mass ratio $\mathring\varepsilon\coloneqq m_p/\mathring M$ and the symmetric mass ratio $\check\nu\coloneqq m_p \mathring M/\mathring M_\text{tot}^2$, where $\mathring M_\text{tot}\coloneqq \mathring M + m_p$ is the total reference mass of the binary. It will also be useful to introduce the large mass ratio $\mathring q\coloneqq1/\mathring\varepsilon$. Throughout this paper, we adopt geometric units where the gravitational constant $G$ and the speed of light $c$ are set to unity.

Due to the presence of the perturbing secondary, the full spacetime metric $\mathbf{g}_{\mu\nu}$ decomposes into a background $g_{\mu\nu}$ and a perturbation $h_{\mu\nu}$,
\begin{equation}
    \mathbf{g}_{\mu\nu} = g_{\mu\nu} + h_{\mu\nu}.\label{metricfull}
\end{equation}
The background metric $g_{\mu\nu}$ is taken to be the metric of the primary Kerr black hole in isolation. We work in dimensionless Boyer-Lindquist (BL) coordinates that we denote as $(t, x^i)=(t, r, \theta, \phi)$, normalized by the primary mass $\mathring M$. Therefore, the standard dimensionful BL coordinates are given by $(\mathring M t, \mathring M r, \theta, \phi)$. In dimensionless BL coordinates, the line element of the Kerr metric is given by 
\begin{align}
    {\mathring M}^{-2}ds^2 = -&\left(1 - \frac{2 r}{\rho^2}\right)dt^2 - 2\frac{2  \mathring\chi r\sin^2 \theta}{\rho^2} dt d\phi\nonumber\\
    &+\frac{\Sigma}{\rho^2}\sin^2 \theta d\phi^2 + \frac{\rho^2}{\Delta} dr^2 + \rho^2 d\theta^2,
\end{align}
where
\begin{subequations}
\begin{align}
    \rho^2 &\coloneqq r^2 + \mathring\chi^2 \cos^2 \theta,\\
    \Delta &\coloneqq r^2 - 2 r + \mathring\chi^2,\\
    \Sigma &\coloneqq \big(r^2 + \mathring\chi^2\big)^2 - \mathring\chi^2 \Delta \sin^2 \theta.
\end{align}
\end{subequations}
The perturbation $h_{\mu\nu}\sim\mathring\varepsilon$ contains all corrections due to the orbiting secondary and nonlinear effects due to the interaction of the two bodies. This includes the time-dependent changes $\delta M$ and $\delta \chi$ of, respectively, the mass and spin of the primary black hole, which evolve due to the absorption of energy and angular-momentum fluxes.

The Einstein equations can be perturbatively expanded in $h_{\mu\nu}$ as
\begin{align}\label{eq:EFE self-consistent}
    \delta G_{\mu\nu}[h] + \delta^2G_{\mu\nu}[h,h] + {\cal O}(\mathring\varepsilon^3) = 8\pi T_{\mu\nu},
\end{align}
where $\delta G_{\mu\nu}$ is the linearized Einstein tensor on the background $g_{\mu\nu}$, while $\delta^2G_{\mu\nu}$ is quadratic in $h_{\mu\nu}$. The right-hand side is given by the point-particle Detweiler stress-energy tensor~\cite{Detweiler:2011tt,Upton:2021oxf},
\begin{equation}\label{eq:TDet}
    T_{\mu\nu} = m_p \int \tilde u_\mu \tilde u_\nu \frac{\delta^4(x^\alpha-z^\alpha(\tilde\tau))}{\sqrt{-\tilde g}}d\tilde\tau,
\end{equation}
where $z^\mu(\tilde\tau)$ is the particle's worldline. Here we have split the metric perturbation as $h_{\mu\nu} = h^{\cal P}_{\mu\nu} + h^{\mathcal R}_{\mu\nu}$, where $h^{\cal P}_{\mu\nu}$ is an analytically known ``puncture'', which is singular at the particle's position, and $h^{\mathcal R}_{\mu\nu}$ is the regular residual field, as defined in Ref.~\cite{Pound:2014xva}. We can then define an effective metric, $\tilde g_{\mu\nu}=g_{\mu\nu}+h^{\mathcal R}_{\mu\nu}$, which is regular at the particle's position. The parameter $\tilde\tau$ is proper time in the effective metric, and the particle's 4-velocity is $\tilde u^\mu \coloneqq dz^\mu/d\tilde\tau$ (with $\tilde u_\mu \coloneqq\tilde g_{\mu\nu}\tilde u^\nu$).

In the effective metric $\tilde{g}_{\mu\nu}$, the secondary object follows a geodesic~\cite{Detweiler:2000gt,Pound_2012,Pound:2017psq}. In terms of the background metric $g_{\mu\nu}$, the particle obeys the self-forced equations of motion
\begin{equation}\label{eom tau}
    \frac{D^2z^\mu}{d\tau^2} = f^\mu.
\end{equation}
Here $\tau$ is proper time in $g_{\mu\nu}$ (normalized with respect to the primary mass $\mathring M$) and $D/d\tau \coloneqq u^\mu\nabla_\mu$, with the 4-velocity $u^\mu\coloneqq dz^\mu/d\tau $ and covariant derivative $\nabla_\mu$ compatible with $g_{\alpha\beta}$. The (gravitational) self-force per unit mass of the secondary, $f^\mu$, can be expressed in terms of the residual metric perturbations as in Refs.~\cite{Pound:2012nt,Pound_2015,Pound:2017psq}.

In order to describe the evolution of the binary we foliate the spacetime with surfaces of constant hyperboloidal time $s$ that extend from the primary's future horizon to future null infinity, following Refs.~\cite{Miller:2020bft,Miller:2023ers}. We assume that $s$ reduces to BL coordinate time $t$ at the particle, which we also use to parametrize the worldline $z^\mu=z^\mu(t,\mathring\varepsilon)$:
\begin{equation}\label{eq:worldline}
    z^\mu(t,\mathring\varepsilon) = \big(t, x_p^i(t,\mathring\varepsilon)\big) = \big(t, r_p(t,\mathring\varepsilon), \pi/2, \phi_p(t, \mathring\varepsilon)\big).
\end{equation}
A subscript $p$ indicates  evaluation at the particle's position, and we have restricted to equatorial orbits ($\theta_p=\pi/2$). Without loss of generality, we take the the orbital angular momentum to be in the $+z$ direction. The dimensionless spin of the primary is either anti-aligned with the $z$-axis for $\mathring\chi\in[-1,0]$ or aligned with the $z$-axis for $\mathring\chi\in[0,1]$. As a consequence, the orbital phase $\phi_p$ is a monotonically increasing function of $t$, while negative and positive spin $\mathring\chi$ corresponds to retrograde and prograde orbits, respectively. This choice of orientation can always be achieved without ambiguity during the inspiral and transition.\footnote{This is no longer necessarily the case during the plunge: for a particle plunging from a retrograde ISCO, the orbital frequency $\Omega(t)\coloneqq d\phi_p/dt$ changes sign before reaching the ergosphere~\cite{PhysRevLett.129.161101}, and the orbital phase is no longer monotonic. The orientation of the coordinate system is then inherited from the inspiral and transition.}

The binary's mechanical phase space is spanned by $(x_p^i, \dot x_p^i, \delta M, \delta \chi)$, where we have used an index $i$ for spatial coordinates and an overdot to denote $d/dt$. For quasicircular binaries, the orbital radius is not independent and can be expressed in terms of other phase-space coordinates. More precisely, the set of mechanical variables describing the binary can be reduced to the azimuthal orbital phase $\phi_p$ and the set of evolving parameters $(\Delta)J^a=\{(\Delta)\Omega, \delta M, \delta \chi\}$. Here we have used the notation $(\Delta)\Omega$ from Ref.~\cite{Kuchler:2024esj} to denote the orbital frequency $\Omega\coloneqq d\phi_p/dt$ during the inspiral or $\Delta\Omega\coloneqq(\Omega-\Omega_\star)/\mathring\varepsilon^{2/5}$ during the transition to plunge, where $\Omega_\star=\Omega_\star(\mathring\chi)$ is the ISCO frequency around a Kerr black hole with spin $\mathring\chi$. As introduced above, $\delta M$ and $\delta\chi$ refer to the evolving corrections to the primary mass $\mathring M$ and dimensionless spin $\mathring\chi$, respectively. They are related to the dynamical mass $M(s)$ and spin $\chi(s)$ of the primary black hole as
\begin{subequations}
\begin{align}
    M(s) &= \mathring M\left(1 + \mathring\varepsilon \, \delta M(s)\right),\\
    \chi(s) &= \mathring\chi\left(1 + \mathring\varepsilon \, \delta  \chi(s)\right).
\end{align}
\end{subequations}
The mechanical variables are defined as functions of hyperboloidal time: on a slice of constant $s$, $(\Delta)\Omega$ is equal to its value at the point where the slice intersects the worldline (this extends to any function defined on the worldline, such as $\phi_p$), while $\delta M$ and $\delta \chi$ are equal to their values where the slice intersects the horizon of the primary. The rescaled variable $\Delta\Omega$ is chosen such that it is order unity during the transition across the ISCO, which occurs over a frequency window $(\Omega-\Omega_\star)=\mathcal{O}(\mathring\varepsilon^{2/5})$.

The crucial feature of our formulation of the problem is that the metric perturbations depend on $s$ only through $\phi_p$ and $(\Delta)J^a$. In addition, since $\phi_p$ is a $2\pi$-periodic variable, we can express $h_{\mu\nu}$ in terms of Fourier series as
\begin{equation}\label{hmunu Fourier}
    h_{\mu\nu}(s, x^i, \mathring\varepsilon) =\! \sum_{m=-\infty}^{+\infty} \! h_{\mu\nu}^m((\Delta)J^a(s, \mathring\varepsilon), x^i, \mathring\varepsilon) e^{-im\phi_p(s, \mathring\varepsilon)}.
\end{equation}
The Fourier coefficients $h_{\mu\nu}^m$ will be expanded in series for small $\mathring\varepsilon$, though the form of that expansion depends on the stage of the binaries evolution, see Refs.~\cite{Kuchler:2024esj,Kuchler:2025hwx}. The form~\eqref{hmunu Fourier} can be derived from the more general self-consistent formulation~\cite{Pound:2009sm} of GSF theory following the arguments in Ref.~\cite{Lewis:2025ydo}.

The system's evolution through phase space is then governed by the following set of ordinary differential equations:
\begin{align}
    \frac{d\phi_p}{ds} &= \Omega,
    \\\label{eq:FJa}
    \frac{d(\Delta)J^a}{ds} &= F^{(\Delta)J^a}((\Delta)J^b; \mathring\varepsilon,\mathring\chi).
\end{align}
Here and below, we will often leave the parametric dependencies implicit and write $F^{(\Delta) J^a}\left((\Delta) J^b;\mathring\varepsilon, {\mathring\chi}\right)$ as $ F^{(\Delta) J^a}\left((\Delta) J^b\right)$, or simply $F^{(\Delta) J^a}$. The forcing functions $F^{(\Delta)\Omega}$ can be expressed in terms of the self-force through the equations of motion~\eqref{eom tau}, while $F^{\delta M}$ and $F^{\delta\chi}$ are determined from the horizon fluxes of energy and angular momentum. We note that the forcing functions do not depend on $\phi_p$ due to the axisymmetry of the problem: the metric perturbations~\eqref{hmunu Fourier} depend on $\phi_p$ only in the combination $\phi-\phi_p$, becoming independent of $\phi_p$ when evaluated along the worldline. As a consequence, the self-force and other dynamical quantities derived from it are also independent of $\phi_p$~\cite{Pound:2021qin,Lewis:2025ydo}.

Our basic approach, as employed in Refs.~\cite{Kuchler:2023jbu,Kuchler:2025hwx}, is the following: 
\begin{enumerate}
    \item In each regime (the inspiral or the transition), we expand the forcing functions $F^{(\Delta)J^a}$ and the metric amplitudes $h^m_{\mu\nu}$ in a series in small $\mathring\varepsilon$ at fixed phase-space coordinates $(\phi_p,(\Delta) J^a)$.
    \item From the EFE and the self-forced equations of motion, we obtain differential equations for the coefficients in each of these expansions.
    \item We impose retarded boundary conditions at the horizon and infinity. 
    \item We fix the remaining freedom in the  solution in each regime through an asymptotic matching condition: the expansions of $F^{(\Delta)J^a}$ and  $h^m_{\mu\nu}$ in the transition must agree with those in the inspiral when both are re-expanded in a common form. Concretely, suppose a function $f$ is expanded for small $\mathring \varepsilon$ at fixed $\Delta J^a$ in the transition and then re-expanded at ``early times'' in the transition  ($\Delta\Omega\ll-\Omega_\star$, equivalent to $\mathring\varepsilon\ll 1$ at fixed $\Omega<\Omega_\star$). This must agree, order by order in $\mathring\varepsilon$ and $\Delta\Omega$, with the result of first expanding $f$ for small $\mathring\varepsilon$ at fixed $\Omega$, then re-expanding near the ISCO ($|\Omega-\Omega_\star|\ll \Omega_\star$), and finally rewriting in terms of~$\Delta\Omega$.   
\end{enumerate}
Another way of stating the matching condition is simply that the two expansions must commute.

\subsection{Orbital motion}\label{subsec:nonperturbEOM}

The trajectory of the secondary~\eqref{eq:worldline} can be written in terms of the mechanical variables $(\Delta)J^a$ as
\begin{equation}\label{eq:worldline Ja}
    z^\mu(t, \mathring\varepsilon) = \big(t, r_p\left((\Delta)J^a(t),\mathring\varepsilon\right), \pi/2, \phi_p(t, \mathring\varepsilon)\big).
\end{equation}
Time derivatives are then evaluated using the chain rule
\begin{equation}\label{eq:chainrule}
    \frac{d}{d t}= \Omega\,\partial_{ \phi_p}+F^{(\Delta)  J^a}\partial_{(\Delta)  J^a}
\end{equation}
and Eq.~\eqref{eq:FJa}. To streamline the resulting expressions, we adopt the notation of Ref.~\cite{Miller:2023ers} for the directional parametric derivative:
\begin{equation}
    \paramD \coloneqq F^{(\Delta)  J^a}\partial_{(\Delta)  J^a}\coloneqq\mathcal{V}^a\partial_{(\Delta)J^a},
\end{equation}
where $\mathcal{V}^a\coloneqq F^{(\Delta)J^a}$ represents the velocity through parameter space. The tangent vector to the trajectory in the non-affine $t$ parametrization is hence given by
\begin{equation}\label{eq:dz/dt}
   \frac{d z^\mu}{dt} \!=\! \left(1, \paramD r_p, 
   0, \Omega\right).
\end{equation}

Written in terms of the non-affine parameter $t$, the self-forced equations of motion~\eqref{eom tau} become
\begin{equation}\label{eq:eomNP}
    \frac{d^2 z^\mu}{d t^2} + U^{-1}\frac{d U}{d t}\frac{d z^\mu}{d t} + \Gamma^\mu_{\alpha\beta}\frac{d z^\alpha}{d t}\frac{d z^\beta}{d t} = U^{-2} f^\mu
\end{equation}
with $\Gamma^\mu_{\alpha\beta}$ the Christoffel symbols of the background Kerr metric $g_{\mu\nu}$, evaluated on the worldline~\eqref{eq:worldline Ja}. Here the redshift factor $U$ is defined as
\begin{equation}\label{eq:redshiftNP}
     U\coloneqq\frac{dt}{d\tau}=\frac{1}{\sqrt{-g_{\mu\nu} (d z^\mu/d t)(dz^\nu/d t)}}
\end{equation}
from the normalization condition for timelike curves, $u_\mu u^\mu=-1$.

Plugging Eqs.~\eqref{eq:worldline Ja} and \eqref{eq:dz/dt} into Eqs.~\eqref{eq:redshiftNP} and \eqref{eq:eomNP} yields
\begin{subequations}\label{eq:eom}
\begin{align}
    U^{-2} =&\; \mathcal{R}^{-2}(r_p, \Omega) - \frac{r_p^2}{\Delta_p} (\paramD r_p)^2,
    \label{Reom}
    \\
    U^{-2} f^t =&\; U^{-1} \paramD U + \frac{\mathcal{C}\left(r_p, \Omega\right)}{r_p^2 \Delta_p} \paramD r_p,\label{teom}
    \\
    U^{-2} f^r=&\; \paramD^2 r_p + U^{-1} \paramD U \, \paramD r_p + \frac{\Delta_p}{r_p^4} \mathcal{G}(r_p, \Omega) \nonumber\\
    & + \frac{\mathcal{E}(r_p)}{r_p \Delta_p} (\paramD r_p)^2,\label{reom}
\end{align}
\end{subequations}
where we have defined $\Delta_p\coloneqq \Delta(r_p)$,
\begin{subequations}\label{eq:fctdef}
\begin{align}
    \mathcal{C}(r, \Omega; \mathring\chi) &\coloneqq 2(\mathring\chi^2 + r^2 - \mathring\chi^3 \Omega - 3 \mathring\chi r^2 \Omega)\label{Cdef},
    \\
    \mathcal{G}(r, \Omega; \mathring\chi) &\coloneqq 1 - 2 \mathring\chi \Omega + \mathring\chi^2 \Omega^2 - r^3 \Omega^2\label{Gdef},
    \\
    \mathcal{R}(r, \Omega; \mathring\chi) &\coloneqq \sqrt{\frac{r}{r - \mathring\chi^2 r \Omega^2 - r^3 \Omega^2 - 2(1 - \mathring\chi \Omega)^2}} \label{Rdef},
    \\
    \mathcal{E}(r; \mathring\chi) &\coloneqq \mathring\chi^2 - r.
\end{align}
\end{subequations}
In the following, we will often drop the parametric dependence in the functions~\eqref{eq:fctdef} and write, for example, $\mathcal{C}(r, \Omega; \mathring\chi)$ as $\mathcal{C}(r, \Omega)$.

\subsection{Study of circular geodesics}\label{subsec:geodesicstudy}

For later use in formulating the multiscale expansion of the motion and field equations, let us first turn off the self-force contributions to the equations of motion~\eqref{eq:eom} and briefly analyze the circular geodesic solutions. 

First note that with $f^\mu=0$, the $1$-parameter family of circular geodesics corresponds to the level set $\mathcal{G}=0$. Parameterizing that curve by the orbital frequency $\Omega$ gives
\begin{equation}\label{Ggeo}
     r_{(0)}(\Omega) = (1 - \mathring\chi\,\Omega)^{2/3}/\Omega^{2/3},
\end{equation}
and one can indeed check that ${\mathcal{G}}(r_{(0)}(\Omega), \Omega)=0$.

Second, one can readily interpret ${\mathcal{R}}( r, \Omega)$ as the redshift factor of a particle following a (possibly non-geodesic) circular motion of radius $r$ and frequency $\Omega$ on the equatorial plane of the Kerr primary. Setting again $f^\mu=0$ and evaluating $\mathcal{R}$ on the level set $\mathcal{G}=0$, we find $\mathcal{R}$ reduces to the redshift factor $U_{(0)}(\Omega)$ of a particle following a circular geodesic of orbital frequency $\Omega$,
 \begin{equation}\label{U0def}
     \mathcal{R}(r_{(0)}(\Omega), \Omega) = U_{(0)}(\Omega),
 \end{equation}
where
\begin{equation}\label{U0PA}
    U_{(0)}^{-2}(\Omega) = (1 - \mathring\chi \Omega) \left[1 + \mathring\chi \Omega - 3\Omega^{2/3} ({1 - \mathring\chi \Omega})^{1/3}\right]. 
\end{equation}

Moreover, the directional derivative of ${\mathcal{R}}$ along $r$ is proportional to $\mathcal{G}$,
\begin{equation}
    \partial_r\mathcal{R}(r, \Omega) = -\frac{\mathcal{G}(r, \Omega)\mathcal{R}^3(r, \Omega)}{r^2},
\end{equation}
and therefore vanishes when evaluated on a circular geodesic. When evaluated on the level set $\mathcal{G}=0$, the normal of $\mathcal{R}$ is directed along $d\Omega$ 
\begin{equation}\label{dRedshift}
    \frac{d\mathcal{R}(r_{(0)}(\Omega), \Omega)}{d\Omega} = \Omega U_{(0)}^3(\Omega)B(\Omega),
\end{equation}
while from Eq.~\eqref{Cdef}, we have
\begin{equation}\label{Cgeo}
    \mathcal{C}(r_{(0)}(\Omega), \Omega) = 2 \Omega r_{(0)}^{3/2} B(\Omega).
\end{equation}
In Eqs.~\eqref{dRedshift} and \eqref{Cgeo}, the function $ B( \Omega)$ is given as in Ref.~\cite{Compere:2021zfj} by
\begin{align}
    B(\Omega) &\coloneqq B(r_{(0)}(\Omega)) \nonumber\\
    &= r_{(0)}^2 - 2\mathring\chi r_{(0)}^{1/2} + \mathring\chi^2\nonumber\\
    &=\mathring\chi^2 - 3 \mathring\chi (\Omega^{-1} - \mathring\chi)^{1/3} + \Omega^{-1}(\Omega^{-1} - \mathring\chi)^{1/3}.\label{eq:B}
\end{align}

Finally, for any $n \geq 0$ 
\begin{subequations}
\begin{align}\label{dGgeo}
    \frac{d^n{\mathcal{G}}}{d \Omega^n}(r_{(0)}( \Omega), \Omega) &= 0,\\
    \frac{d^n{\mathcal{R}}}{d \Omega^n}(r_{(0)}( \Omega), \Omega) &= \frac{d^n  U_{(0)}}{d \Omega^n},
\end{align}
\end{subequations}
from which we can extract, for example,
\begin{equation}\label{dr0geo}
    \frac{d r_{(0)}}{d \Omega} = -\frac{\partial_\Omega \mathcal{G}}{\partial_r \mathcal{G}} = -\frac{2}{3}\frac{1}{r^{1/2}_{(0)}(\Omega)\Omega^2}.
\end{equation}

\subsection{Einstein field equations}

As anticipated below Eq.~\eqref{hmunu Fourier}, the metric perturbations can be expanded for small $\mathring\varepsilon$: during the inspiral, the expansion reads~\cite{Miller:2020bft}
\begin{equation}\label{eq:h inspiral}
    h_{\mu\nu} = \sum_{n\geq1}\mathring\varepsilon^n\sum_{m=-\infty}^\infty h^{(n),m}_{\mu\nu}(J^a, x^i)e^{-im\phi_p},
\end{equation}
where an index $(n)$ indicates the coefficient of $\mathring\varepsilon^n$, while in the transition regime we have~\cite{Kuchler:2024esj}
\begin{equation}\label{eq:h transition}
    h_{\mu\nu} = \sum_{n\geq5}\mathring\varepsilon^{n/5}\sum_{m=-\infty}^\infty h^{[n],m}_{\mu\nu}(\Delta J^a, x^i)e^{-im\phi_p},
\end{equation}
where an index $[n]$ labels the coefficient of $\mathring\varepsilon^{n/5}$.

In Ref.~\cite{Kuchler:2024esj} the structure of the Schwarzschild metric perturbations was derived from the field equations as follows: during the inspiral, the metric perturbations can be expressed as a sum of terms factored into a piece that is smooth and a piece that is singular at the ISCO frequency. In particular, the first-order metric perturbation $h_{\mu\nu}^{(1)}$ was shown to be a smooth function of the orbital frequency. Similarly, in the transition regime, the metric perturbations at each perturbative order can be written as a sum of terms factored into a $\Delta\Omega$-dependent and a $\Delta\Omega$-independent piece. At the orders we will be interested in this paper, it was found that $h_{\mu\nu}^{[5]}$ does not depend on $\Delta\Omega$, while $h_{\mu\nu}^{[6]}(\Delta J^a, x^i)=0$ and $h_{\mu\nu}^{[7]}(\Delta J^a, x^i)=\Delta\Omega\,h_{\mu\nu}^{[7]A}(x^i)$. These structures are dictated by the field equations, in particular by the order of the highest-order time derivative. It follows that the structure of the metric perturbations remains unaltered when considering a Kerr background.

In Ref.~\cite{Kuchler:2024esj} it was also shown that the metric perturbations to next-to-leading order in the transition expansion can be obtained from the inspiral's first-order metric perturbation as
\begin{subequations}
\begin{align}
    h_{\mu\nu}^{[5]}(\delta M, \delta\chi, x^i) &= h_{\mu\nu}^{(1)}(\Omega=\Omega_\star, \delta M, \delta\chi, x^i),
    \\
    h_{\mu\nu}^{[7]A}(x^i) &= \partial_\Omega h_{\mu\nu}^{(1)}(\Omega=\Omega_\star, x^i).
\end{align}
\end{subequations}
When working at leading order in the inspiral and next-to-next-to-leading order in the transition expansions, it is therefore sufficient to solve for $h_{\mu\nu}^{(1)}$, rather than directly solving field equations in the transition regime. 

In addition, $\delta M$ and $\delta\chi$ only enter $h^{(1)}_{\mu\nu}$ in non-radiative modes~\cite{Miller:2020bft}, meaning we can neglect them at our orders of interest in this paper. We can then obtain the waveform's first-order mode amplitudes by solving the Teukolsky equations for the fourth Weyl scalar $\Psi_4$ (for a review, see Ref.~\cite{Pound:2021qin}). At zeroth order, $\mathcal{O}(\mathring\varepsilon^0)$, the background Weyl scalar vanishes, and we can expand $\Psi_4$ in powers of the mass ratio as
\begin{equation}
\zeta^4\Psi_4=\mathring\varepsilon\psi^{(1)}_4+ \mathring\varepsilon^2\psi^{(2)}_4+\mathcal{O}(\mathring\varepsilon^3),
\end{equation}
with $\zeta=r-i\mathring\chi\cos\theta$. The linearized scalar $\psi^{(1)}_4$ satisfies the Teukolsky equation. On phase space, time derivatives are substituted using the chain rule~\eqref{eq:chainrule}. At first order in $\mathring\varepsilon$ and with the ansatz $\psi_4^{(1)}\sim e^{-im\phi_p}$, this reduces to $\partial_t\rightarrow-im\Omega$. The Teukolsky equation therefore decouples into an angular part and a radial part in the usual way~\cite{Teukolsky:1974yv}. Therefore, we can decompose the linearized Weyl scalar $\psi_4^{(1)}$ in the basis of spin-weighted spheroidal harmonics as
\begin{equation}\label{eq:psi4decomposition}
    \psi_4^{(1)}=\sum_{\ell =2}^\infty\sum_{m=-\ell}^{\ell} \psi^{(1)}_{\ell m}(\Omega, r)\prescript{}{-2}{S}_{\ell m}(\theta,\phi)e^{-im\phi_p},
\end{equation}
with $\prescript{}{-2}{S}_{\ell m}$ satisfying the spin$-2$-weighted spheroidal harmonic equation and $\psi^{(1)}_{\ell m}$ the sourced  Teukolsky radial equation depending upon $\Omega$. It is readily solved using the Black Hole Perturbation Toolkit's \texttt{Teukolsky} package~\cite{TeukolskyPackage}, for example, through the method of variation of constants: two linearly independent solutions to the homogeneous radial Teukolsky equation $\prescript{}{-2}R^{\rm in}_{\ell m}( r)$ and $\prescript{}{-2}R^{\rm up}_{\ell m}( r)$ are found, whose asymptotic behaviour are given in Ref.~\cite{Pound:2021qin}, and the solution is
\begin{align}
    \psi^{(1)}_{\ell m}(\Omega,r) &= \prescript{}{-2}{C}^{\rm in}_{\ell m}(r;\Omega)\prescript{}{-2}R^{\rm in}_{\ell m}(r) \nonumber\\
    &+\prescript{}{-2}{C}^{\rm up}_{\ell m}(r;\Omega)\prescript{}{-2}R^{\rm up}_{\ell m}(r).\label{eq:varconstant}
\end{align}

Let us define the asymptotic tetrad leg $\bar m^\alpha=\frac{1}{\sqrt{2}\rho}(0,0,1,-i\, \text{csc}\, \theta)$ which matches with both the Kinnersley and the Carter tetrad up to  irrelevant subleading terms in the radial component~\cite{Pound:2021qin}. The projected metric perturbation $h_{\bar m\bar m}=\bar{m}^\mu\bar{m}^\nu h_{\mu\nu}$ is then related to $\Psi_4$ as 
\begin{equation}\label{eq:psitoh}
    \zeta^4\Psi_4=-\frac{1}{2}\partial_t^2h_{\bar m\bar m},
\end{equation}
while the wavestrain reads
\begin{equation}
    h \coloneqq   \lim_{r\rightarrow \infty}r(h_+-ih_\times) =\lim_{r\rightarrow \infty}r\, h_{\bar m\bar m}.
\end{equation}
Decomposing $\, h$ into spin-weighted spheroidal harmonics, as in
\begin{align}
    h = \sum_{\ell =2}^\infty \sum_{m=-\ell}^\ell  \, h_{\ell m}(\phi_p,(\Delta) J^a)\,  \mbox{}_{-2}S_{\ell m}(\theta,\phi),
\end{align}
with
\begin{equation}
    h_{\ell m}(\phi_p,(\Delta) J^a)=H_{\ell m}((\Delta) J^a)e^{-im\phi_p},
\end{equation}
and using the relation~\eqref{eq:psitoh}, one can readily express the $(\ell,m)$-mode amplitudes $H_{\ell m}$ in terms of the Teukolsky amplitudes as
\begin{equation}\label{eq:HlmTeukMode}
    H_{\ell m}(\Omega)=\mathring\varepsilon\frac{2}{(m \Omega)^2} \lim_{r\rightarrow\infty}\prescript{}{-2}{C}^{\rm up}_{\ell m}(r; \Omega)+\mathcal{O}(\mathring\varepsilon^{8/5}).
\end{equation}
Here the omitted terms are $\mathcal{O}(\mathring\varepsilon^{8/5})$ in the transition (corresponding to $h^{[8],m}_{\mu\nu}$) but reduce to  $\mathcal{O}(\mathring\varepsilon^2)$ in the inspiral (corresponding to $h^{(2),m}_{\mu\nu}$).

\section{The inspiral's post-adiabatic expansion}\label{sec:PAexpansion}

In this section we derive the post-adiabatic expansion of the quasi-circular inspiral, extending the results in Appendix~A of Ref.~\cite{Miller:2020bft} (see also the summary in Ref.~\cite{Miller:2023ers}) to the case of a Kerr primary.

\subsection{Perturbative expansion}

The inspiral motion is driven by the loss of energy and angular momentum through the emission of GWs, which occurs on the radiation-reaction timescale $1/\mathring\varepsilon$. The mechanical variables describing the inspiral are the orbital phase $\phi_p$, which varies on a ``fast'' orbital timescale ${\cal O}(1)$, and the parameters $J^a=\{\Omega, \delta M, \delta\chi\}$, which evolve on the radiation-reaction timescale.

The inspiral's post-adiabatic expansion consist in writing all quantities of interest as functions of $(\phi_p, J^a, \varepsilon)$, which are then expanded in integer powers of $\mathring\varepsilon$ while keeping the mechanical variables $(\phi_p, J^a)$ fixed. For the mode amplitudes $H_{\ell m}$, we obtain
\begin{equation}\label{eq:HlmPA}
    H_{\ell m}(J^a; \mathring\varepsilon) = \sum_{n=1}^\infty {\mathring\varepsilon}^n H^{(n)}_{\ell m}(J^a).
\end{equation}
We refer to $H^{(n)}_{\ell m}$ as the $n$th-order mode amplitudes. The first-order mode amplitudes, $H^{(1)}_{\ell m}$, are obtained from the Teukolsky formalism as described in the previous section. Similarly, we expand the orbital quantities such as the orbital radius $r_p$, the redshift factor $U$, the forcing functions $F^{J^a}$ and the self-force $f^\mu$ as power series in $\mathring\varepsilon$:
\begin{subequations}\label{PAExpansionOrbital}
\begin{align}
    r_p(J^a; \mathring\varepsilon) &= \sum_{n=0}^\infty \mathring\varepsilon^n r_{(n)}(J^a),
    \\
    U(J^a; \mathring\varepsilon) &=\sum_{n=0}^\infty \mathring\varepsilon^n U_{(n)}(J^a),
    \\\label{FOmega}
    F^{\Omega}(J^a; \mathring\varepsilon) &= \mathring\varepsilon \sum_{n=0}^\infty \mathring\varepsilon^n F_{(n)}^{\Omega}(J^a),
    \\
    F^{\delta M}(J^a; \mathring\varepsilon) &= \sum_{n=1}^\infty \mathring\varepsilon^n F_{(n)}^{\delta M}(J^a),
    \\
    F^{\delta \chi}(J^a; \mathring\varepsilon) &= \sum_{n=1}^\infty \mathring\varepsilon^n F_{(n)}^{\delta \chi}(J^a),
    \\
    f^\mu(J^a; \mathring\varepsilon) &= \sum_{n=1}^\infty{\mathring\varepsilon}^n f^\mu_{(n)}(J^a).
\end{align}
\end{subequations}
The term $F_{(n)}^{J^a}$ is referred to as the $n$th post-adiabatic (or $n$PA) forcing function of $J^a$, following the nomenclature first introduced in Ref.~\cite{Hinderer:2008dm} for the expansion of the orbital phase on the radiation-reaction timescale.

Using the expansions above, we perform the post-adiabatic expansion of Eq.~\eqref{eq:eom}. The corrections $U_{(n)}$ and $r_{(n)}$ are obtained at order $\mathcal{O}(\mathring\varepsilon^n)$ from Eqs.~\eqref{Reom} and~\eqref{reom}, respectively. The forcing function $F^\Omega_{(n)}$ is instead obtained from Eq.~\eqref{teom} at order $\mathcal{O}(\mathring\varepsilon^{n+1})$. At adiabatic order, we get
\begin{subequations}\label{eq:0PA}
\begin{align}
    r_{(0)}(\Omega) &= \frac{(1 - \mathring\chi \Omega)^{2/3}}{\Omega^{2/3}},
    \\
    U^{-2}_{(0)}(\Omega) &= (1 - \mathring\chi \Omega) \! \left[1 \!+ \mathring\chi \Omega - 3 \Omega^{2/3} ({1 \!- \mathring\chi \Omega})^{1/3}\!\right]\!,\label{U0}
    \\
    F_{(0)}^{\Omega}(\Omega) &= -\frac{3 \Delta_{(0)}(\Omega)}{\Omega U_{(0)}^4(\Omega) D(\Omega) B(\Omega)} f_{(1)}^t(\Omega),\label{F01}
\end{align}
\end{subequations}
where the function $D(\Omega)$ is defined as in Ref.~\cite{Compere:2021zfj} to be
\begin{align}
    D(\Omega) \coloneqq&\; D(r_{(0)}(\Omega)) \nonumber\\
    =&\; r_{(0)}^2 - 6 r_{(0)} + 8 \mathring\chi r_{(0)}^{1/2} - 3\mathring\chi^2 \nonumber
    \\
    =&\; -3\mathring\chi^2 - 6(\Omega^{-1} -\mathring\chi)^{2/3}\nonumber
    \\
    &\;+ 7\mathring\chi (\Omega^{-1} - \mathring\chi)^{1/3} + \Omega^{-1}(\Omega^{-1} - \mathring\chi)^{1/3},\label{eq:D}
\end{align}
and the function $\Delta_{(0)}(\Omega)$ is to be understood as $\Delta_{(0)}(\Omega) \coloneqq \Delta(r_{(0)}(\Omega))$. We note that the expressions for $r_{(0)}(\Omega)$ and $U_{(0)}(\Omega)$ are simply the geodesic relationships obtained in Sec.~\ref{subsec:geodesicstudy}.

The first-order dissipative self-force $f_{(1)}^t$ can be expressed in terms of the Teukolsky fluxes through the energy flux balance law. To do so, we first write the specific orbital energy as $E = E_{(0)} + \mathring\varepsilon \, E_{(1)} + \mathcal{O}(\mathring\varepsilon^2) $ with $E_{(0)}$ the adiabatic energy defined from circular geodesics. We can then relate the first-order dissipative self-forces to the rate of change of $E_{(0)}$ using $d E/d \tau = \nabla_{u} (-u_t)$ at leading order:
\begin{equation}\label{eq:dE0dt1}
    \frac{1}{\mathring\varepsilon} \frac{d E_{(0)}}{dt} =- U_{(0)}^{-1}\left(g_{tt} f_{(1)}^t + g_{t\phi}f_{(1)}^\phi\right). 
\end{equation}
Secondly, solving the orthogonality condition $f^\mu u_\mu=0$ at leading order, yields
\begin{equation}\label{eq:orthocond}
    f_{(1)}^\phi = -\frac{g_{tt} + g_{t\phi} \Omega}{g_{t\phi} + g_{\phi\phi} \Omega} f_{(1)}^t.
\end{equation}
Finally, the energy flux balance law reads
\begin{equation}\label{eq:fluxbalancelaw}
    \frac{1}{\mathring{\varepsilon}}\frac{dE_{(0)}}{dt} = -\left(\mathcal{F}^\infty_{(0)} + \mathcal{F}^\mathcal{H}_{(0)}\right),
\end{equation}
where ${\mathcal{F}}^\infty={\mathring\varepsilon}^2{\mathcal{F}}^\infty_{(0)}+\mathcal O({\mathring\varepsilon}^3)$ and ${\mathcal{F}}^\mathcal{H}={\mathring\varepsilon}^2{\mathcal{F}}^\mathcal{H}_{(0)}+\mathcal O({\mathring\varepsilon}^3)$ are the energy fluxes to null infinity and down the horizon of the primary, respectively. The leading-order quantities $\mathcal{F}^\infty_{(0)}$ and $\mathcal{F}^\mathcal{H}_{(0)}$ are the 0PA Teukolsky fluxes, which we obtain from the BHPToolkit's \texttt{Teukolsky} package~\cite{TeukolskyPackage}. Combining Eqs.~\eqref{eq:dE0dt1}, \eqref{eq:orthocond}, and \eqref{eq:fluxbalancelaw} allows us to write the adiabatic forcing function of the orbital frequency~\eqref{F01} as
\begin{equation}\label{FOmega0PA}
    F_{(0)}^{\Omega}(\Omega) = \frac{3}{\Omega U_{(0)}^3(\Omega) D(\Omega)}\left(\mathcal{F}^\infty_{(0)}(\Omega) + \mathcal{F}^\mathcal{H}_{(0)}(\Omega)\right).
\end{equation}

\begin{figure*}
\includegraphics[width=.47\textwidth]{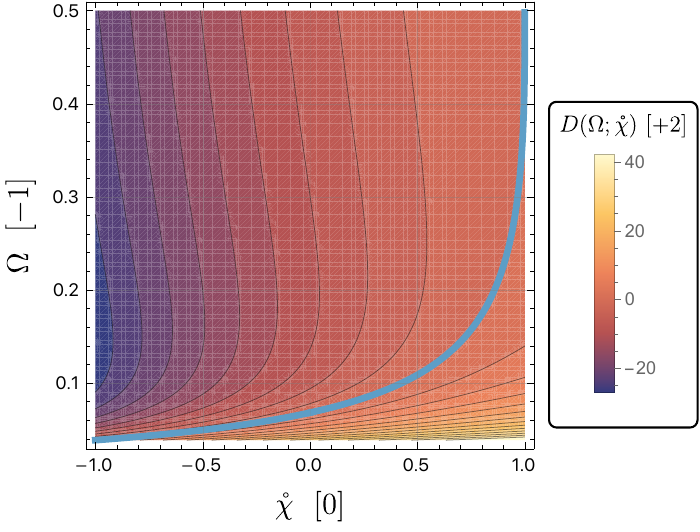}
\includegraphics[width=.43\textwidth]{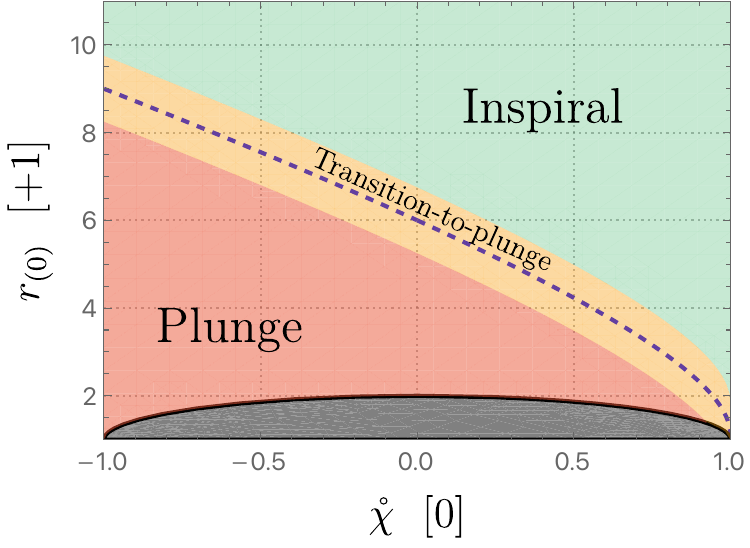}
\caption{\label{fig:paramspace} Left panel: the function $D(\Omega; \mathring\chi)$ in Eq.~\eqref{eq:D} plotted in terms of the orbital frequency $\Omega$ and the primary's spin $\mathring\chi$. The blue level set corresponds to $ D(\Omega_\star(\mathring\chi); \mathring\chi)=0$, where $ \Omega_\star(\mathring\chi)$ is the ISCO frequency. The region below the blue curve corresponds to $D>0$ (inspiral), while the region above it corresponds to $ D<0$ (plunge). Right panel: phase diagram of quasicircular motion around a primary Kerr black hole of spin $\mathring\chi$. The dashed blue curve corresponds to the ISCO radius $r_\star(\mathring\chi)$. Above that curve, circular orbits are stable, while below it they are unstable. We qualitatively depict the region of validity of the inspiral (green region above the ISCO), the transition to plunge (orange band around the ISCO) as well as the plunge (red region below the ISCO). The gray region corresponds to the primary black hole's interior. All quantities have been adimensionalized using the primary mass $\mathring M$. The numbers in square brackets next to a quantity correspond to its mass dimension as defined in Appendix~\ref{app:massdimension}.}
\end{figure*}

The function $D(\Omega)$ has a root at the ISCO frequency 
\begin{align}
&\Omega_\star(\mathring\chi) =\frac{1}{r_\star^{3/2} + \mathring\chi},\\
&r_\star = 3+Z_2-\text{sign}(\mathring\chi) \sqrt{(3-Z_1)(3+Z_1+2Z_2)}, \label{rstar} \\
&Z_1 =1+\sqrt[3]{1-\mathring\chi^2}\left(\sqrt[3]{1+\mathring\chi}+\sqrt[3]{1-\mathring\chi}\right), \\  &Z_2 =\sqrt{3\mathring\chi^2+Z_1^2}, 
\end{align}
marking the breakdown of the two-timescale expansion in place during the inspiral. In Eq.~\eqref{rstar}, $r_\star$ is the ISCO radius. The left panel of Fig.~\ref{fig:paramspace} shows the behavior of the function $D(\Omega;\mathring\chi)$ in terms of the orbital frequency and the primary black hole's spin. The level set $D=0$, which locates the ISCO frequency $\Omega_\star(\mathring\chi)$, splits the parameter space into the inspiral regime ($ D>0$) and the plunge regime ($ D<0$), while the region close to $D=0$ corresponds to the transition-to-plunge regime. These three regions in the parameter space are schematically depicted in Fig.~\ref{fig:paramspace}.

At next-to-leading order, the first post-adiabatic (1PA) corrections to the orbital radius and the redshift factor are given by
\begin{subequations}\label{eq:r1U1}
\begin{align}
    r_{(1)}(J^a) &= -\frac{r_{(0)}^2(\Omega)}{3 \Omega^2 U_{(0)}^2(\Omega) \Delta(\Omega)}f_{(1)}^r(J^a),\label{r1}
    \\
    U_{(1)}(J^a) &= 0.\label{U1}
\end{align}
\end{subequations}
Moreover, solving Eq.~\eqref{teom} at order $\mathcal{O}(\mathring\varepsilon^2)$ gives the 1PA forcing function of the orbital frequency:
\begin{align}\label{F1PA}
    F_{(1)}^{\Omega}(J^{a}) =& -\frac{3\Delta_{(0)}(\Omega)}{\Omega U_{(0)}^4(\Omega) D(\Omega) B(\Omega)} f_{(2)}^t(J^a)\nonumber\\
    &-\frac{2 r_{(0)}^{3/2}(\Omega)F_{(0)}^{J^a} (J^b)\partial_{J^a} f_{(1)}^r(J^b)}{\Omega^2 U_{(0)}^4(\Omega)  D(\Omega) \Delta_{(0)}(\Omega)}\nonumber\\
    &-\frac{4 P(\Omega)f_{(1)}^r(J^a) f_{(1)}^t(\Omega)}{\Omega^6 U_{(0)}^8(\Omega) r_{(0)}^{1/2}(\Omega) B^2(\Omega) D^2(\Omega) \Delta_{(0)}(\Omega)},
\end{align}
with
\begin{align}
    P(\Omega) =&\; r_{(0)}(\Omega) \Delta_{(0)}(\Omega)\left[-4 + 3 \Omega^2 B(\Omega) r_{(0)}(\Omega)\right.\nonumber\\
    &\left.+ 6 \Omega r_{(0)}^{3/2}(\Omega) + 3 \Omega^4 B^2(\Omega) r_{(0)}(\Omega) U_{(0)}^2(\Omega)\right]\nonumber\\
    &-2 \Omega B(\Omega) r_{(0)}^{3/2}(\Omega)(d\Delta_{(0)}/d r_{(0)})(\Omega).
\end{align}

For nonspinning binaries (meaning $\mathring\chi=0$), Eqs.~\eqref{eq:0PA}, \eqref{eq:r1U1}, and \eqref{F1PA} all reduce to the expressions in Appendix~A of Ref.~\cite{Miller:2020bft} and Sec.~III of Ref.~\cite{Kuchler:2024esj}.

\subsection{Re-expansion in the symmetric mass ratio}

All the expansions above are series in integer powers of $\mathring\varepsilon$, while keeping the phase-space coordinates $J^a=\{\Omega,\delta M,\delta\chi\}$ fixed and with the overall mass scale set by the mass $\mathring M$ of the primary. This setup is not well suited for comparisons with NR: the small-mass-ratio expansion explicitly breaks the symmetry arising from the exchange of the primary and the secondary compact objects, which is particularly important for the typical range of mass ratios covered by NR~\cite{Tiec:2014lba}. Moreover, NR simulations use the total mass as the mass scale. We therefore proceed to re-expand all quantities as power series in the symmetric mass ratio $\check\nu=m_p {\mathring M}/{\mathring M_\text{tot}}^2 = \mathring\varepsilon/(1+\mathring\varepsilon)^2$ and we set the overall mass scale as the total reference mass $\mathring M_\text{tot} = \mathring{M} + m_p = \mathring{M}(1+\mathring{\varepsilon})$. Historically, this re-expansion has been found to yield the most accurate comparisons in the regime of comparable-mass binaries~\cite{LeTiec:2011bk,LeTiec:2011dp,LeTiec:2013uey,Nagar:2013sga,LeTiec:2017ebm,Rifat:2019ltp,vandeMeent:2020xgc,Warburton:2021kwk,Wardell:2021fyy,Kuchler:2024esj,Kuchler:2025hwx}.

Before performing the re-expansion, let us introduce a handy notation: a quantity $f$ of mass dimension $n$ is adimensionalized with respect to the primary mass, coherently with all notations introduced in the previous sections. Now, that same quantity but with a check overhead, $\check f$, is adimensionalized with respect to the total reference mass $\mathring M_\text{tot}$. Both quantities are simply related by
\begin{align}\label{eq:massdimension}
    \check f=\left(\frac{{\mathring M}}{{\mathring M_\text{tot}}}\right)^n f.
\end{align}
Therefore, the mass dimension $n$ is the only number we need to keep track of when changing the mass scale of the system. Appendix~\ref{app:massdimension} provides a list of quantities together with their mass dimension. In particular, we denote the BL and hyperboloidal times normalized by the total mass as $\check t$ and $\check s$, respectively. 

As a summary, we expand all quantities in terms of the symmetric mass ratio $\check \nu$ and total reference mass $\mathring M_\text{tot}$. We furthermore define the phase-space coordinates with a check overhead, $\check J^a=\{\check \Omega,\delta \check M_\text{tot},\delta\check \chi\}$, as
\begin{subequations}
\begin{align}
    \check \Omega(\check s) &\coloneqq\, \frac{{\mathring M_\text{tot}}}{{\mathring M}}\Omega(s),\\
    \chi(\check s) &\coloneqq\, {\mathring\chi}\left(1+\check\nu\,  \delta\check\chi(s)\right).\\
    M_\text{tot}(\check s) &\coloneqq\, {\mathring M_\text{tot}}\big(1+\check\nu\, \delta\check M_\text{tot}(s)\big).
\end{align}
\end{subequations}

To first subleading order in the symmetric mass ratio, re-expanding the old phase-space coordinates $J^a$ into the new ones $\check J^a$, yields
\begin{subequations}
\begin{align}
    \Omega &= \left[1-\check\nu+\mathcal{O}(\check\nu^2)\right]\check\Omega,\\
    \delta \chi &= \left[1-2\check\nu+\mathcal{O}(\check\nu^2)\right] \delta\check\chi,\\
    \delta M &= \left[1-\check\nu+\mathcal{O}(\check\nu^2)\right] \delta \check M_\text{tot}.
\end{align}
\end{subequations}

Similarly, the forcing function $F^\Omega$~\eqref{FOmega} is re-expanded as
\begin{align}\label{eq:F0PAreexp}
    \check F^{\check\Omega}(\check J^a) =&\; \check\nu F_{(0)}^{\Omega}(\check\Omega) + \check\nu^2 \left[F_{(1)}^{\Omega}(\check\Omega,\delta \check M_\text{tot},\delta\check \chi)\right.\nonumber\\
    &\left.+4 F_{(0)}^{\Omega}(\check\Omega)-\check\Omega\,\partial_{\check\Omega} F_{(0)}^{\Omega}(\check\Omega)\right]+\mathcal{O}(\check\nu^3),
\end{align}
with $\check F^{\check\Omega}=d\check\Omega/d\check t$, which is consistent with the notation introduced above. This defines $\check F_{(0)}^{\check\Omega}(\check \Omega)$ and $\check F_{(1)}^{\check\Omega}(\check J^a)$ as the coefficients of $\check\nu^1$ and $\check \nu^2$ in Eq.~\eqref{eq:F0PAreexp}, respectively.

Finally, the mode amplitudes $H_{\ell m}$ from Eq.~\eqref{eq:HlmPA} are re-expanded as 
\begin{align}\label{eq:HlmPAreexp}
    \check H_{\ell m}(\check J^a) =&\, \check\nu H^{(1)}_{\ell m}(\check J^a) + \check\nu^2 \left[H^{(2)}_{\ell m}(\check J^a) + H^{(1)}_{\ell m}(\check J^a)\right.\nonumber\\
    &-\check\Omega\,\partial_{\check\Omega} H^{(1)}_{\ell m}(\check J^a) - \delta\check M_\text{tot}\partial_{\delta\check M_\text{tot}} H^{(1)}_{\ell m}(\check J^a)\nonumber\\
    &\left.- 2\delta\check\chi\partial_{\delta\check \chi} H^{(1)}_{\ell m}(\check J^a)\right] + \mathcal{O}(\check\nu^3).
\end{align}
Again, the first- and second-order mode amplitudes $\check H^{(1)}_{\ell m}$ and $\check H^{(2)}_{\ell m}$ are defined as the coefficients of, respectively, $\check\nu^1$ and $\check \nu^2$ in the equation above.

\section{The transition-to-plunge expansion}\label{sec:PLTexpansion}

In this section we describe the perturbative expansion of the transition-to-plunge dynamics in the phase-space approach. This generalizes results obtained in Refs.~\cite{Kuchler:2023jbu,Kuchler:2024esj} to the case of a spinning primary.

\subsection{Perturbative expansion}\label{subsec:PLTexpansion}

In the transition regime, the orbital frequency no longer evolves on the radiation-reaction timescale but on a new timescale $\sim1/{\mathring\varepsilon}^{1/5}$~\cite{Buonanno:2000ef,Ori:2000zn}, which we refer to as the ISCO-crossing timescale. We write all quantities of interest as functions of $(\phi_p,\Delta J^a, \mathring\varepsilon)$, where $\Delta J^a=\{\Delta\Omega,\delta M,\delta\chi\}$. We recall that $\Delta\Omega$ is related to the orbital frequency through $\Omega=\Omega_\star+{\mathring\varepsilon}^{2/5}\Delta\Omega$, with $\Omega_\star$ the geodesic orbital frequency at the (Kerr background) ISCO crossing. The transition expansion then consists of expanding all functions in integer powers of $\mathring\lambda\coloneqq\mathring\varepsilon^{1/5}$ at fixed $(\phi_p, \Delta J^a)$~\cite{Kuchler:2023jbu,Kuchler:2024esj}. 

For the mode amplitudes of the metric perturbations we obtain
\begin{align}
    H_{\ell m}(\Delta J^a,\mathring \lambda) =&\; \mathring\lambda^5 H^{[5]}_{\ell m}(\Delta J^a) + \sum_{n=7}^9\mathring\lambda^n H^{[n]}_{\ell m}(\Delta\Omega)\nonumber\\
    &+ \sum_{n=10}^{\infty}\mathring\lambda^n H^{[n]}_{\ell m}(\Delta J^a).\label{Hlm transition}
\end{align}
The facts that $H^{[6]}_{\ell m}=0$ and that $H^{[n]}_{\ell m}$ for $n=7,8,9$ only depend on $\Delta\Omega$ follow from the same arguments as in Ref.~\cite{Kuchler:2024esj}. There, it was also shown that the leading-order amplitude $H^{[5]}_{\ell m}$ does not depend on $\Delta\Omega$ and that the 2PLT term can be written as $H^{[7]}_{\ell m} = \Delta\Omega H^{[7]A}_{\ell m}$, where $H^{[7]A}_{\ell m}$ is a constant.

The forcing functions are defined as $d\Delta\Omega/dt \coloneqq  F^{\Delta\Omega}={\cal O}(\mathring\lambda)$, $d\delta M/dt \coloneqq  F^{\delta M}={\cal O}(\mathring\lambda^5)$ and $d\delta \chi/dt \coloneqq F^{\delta \chi}={\cal O}(\mathring\lambda^5)$~\cite{Kuchler:2024esj}. Their transition expansions read
\begin{subequations}
\begin{align}
    F^{\Delta\Omega}(\Delta J^a,{\mathring\lambda}) &=\, {\mathring\lambda} \sum_{n = 0}^\infty {\mathring\lambda}^nF_{[n]}^{\Delta\Omega}(\Delta J^{a}),\\
    F^{\delta M}(\Delta J^a,{\mathring\lambda}) &=\, {\mathring\lambda}^5 \sum_{n = 0}^\infty {\mathring\lambda}^nF_{[3+n]}^{\delta M}(\Delta J^{a}),\\
    F^{\delta \chi}(\Delta J^a,{\mathring\lambda}) &=\, {\mathring\lambda}^5 \sum_{n = 0}^\infty {\mathring\lambda}^nF_{[3+n]}^{\delta \chi}(\Delta J^{a}).\label{FPLT}    
\end{align}
\end{subequations}
Here, we use the same labeling convention as Ref.~\cite{Kuchler:2025hwx}, where the index $[n]$ corresponds to the $n$th post-leading transition ($n$PLT) order at which a given forcing function contributes to the orbital phase. The PLT expansions of $r_p$, $U$, and $f^\mu$ read
\begin{subequations}
\begin{align}
     r_p(\Delta J^a,{\mathring\lambda}) &=\, r_\star+ {\mathring\lambda}^{2} \sum_{n=0}^\infty {\mathring\lambda}^n r_{[n]}(\Delta J^a)\label{rPLT} ,\\
    U(\Delta J^a,{\mathring\lambda}) &=\, U_\star+{\mathring\lambda}^2\sum_{n=0}^\infty {\mathring\lambda}^n U_{[n]}(\Delta J^a)\label{UPLT}, \\
    f^\mu(\Delta J^a,{\mathring\lambda}) &=\, {\mathring\lambda}^5 \sum_{n=0}^\infty {\mathring\lambda}^n f^\mu_{[5+n]}(\Delta J^a),\label{fPLT}
\end{align}
\end{subequations}
where we use a star to indicate evaluation on the geodesic background ISCO~\eqref{rstar}.

We use the two sets of expansions above to solve the equations of motion~\eqref{eq:eom} order by order in the expansion parameter $\mathring\lambda$. In general, the hierarchical structure is the following: at order ${\cal O}(\mathring\lambda^{2+n})$, Eqs.~\eqref{reom} and \eqref{Reom} give two algebraic equations for $r_{[n]}$ and $U_{[n]}$, respectively, while Eq.~\eqref{teom} at order ${\cal O}(\mathring\lambda^{5+n})$ yields a differential equation for the forcing function $F_{[n]}^{\Delta\Omega}$.

At leading order, we trivially recover the geodesic relations evaluated on the marginal circular geodesic: $r_\star=r_{(0)}(\Omega_\star)$, $U_\star=U_{(0)}(\Omega_\star)$. The corrections to the orbital radius and the redshift are given up to 2PLT order by
\begin{subequations}\label{eq:0PLTOrb}
\begin{align}
    r_{[0]}\left(\Delta\Omega\right) &= \left(\frac{d r_{(0)}}{d\Omega}\right)_\star \Delta\Omega,\label{eq:0PLTRad}\\
    U_{[0]}\left(\Delta\Omega\right) &= \left(\frac{dU_{(0)}}{d\Omega}\right)_\star \Delta\Omega,\label{eq:0PLTRed}
\end{align}
\end{subequations}
\begin{subequations}\label{eq:1PLTOrb}
\begin{align}
    r_{[1]}\left(\Delta\Omega\right) &= 0,\label{eq:1PLTRad}\\
    U_{[1]}\left(\Delta\Omega\right) &= 0\label{eq:1PLTRed},
\end{align}
\end{subequations}
and
\begin{subequations}\label{eq:2PLTOrb}
\begin{align}
    r_{[2]}(\Delta\Omega) =&\; \frac{1}{2} \left(\frac{d^2r_{(0)}}{d\Omega^2}\right)_\star \Delta\Omega^2\nonumber\\
    &- \frac{r_\star^{5/2}(dr_{(0)}/d\Omega)_\star^2}{2\Delta_\star}F_{[0]}^{\Delta\Omega}\partial_{\Delta\Omega}F_{[0]}^{\Delta\Omega},\label{eq:2PLTRad}\\
    U_{[2]}(\Delta\Omega) =&\; \frac{1}{2}\left(\frac{d^2U_{(0)}}{d\Omega^2}\right)_\star\Delta\Omega^2. \label{eq:2PLTRed}
\end{align}
\end{subequations}

The 0PLT corrections to the orbital radius~\eqref{eq:0PLTRad} and the redshift~\eqref{eq:0PLTRed} solely come from evaluating the adiabatic quantities $r_{(0)}$ and $U_{(0)}$ at the frequency $\Omega_\star+{\mathring\lambda}^2\Delta\Omega$. The first term of the 2PLT orbital radius, given in Eq.~\eqref{eq:2PLTRad}, as well as the entire 2PLT correction to the redshift, given in Eq.~\eqref{eq:2PLTRed}, also have the same origin. The 3PLT and 4PLT corrections are given in Appendix~\ref{app:ttp functions}.

Using Eqs.~\eqref{eq:0PLTOrb}, \eqref{eq:1PLTOrb}, and \eqref{eq:2PLTOrb} in the ${\cal O}(\mathring\lambda^5)$ term of Eq.~\eqref{teom} gives the following differential equation for~$F^{\Delta\Omega}_{[0]}$:%
\begin{multline}\label{eq:F0PLT}
\left(F_{[0]}^{\Delta\Omega}\right)^2\frac{d^2F_{[0]}^{\Delta\Omega}}{d\Delta\Omega^2}
+F_{[0]}^{\Delta\Omega}\left(\frac{dF_{[0]}^{\Delta\Omega}}{d\Delta\Omega}\right)^2
\\+\zeta_1(\Omega_\star)\Delta\Omega F_{[0]}^{\Delta\Omega}
=\zeta_2(\Omega_\star) f_{[5]}^t. 
\end{multline}
The two coefficients $\zeta_{1}$ and $\zeta_{2}$ only depend on the primary black hole's spin, through the ISCO frequency $\Omega_\star(\mathring\chi)$. They are given explicitly by
\begin{subequations}\label{zeta12}
\begin{align}
   \zeta_1(\Omega_\star)&= -\frac{U_\star^2\Omega_\star^2\Delta_\star (dD/d\Omega)_\star}{2r_\star^{3/2}(dr_{(0)}/d\Omega)_\star}\label{zeta1},\\
   \zeta_2(\Omega_\star)&= -\frac{\Delta_\star^2}{\Omega_\star r_\star^2(dr_{(0)}/d\Omega)_\star^2 U_\star^2 B_\star}.\label{zeta2}
\end{align}
\end{subequations}
As prescribed in Ref.~\cite{Kuchler:2024esj}, defining $s_{[0]}=\int d\Delta\Omega/F_{[0]}^{\Delta\Omega}$, Eq.~\eqref{eq:F0PLT} can be written as
\begin{equation}\label{Painleve}
    \frac{d^2\Delta\Omega}{ds_{[0]}^2}+\frac{\zeta_1}{2}\Delta\Omega^2=\zeta_2f^t_{[5]}s_{[0]},
\end{equation}
which can be recognized as a Painlev\'e transcendental equation of the first kind; see Refs.~\cite{Compere:2019cqe,Compere:2021zfj,Kuchler:2024esj}.

We checked that the prograde near-extremal expansion of the coefficients $\zeta_{1}$ in Eq.~\eqref{zeta1} and $\zeta_{2}$ in Eq.~\eqref{zeta2} are consistent with the results of Ref.~\cite{Compere:2019cqe}. Indeed, close to ${\mathring\chi}=+1$, the near-extremal behavior of these coefficients is easily found to be
\begin{subequations}\label{eq:zeta12nearextremal}
\begin{align}
    \zeta_1 &= -2^{4/3}(1-{\mathring\chi})^{2/3}+\mathcal{O}\bigl[(1-{\mathring\chi})^{}\bigr],\\
    \zeta_2 &= \frac{1}{2^{2/3}}\frac{27}{32}(1-{\mathring\chi})^{5/3}+\mathcal{O}\bigl[(1-{\mathring\chi})^{2}\bigr].
\end{align}
\end{subequations}
With the change of variables $r_{[0]}=(dr_{(0)}/d\Omega)_\star\Delta\Omega$ and $s_{[0]}={\mathring\varepsilon}^{1/5} U_\star( \tau-\tau_\star)$ as well as the near-extremal behaviors~\eqref{eq:zeta12nearextremal}, Eq.~\eqref{Painleve} reproduces Eq.~(26) of Ref.~\cite{Compere:2019cqe}.

We find that the 1PLT forcing function $F_{[1]}^{\Delta\Omega}$ satisfies the homogeneous equation
\begin{align}\label{eq:F1PLT}
    &\left(F_{[0]}^{\Delta\Omega}\right)^2\frac{d^2F_{[1]}^{\Delta\Omega}}{d\Delta\Omega^2}
    +2F_{[0]}^{\Delta\Omega}F_{[1]}^{\Delta\Omega}\frac{d^2F_{[0]}^{\Delta\Omega}}{d\Delta\Omega^2}
    +F_{[1]}^{\Delta\Omega}\left(\frac{dF_{[0]}^{\Delta\Omega}}{d\Delta\Omega}\right)^2 \nonumber\\
&    +2F_{[0]}^{\Delta\Omega}\frac{dF_{[0]}^{\Delta\Omega}}{d\Delta\Omega}\frac{dF_{[1]}^{\Delta\Omega}}{d\Delta\Omega}
    +\zeta_1(\Omega_\star)\Delta\Omega F_{[1]}^{\Delta\Omega}=0.
\end{align}
We shall see in Sec.~\ref{sec:matching} that the asymptotic match of the 1PLT forcing function with the inspiral forces the solution $F_{[1]}^{\Delta\Omega}$ to identically vanish, a straightforward generalization of the results in Refs.~\cite{Kuchler:2023jbu,Kuchler:2024esj}. The 2PLT forcing function $F_{[2]}^{\Delta\Omega}$ satisfies
\begin{widetext}
\begin{align}
    \left(F_{[0]}^{\Delta\Omega}\right)^2\frac{d^2F_{[2]}^{\Delta\Omega}}{d\Delta\Omega^2}
    +2F_{[0]}^{\Delta\Omega}F_{[2]}^{\Delta\Omega}&\frac{d^2F_{[0]}^{\Delta\Omega}}{d\Delta\Omega^2}
    +F_{[2]}^{\Delta\Omega}\left(\frac{dF_{[0]}^{\Delta\Omega}}{d\Delta\Omega}\right)^2
    +2F_{[0]}^{\Delta\Omega}\frac{dF_{[0]}^{\Delta\Omega}}{d\Delta\Omega}\frac{dF_{[2]}^{\Delta\Omega}}{d\Delta\Omega}
    +\zeta_1(\Omega_\star)\Delta\Omega F_{[2]}^{\Delta\Omega}\nonumber\\
    &=\zeta_2(\Omega_\star)f_{[7]}^t
    +\zeta_3(\Omega_\star)f_{[5]}^t\Delta\Omega
    +\zeta_4(\Omega_\star)F_{[0]}^{\Delta\Omega}\Delta\Omega^2
    +\zeta_5(\Omega_\star)\left(F_{[0]}^{\Delta\Omega}\right)^2\frac{dF_{[0]}^{\Delta\Omega}}{d\Delta\Omega}.\label{eq:F2PLT} 
\end{align}
\end{widetext}
Just as for $\zeta_1$ and $\zeta_2$, the coefficients $\zeta_3$, $\zeta_4$, and $\zeta_5$ appearing in this equation depend on the primary spin $\mathring\chi$ through the ISCO frequency. Their analytical expressions are given by%
\begingroup%
\allowdisplaybreaks%
\begin{align}
    \zeta_3(\Omega_\star) &= -\frac{\Delta_\star^2}{3r_\star^{7/2}\Omega_\star^3B_\star^2U_\star^5(dr_{(0)}/d\Omega)_\star^2}\left[4B_\star U_\star^3\phantom{\frac{}{}}\right.\nonumber\\*
    &+6r_\star^{3/2}\Omega_\star B_\star U_\star^3-3r_\star^2(r_\star-1)\Omega_\star^2 U_\star^5 B_\star\nonumber\\*
    &-\left.3r_\star^{3/2}\Omega_\star^3U_\star^5B_\star^2-3r_\star^{3/2}\Omega_\star\big(d^2U_{(0)}/d\Omega^2\big)_\star\right],\\
    \zeta_4(\Omega_\star) &= -\frac{U_\star^4\Delta_\star^2}{24r_\star^{5/2}(dr_{(0)}/d\Omega)_\star^2}\left[4(2+3r_\star^{3/2}\Omega_\star)(dD/d\Omega)_\star\right.\nonumber\\*
    &+\left.3r_\star^{3/2}\Omega_\star^2(d^2D/d\Omega^2)_\star\right],\\
    \zeta_5(\Omega_\star)&=-\frac{U_\star^4\Delta_\star^2}{32\Omega_\star B_\star(dr_{(0)}/d\Omega)_\star}\Bigl\{16\Omega_\star B_\star-16r_\star^{1/2}\nonumber\\*
    &+3r_\star^2\Omega_\star^5B_\star\!\left[r_\star^{1/2}U_\star^2(dD/d\Omega)_\star+18(d^2r_{(0)}/d\Omega^2)_\star\right]\nonumber\\*
    &+48r_\star^2\Omega_\star-24r_\star^{3/2}\Omega^2B_\star(1+3B_\star U_\star^2\Omega_\star^2)\nonumber\\*
    &+12r_\star\Omega_\star^3U_\star^2B_\star\mathcal{E}_\star\Bigr\}.
\end{align}
\endgroup%

We recognize the left-hand side of both Eqs.~\eqref{eq:F1PLT} and~\eqref{eq:F2PLT} to be the linearization of the Painlev\'e transcendental equation~\eqref{eq:F0PLT} around the solution $F_{[0]}^{\Delta\Omega}$~\cite{Kuchler:2023jbu,Kuchler:2024esj}.

\subsection{Re-expansion in the symmetric mass ratio}\label{subsec:PLTreexp}

Analogously to the inspiral, we re-expand the forcing functions $F^{\Delta\Omega}$ and the mode amplitudes $H_{\ell m}$ as power series in the fractional power of the symmetric mass ratio, 
\begin{align}
\check\sigma\coloneqq\check\nu^{1/5}, 
\end{align}
while keeping the new set of phase-space coordinates $\Delta\check J^a=\{\Delta\check\Omega,\delta\check M_\text{tot}, \delta\check \chi\}$ fixed and using the total mass of the binary as the mass scale. Here, $\Delta\check\Omega$ has been defined through
\begin{equation}\label{eq:check DeltaOmega}
    \check\Omega = \check\Omega_\star + \check\sigma^2\Delta\check\Omega,
\end{equation}
which relates to $\Delta\Omega=(\Omega-\Omega_\star)/\mathring\lambda^2$ as
\begin{align}
    \Delta\Omega=\left[1-\frac{9}{5}\check\sigma^5+\mathcal{O}(\check\sigma^{10})\right]\Delta\check\Omega .
\end{align}

Up to 2PLT order, the forcing functions $\check F_{[n]}^{\Delta\check\Omega}$ (i.e., the coefficients of $\check\sigma^n$ in a transition expansion of $d\Delta\check\Omega/d\check t$ at fixed $\Delta\check J^a$) are simply given by
\begin{equation}\label{eq:PLTReexpansion1}
    \check F_{[n]}^{\Delta\check\Omega}(\Delta\check\Omega) = F_{[n]}^{\Delta\Omega}(\Delta\check\Omega), \quad 0\leq n\leq2.
\end{equation}
The functions start to differ at 5PLT order; see Appendix~\ref{app:re-expansions}. Similarly, the mode amplitudes of the metric perturbations~\eqref{Hlm transition} get re-expanded to give
\begin{subequations}
\begin{align}
    \check H_{\ell m}^{[5]}(\Delta\check J^a) &=\, H_{\ell m}^{[5]}(\Delta\check J^a),\\
    \check H_{\ell m}^{[6]}(\Delta\check\Omega) &=\, H_{\ell m}^{[6]}(\Delta\check\Omega)=0,\\
    \check H_{\ell m}^{[7]}(\Delta\check\Omega) &=\, H_{\ell m}^{[7]}(\Delta\check\Omega).
\end{align}
\end{subequations}
Again, the functions start to differ at 5PLT order.

\section{Asymptotically matching the inspiral with the transition}\label{sec:matching}

We now proceed to verify the asymptotic match between inspiral and transition-to-plunge quantities through 0PA and 2PLT order, respectively. Following Ref.~\cite{Kuchler:2024esj}, we perform a near-ISCO expansion $\Omega\to\Omega_\star+\mathring\lambda^2\Delta\Omega$ of the inspiral quantities and compare it with the early-time asymptotics as $\Delta\Omega\rightarrow-\infty$ of the transition quantities.

\subsection{Match of the mode amplitudes and self-force}

The inspiral's first-order mode amplitudes are smooth functions of the orbital frequency at the ISCO and can therefore by Taylor expanded around it. The near-ISCO expansion of the inspiral's mode amplitudes then reads
\begin{equation}\label{eq:Hlm inspiral nearISCO}
    \check H_{\ell m} \!=\! \check\nu \left[\!\left.\check H_{\ell m}^{(1)}\right\vert_\star \!+\! \check\sigma^2\Delta\check\Omega\left.\partial_{\check\Omega}\check H_{\ell m}^{(1)}\right\vert_\star \!+\! \mathcal{O}(\Delta\check\Omega^2) \right] + \mathcal{O}(\check\nu^2),
\end{equation}
where we have used the definition~\eqref{eq:check DeltaOmega} of $\Delta\check\Omega$ to rewrite the differences $(\check\Omega-\check\Omega_\star)$. The early-time solution of the transition mode amplitudes to 2PLT order is simply given by
\begin{equation}\label{eq:Hlm transition early}
    \check H_{\ell m} = \check\sigma^5\check H^{[5]}_{\ell m} + \check\sigma^7 \Delta\check\Omega \, \check H^{[7]A}_{\ell m} + \mathcal{O}(\check\sigma^8).
\end{equation}
Comparing the coefficients of equal powers of $\check\sigma$ and $\Delta\check\Omega$ in Eqs.~\eqref{eq:Hlm inspiral nearISCO} and \eqref{eq:Hlm transition early} yields the following matching conditions
\begin{subequations}\label{PAPLTMatchingHlm}
\begin{align}
    \check H_{\ell m}^{[5]} &= \check H_{\ell m}^{(1)}(\check\Omega_\star),\\
    \check H_{\ell m}^{[7]} &= \Delta\check\Omega \, \check H_{\ell m}^{[7]A} =\Delta\check\Omega\,\partial_{\check\Omega} \check H_{\ell m}^{(1)}(\check\Omega_\star).
\end{align}
\end{subequations}
In an analogous way, the match of the self-force reads
\begin{subequations}\label{PAPLTMatchingfmu}
\begin{align}
    f^\mu_{[5]} &= f^\mu_{(1)}(\check\Omega_\star),\\
    f^\mu_{[7]} &= \Delta\Omega\,\partial_\Omega f^\mu_{(1)}(\Omega_\star).
\end{align}
\end{subequations}

\subsection{Match of the forcing functions of the orbital frequency}

We obtain the generic matching to all PA and PLT orders in the following way: the inspiral's post-adiabatic expansion of the rate of change of the orbital frequency is given by $d\check\Omega/d\check t = \check\nu \sum_{i\geq0}\check\nu^i \check F^{\check\Omega}_{(i)}(\check\Omega)$. For simplicity we suppress the dependence on $\delta\check M$ and $\delta\check\chi$. Re-expanding close to the ISCO and recalling Eq.~\eqref{eq:check DeltaOmega}, we have
\begin{equation}\label{eq:inspiral nearISCO sigma}
    \frac{d\check\Omega}{d\check t} = \sum_{i\geq0}\sum_j \check F^{(5+5i+2j,j)}_{(i)} \check\sigma^{5+5i+2j} \Delta\check\Omega^j,
\end{equation}
where $\check F^{(a,b)}_{(i)}$ are the (in general $\delta\check M$- and $\delta\check\chi$-dependent) coefficients of $\check\sigma^a$ and $\Delta\check\Omega^b$ in the $i$PA order's near-ISCO expansion. The range over which the index $j$ runs is determined by the near-ISCO limit of the algebraic expressions for the post-adiabatic forcing functions, see Eqs.~\eqref{eq:F0PAreexp}, \eqref{FOmega0PA} and \eqref{F1PA}. 

In particular, at adiabatic order we obtain
\begin{equation}\label{eq:asymptotic0PAnu}
    \check\nu\check F_{(0)}^{\check\Omega}=\sum_{k=0}^\infty \check F_{(0)}^{(3+2k,k-1)} \check\sigma^{3+2k} \Delta\check\Omega^{k-1}.
\end{equation}
The coefficients in Eq.~\eqref{eq:asymptotic0PAnu} can be expressed alternatively in terms of the same coefficients written in our old set of phase-space parameters $\Delta J ^a$,
\begin{equation}\label{eq:asymptotic0PA}
    {\mathring\varepsilon} F_{(0)}^{\Omega}=\sum_{k=0}^\infty F_{(0)}^{(3+2k,k-1)}{\mathring\lambda}^{3+2k}\Delta\Omega^{k-1}.
\end{equation}
We find at the lowest orders that
\begin{subequations}\label{eq:asymptoticOLDNEW}
\begin{align}
    \check F_{(0)}^{(3,-1)}&= F_{(0)}^{(3,-1)},\\
    \check F_{(0)}^{(5,0)}&= F_{(0)}^{(5,0)}.
\end{align}
\end{subequations}
Finally, the coefficients $F_{(0)}^{(3+2k,k-1)}$ can easily be computed by performing the near-ISCO expansion $\Omega=\Omega_\star+{\mathring\lambda}^2\Delta\Omega$ of Eq. \eqref{FOmega0PA}. This gives, at the first two orders,
\begin{subequations}\label{eq:asymptoticOLD}
\begin{align}
    &\check F_{(0)}^{(3,-1)}= \frac{-3 \Delta_\star f_{(1)\star}^t}{\Omega_\star U_\star^4 B_\star (dD/d\Omega)_\star},\label{eq:asymptoticOLD0PLT}
    \\
    &\check F_{(0)}^{(5,0)}= \Biggl[\frac{3}{2}\frac{(d^2r_{(0)}/d\Omega^2)_\star}{(dr_{(0)}/d\Omega)_\star}\frac{\Delta_\star}{\Omega_\star U_\star^4 B_\star (dD/d\Omega)_\star}\nonumber\\
    &\phantom{00}+\left(\frac{dr_{(0)}}{d\Omega}\right)_\star^{-1}\!\frac{d}{d\Omega}\!\left(\frac{-3\Delta(\Omega)}{\Omega U_{(0)}^4(\Omega)B(\Omega)}\frac{r_{(0)}(\Omega)-r_\star}{D(\Omega)}\right)_{\!\star}\Biggr]f_{(1)\star}^t\nonumber\\
    &\phantom{00}+\frac{-3\Delta_\star}{\Omega_\star U_\star^4 B_\star}\frac{\left(\partial_\Omega f_{(1)}^t\right)_\star}{(dD/d\Omega)_\star}.\label{eq:asymptoticOLD2PLT}
\end{align}
\end{subequations}

In the transition regime the rate of change of the orbital frequency is given by $d\check\Omega/d\check t=\check\sigma^3\sum_{n\geq0}\check\sigma^n F^{\Delta\check\Omega}_{[n]}$. At early times, where $\Delta\check\Omega\to-\infty$, we can re-expand in powers of $\Delta\check\Omega$:
\begin{subequations}
\begin{align}\label{eq:transition early sigma}
    \frac{d\check\Omega}{d\check t} &= \sum_{n\geq0}\sum_m \check F^{(3+n,m)}_{[n]} \check\sigma^{3+n} \Delta\check\Omega^m,\\
    &= \sum_{n\geq0}\sum_m \check F^{(3+n,m)}_{[n]} \check\nu^{(3+n-2m)/5} (\check\Omega-\check\Omega_\star)^m,\label{eq:transition early nu}
\end{align}
\end{subequations}
where $\check F^{(a,b)}_{[n]}$ are the (in general $\delta\check M$- and $\delta\check\chi$-dependent) coefficients of $\check\sigma^a$ and $\Delta\check\Omega^b$ in the $n$PLT order's early-time expansion. The bounds on the values of the index $m$ are readily obtained by the requirement that Eq.~\eqref{eq:transition early nu} only contains integer powers of $\check\nu$, as is required in order to match the inspiral. This yields the condition
\begin{equation}
    m=-1+\frac{n}{2}-\frac{5}{2}i, \quad n,i\geq0.
\end{equation}
In addition $m$ needs to be an integer, since only integer powers of $\Delta\check\Omega$ appear in Eq.~\eqref{eq:inspiral nearISCO sigma}. The 0PLT and 2PLT early-time solutions can then be written as
\begin{subequations}\label{eq:asymptoticPLT}
\begin{align}
    \check\sigma^3\check F_{[0]}^{\check\Omega}&=\sum_{k=0}^\infty \check F_{[0]}^{(3,-1-5k)}\check\sigma^{3}\Delta\check\Omega^{-1-5k},\label{eq:asymptotic0PLT}\\
    \check\sigma^5\check F_{[2]}^{\check\Omega}&=\sum_{k=0}^\infty \check F_{[2]}^{(5,-5k)}\check\sigma^{5}\Delta\check\Omega^{-5k}.\label{eq:asymptotic2PLT}
\end{align}
\end{subequations}

Finally, in order for Eqs.~\eqref{eq:inspiral nearISCO sigma} and \eqref{eq:transition early sigma} to agree term by term, we need to equate the powers of $\check\sigma$ and $\Delta\check\Omega$, which leads to the conditions
\begin{equation}
    n=2+5i+2j, \quad m=j.
\end{equation}
Therefore, the 0PA forcing function $\check F^{\check\Omega}_{(0)}$ ($i=0$) is matched by terms with $n=2+2j$ and $m=j$ in Eq.~\eqref{eq:transition early sigma}. In particular, we have that (i) the leading-order term with $j=-1$, matches a term $\propto\Delta\check\Omega^{-1}$ in the 0PLT forcing function $\check F^{\Delta\check\Omega}_{[0]}$ ($\check F_{(0)}^{(3,-1)}=\check F_{[0]}^{(3,-1)}$), (ii) the first subleading term with $j=0$, matches a term $\propto\Delta\check\Omega^0$ in the 2PLT forcing function $\check F^{\Delta\check\Omega}_{[2]}$ ($\check F_{(0)}^{(5,0)}=\check F_{[2]}^{(5,0)}$), and so on; see Table~I of Ref.~\cite{Albertini:2022rfe} or Table~1 of Ref.~\cite{Kuchler:2024esj}. In the following subsections we show how to compute the asymptotic coefficients in Eq.~\eqref{eq:asymptoticPLT}. We will check that the 0PA-0PLT and 0PA-2PLT matching conditions we have just derived are satisfied.
    
\subsection{0PLT asymptotic matching}

Comparing the late-time asymptotic expansion of the 0PA forcing function \eqref{eq:asymptotic0PA} with the early-time asymptotic expansion of the 0PLT forcing function \eqref{eq:asymptotic0PLT}, we can easily verify the matching condition
\begin{equation}\label{eq:matching0PLT}
    \check F_{[0]}^{(3,-1)}=\check F_{(0)}^{(3,-1)},
\end{equation}
with $\check F_{(0)}^{(3,-1)}$ given in Eq.~\eqref{eq:asymptoticOLD0PLT}.

Let us reformulate the 0PLT dynamics such that the equation of motion has a normalized form. The advantage of this is that the dynamics becomes independent of the primary spin. As is typical in critical phenomena, the behavior of the system near the critical point (in this case, the ISCO) is universal.

To do so, we rescale the forcing function $\check F_{[0]}^{\Delta\check\Omega}$ and its independent variable $\Delta\check\Omega$ to the mass-dimension-$0$ quantities $y$ and $u$ defined by
\begin{subequations}\label{F0PLTrenomalization}
\begin{align}
    \check F_{[0]}^{\Delta\check\Omega} &=\, \frac{\alpha^{3/5}}{\beta^{2/5}}y,\\
    \Delta\check\Omega &=\, \frac{\alpha^{2/5}}{\beta^{3/5}}u,\label{eq:udef}
\end{align}
\end{subequations}
with
\begin{subequations}\label{alphabetadef}
\begin{align}
    \alpha &=\, \zeta_2(\Omega_\star)f_{(1)\star}^t,\\
    \beta &=\, -\zeta_1(\Omega_\star).
\end{align}
\end{subequations}
Under such rescaling, the Painlev\'e equation \eqref{eq:F0PLT} takes the normalized and spin-independent form
\begin{equation}\label{eq:y0PLT}
    y^2\frac{d^2y}{du^2}+y\left(\frac{dy}{du}\right)^2-uy=1.
\end{equation}

There is a 2-parameter family of solutions to Eq.~\eqref{eq:y0PLT}. From the matching with the inspiral~\eqref{eq:asymptoticPLT}, the monotonic solution $y(u)$ admits an asymptotic expansion
\begin{equation}\label{eq:y0PLTansatz}
    y(u)=\sum_{k=0}^\infty {c}_k u^{-1-5k},
\end{equation}
with
\begin{equation}
    \check F_{[0]}^{(3,-1-5k)}=\frac{\alpha^{2k+1}}{\beta^{3k+1}}{c}_k.
\end{equation}
Using Eqs. \eqref{zeta12}, \eqref{eq:asymptoticOLD0PLT}, and \eqref{alphabetadef}, we find the matching condition \eqref{eq:matching0PLT} now simply reads
\begin{equation}\label{0PLTnormalizedmatchingcondition}
    {c}_0=-1.
\end{equation}
We now impose Eq. \eqref{0PLTnormalizedmatchingcondition} as the initial condition of the transition dynamics, which correctly enforces the matching condition \eqref{eq:matching0PLT}. 

\begin{figure}
\includegraphics[width=.48\textwidth]{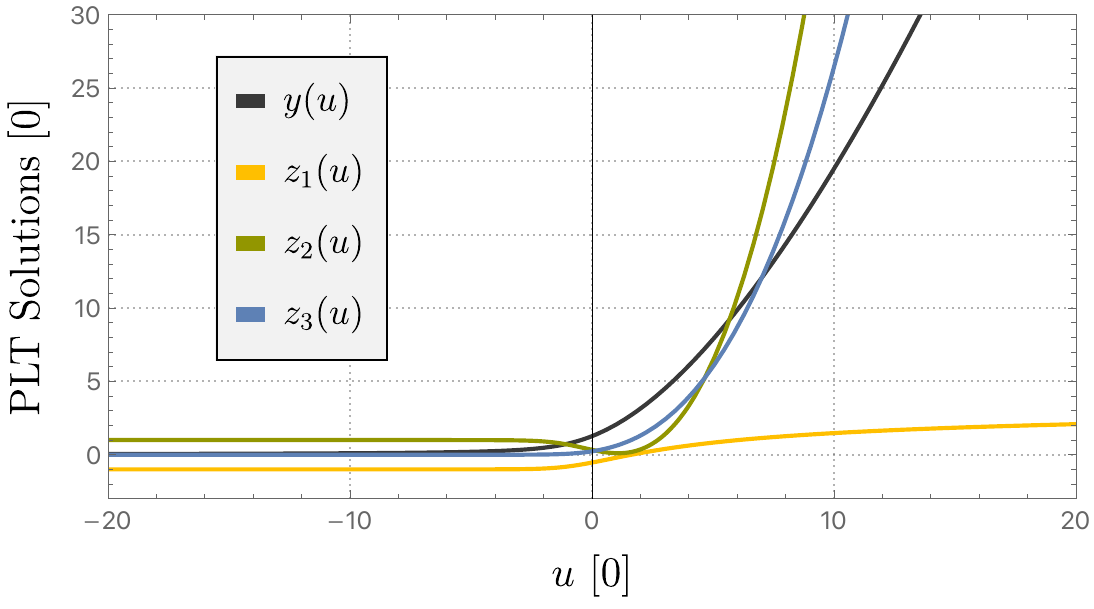}
\caption{\label{fig:y0PLT}Normalized 0PLT and 2PLT forcing functions. The solution to the normalized Painlev\'e equation \eqref{eq:y0PLT} with boundary condition \eqref{0PLTnormalizedmatchingcondition} is shown in black.  The solutions $z_i$ to the sourced linearized Painlev\'e equations~\eqref{eq:zi2PLT} with boundary solutions \eqref{2PLTASmatch} are shown in yellow, green, and blue, respectively. Here $u=0$ corresponds to the location of the ISCO, while $u<0$ ($u>0$) corresponds to an orbital frequency below (above) the ISCO frequency.}
\end{figure}

All the other coefficients ${c}_{i\neq 0}$ are fixed by the requirement that $y(u)$ is a solution to the Painlev\'e equation~\eqref{eq:y0PLT}. Indeed, plugging Eq.~\eqref{eq:y0PLTansatz} into Eq.~\eqref{eq:y0PLT} gives the difference equation
\begin{equation}\left\{\label{0PLTdiffeq}
    \begin{aligned}
        &{c}_0=-1,\\
        &{c}_{i+1}=\sum_{j=0}^i\sum_{k=0}^{i-j}(5k+1)(5k+5j+3){c}_{i-j-k}{c}_{j}{c}_{k}.
    \end{aligned}\right.
\end{equation}
We then pick the unique solution $y(u)$ to Eq.~\eqref{eq:y0PLT} that has the early-time asymptotic expansion~\eqref{eq:y0PLTansatz} with coefficients~\eqref{0PLTdiffeq}. The behavior of this solution is shown in Fig.~\ref{fig:y0PLT}. The solution $y(u)$ can then be rescaled to $\check F_{[0]}^{\Delta\check\Omega}$ for specific values of the primary black hole spin ${\mathring\chi}$ using Eq.~\eqref{F0PLTrenomalization}.

\begin{figure}
\includegraphics[width=.48\textwidth]{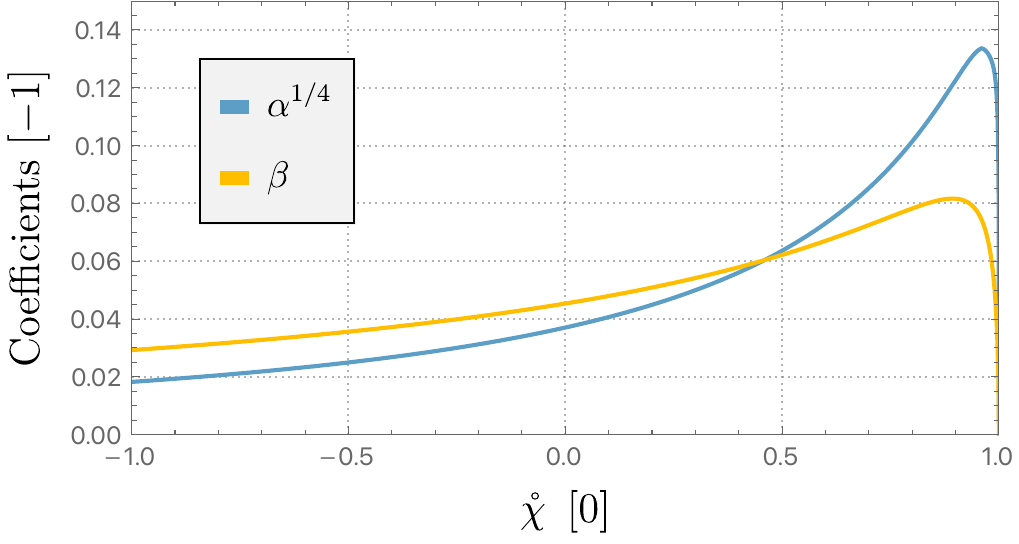}
\caption{\label{fig:alphabeta}Values of the coefficients $\alpha^{1/4}$ and $\beta$, as defined in Eqs.~\eqref{alphabetadef}, as a function of the primary black hole's spin ${\mathring\chi}$.}
\end{figure}

The rescaling coefficients~\eqref{alphabetadef} are both shown in Fig.~\ref{fig:alphabeta} as a function of the primary spin ${\mathring\chi}$. Let us finally prove that both coefficients go to zero in the near-extremal limit ${\mathring\chi \mapsto 1}$. This can be analytically checked by performing a near-extremal Kerr expansion: the $0$PLT dissipative self-force reads
\begin{equation}
    f^t_{[5]}=-\frac{\Omega_\star B_\star}{\Delta_\star}\kappa_\star,
\end{equation}
where $\kappa_\star=-(d u_\phi/d \tau)_\star$, as defined in Ref.~\cite{Compere:2019cqe}, is the specific angular momentum loss per unit of proper time. Reference~\cite{Gralla:2016qfw} showed that $\kappa_\star$ remains finite and non-zero near extremality. Hence, the pole structure of the self-force reads at leading-order as
\begin{equation}
    f^t_{[5]}= -\frac{\kappa_\star}{2^{5/3}}\frac{1}{(1-{\mathring\chi})^{1/3}}+\mathcal{O}\!\left(1\right).
\end{equation}
This shows, together with Eqs. \eqref{eq:zeta12nearextremal} and \eqref{alphabetadef}, that both $\alpha$ and $\beta$ go to zero as ${\mathring\chi}\rightarrow 1$.

\subsection{2PLT asymptotic matching}

Comparing the 0PA late-time asymptotics \eqref{eq:asymptotic0PA} with the 2PLT early-time asymptotics \eqref{eq:asymptotic2PLT} gives the matching condition
\begin{equation}\label{eq:matching2PLT}
    \check F_{[2]}^{(5,0)}=\check F_{(0)}^{(5,0)},
\end{equation}
which can be verified once the transition coefficient has been computed as demonstrated below in Eq.~\eqref{normalizationcoef2PLT}.

Similarly to the matching at 0PLT order, we rescale Eq. \eqref{eq:F2PLT} using the mass-dimension-$0$ quantity $z$ defined by
\begin{equation}\label{F2PLTrenormalization}
    \check F_{[2]}^{\Delta\check\Omega}=\frac{\alpha^{3/5}}{\beta^{2/5}}z,
\end{equation}
as well as $u$ that has been defined in Eq. \eqref{eq:udef}. In these new variables, Eq. \eqref{eq:F2PLT} now reads as
\begin{align}\label{z2PLT}
    y^2\frac{d^2z}{du^2}&+2yz\frac{d^2y}{du^2}+z\left(\frac{dy}{du}\right)^2+2y\frac{dy}{du}\frac{dz}{du}-uz\nonumber\\
    &=\gamma(\Omega_\star)u+\eta(\Omega_\star)u^2y+\tau(\Omega_\star)y^2\frac{dy}{du},
\end{align}
with
\begingroup 
\allowdisplaybreaks%
\begin{subequations}\label{gammaetataudef}
\begin{align}
    \gamma(\Omega_\star) &= \frac{\alpha^{2/5}}{\beta^{3/5}}\left[\frac{\left(\partial_\Omega f_{(1)}^t\right)_\star}{f_{(1)\star}^t}+\frac{\zeta_3(\Omega_\star)}{\zeta_2(\Omega_\star)}\right],\\
    \eta(\Omega_\star) &=\frac{\alpha^{2/5}}{\beta^{8/5}}\zeta_4(\Omega_\star),\\
    \tau(\Omega_\star) &= \frac{\alpha^{2/5}}{\beta^{3/5}}\zeta_5(\Omega_\star).
\end{align}
\end{subequations}
\endgroup

\begin{figure}[t]
\includegraphics[width=\columnwidth]{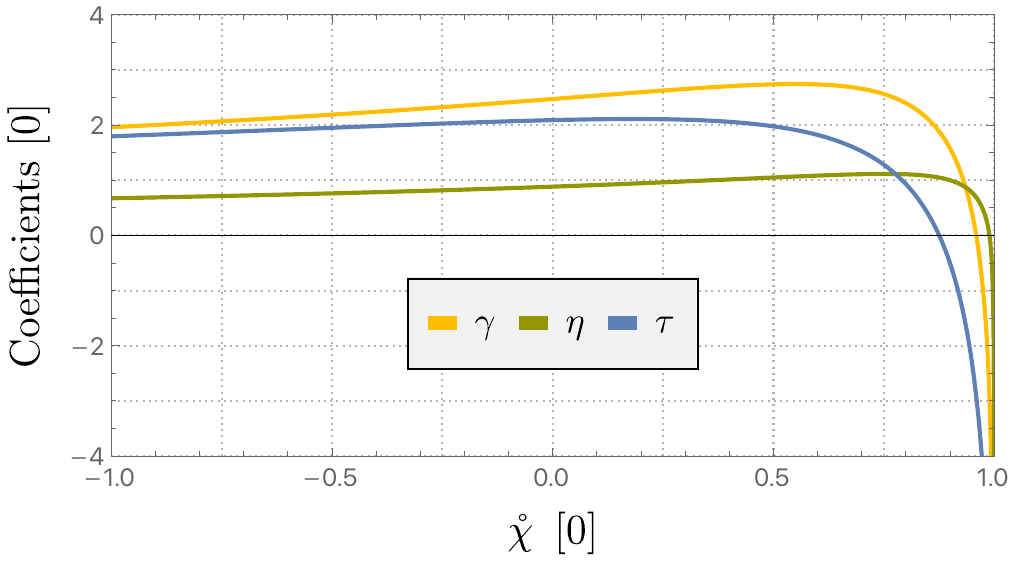}
\caption{\label{fig:gammaetatau} Dependence on the primary's spin ${\mathring\chi}$ of the coefficients 
$\gamma(\Omega_\star)$, $\eta(\Omega_\star)$, and $\tau(\Omega_\star)$  defined in Eqs.~\eqref{gammaetataudef}.}
\end{figure}

In contrast to the 0PLT differential equation~\eqref{eq:y0PLT}, the 2PLT differential equation is no longer universal, as it depends on the primary black hole spin through the coefficients $\gamma$, $\eta$, and $\tau$. However, since $y(u)$ is now known, the linear differential equation~\eqref{z2PLT} for the function $z(u)$ can be solved by first solving the three inhomogeneous solutions for the three sources individually
\begin{equation}\label{eq:zi2PLT}
    \mathcal P_y(z_1)=u, \quad \mathcal P_y(z_2)=u^2y, \quad \mathcal P_y(z_3) = y^2\frac{dy}{du},
\end{equation}
where the linear Painlev\'e operator around the solution $y(u)$ is
\begin{equation}
    \mathcal P_y(z) \coloneqq y^2\frac{d^2z}{du^2}+2yz\frac{d^2y}{du^2}+z\left(\frac{dy}{du}\right)^2+2y\frac{dy}{du}\frac{dz}{du}-uz    . 
\end{equation}
We then perform the linear combination
\begin{equation}\label{ziLC}
    z(u)=\gamma(\Omega_\star)z_1(u)+\eta(\Omega_\star)z_2(u)+\tau(\Omega_\star)z_3(u)
\end{equation}
to obtain the complete solution. The first step can be done once and for all offline while the second step has to be done online. Figure~\ref{fig:gammaetatau} depicts the dependence of the coefficients $\gamma$, $\eta$, and $\tau$ on the primary's spin ${\mathring\chi}$.

The general linear solution admits two parameters, as explained in Sec.~IV.C. of Ref.~\cite{Compere:2021zfj}. The asymptotic behavior \eqref{eq:asymptoticPLT} obtained from matching with the late inspiral selects the monotonic solution. We choose as a basis the three solutions $z_j$ that admit the following asymptotic expansions:
\begin{equation}\label{eq:ziExpansion}
    z_j(u) = \sum_{k=0}^\infty \prescript{\phantom{0}}{j}{{d}_k} u^{-5k},
\end{equation}
where
\begin{equation}\label{normalizationcoef2PLT}
    \check F_{[2]}^{(5,-5k)}=\frac{\alpha^{2k+3/5}}{\beta^{3k+2/5}}\left[\gamma(\Omega_\star)\prescript{\phantom{0}}{1}{{d}_k}+\eta(\Omega_\star)\prescript{\phantom{0}}{2}{{d}_k}+\tau(\Omega_\star)\prescript{\phantom{0}}{3}{{d}_k}\right]\!. 
\end{equation}
The matching condition \eqref{eq:matching2PLT} is found to be satisfied if we impose the following initial conditions:
\begin{align}\label{2PLTASmatch}
    \prescript{\phantom{0}}{1}{{d}_0}=-1,\phantom{0000}\prescript{\phantom{0}}{2}{{d}_0}=1,\phantom{0000}&\prescript{\phantom{0}}{3}{{d}_0}=0.
\end{align}
Furthermore, all the subsequent coefficients $\prescript{\phantom{0}}{j}{{d}_k}$ are fixed by requiring that the functions $z_j$ are solutions to the $2$PLT equations \eqref{eq:zi2PLT}. Indeed, plugging Eq. \eqref{eq:ziExpansion} into Eqs. \eqref{eq:zi2PLT}, we find
\begin{equation}
\label{2PLTdiffeq1}
    \begin{aligned}
        \prescript{}{1}{{d}_{i+1}}=\sum_{j=0}^i\sum_{k=0}^{i-j}\ \,&\bigl[(5k)(5k+1){c}_{i-j-k}{c}_{j}\prescript{\phantom{0}}{1}{{d}_{k}}\\[-1em]
        &+2(5k+1)(5k+2){c}_{i-j-k}\prescript{\phantom{0}}{1}{{d}_{j}}{c}_{k}\\
        &+(5j+1)(5k+1)\prescript{\phantom{0}}{1}{{d}_{i-j-k}}{c}_{j}{c}_{k}\\
        &+2(5k)(5j+1){c}_{i-j-k}{c}_{j}\prescript{\phantom{0}}{1}{{d}_{k}}\bigr],
    \end{aligned}
\end{equation}
\begin{equation}
\label{2PLTdiffeq2}
    \begin{aligned}
        \prescript{}{2}{{d}_{i+1}}=\sum_{j=0}^i\sum_{k=0}^{i-j}\ \,&\bigl[(5k)(5k+1){c}_{i-j-k}{c}_{j}\prescript{\phantom{0}}{2}{{d}_{k}}\\[-1em]
        &+2(5k+1)(5k+2){c}_{i-j-k}\prescript{\phantom{0}}{2}{{d}_{j}}{c}_{k}\\
        &
        +(5j+1)(5k+1)\prescript{\phantom{0}}{2}{{d}_{i-j-k}}{c}_{j}{c}_{k}\\
        &+2(5k)(5j+1){c}_{i-j-k}{c}_{j}\prescript{\phantom{0}}{2}{{d}_{k}}\bigr]\\
        &-{c}_{i+1},
    \end{aligned}
\end{equation}
\begin{equation}
\label{2PLTdiffeq3}
    \begin{aligned}
        \prescript{}{3}{{d}_{i+1}}=\sum_{j=0}^i\sum_{k=0}^{i-j}\ \, &\bigl[(5k)(5k+1){c}_{i-j-k}{c}_{j}\prescript{\phantom{0}}{3}{{d}_{k}}
        \\[-1em]
        &+2(5k+1)(5k+2){c}_{i-j-k}\prescript{\phantom{0}}{3}{{d}_{j}}{c}_{k}\\
        &
        +(5j+1)(5k+1)\prescript{\phantom{0}}{3}{{d}_{i-j-k}}{c}_{j}{c}_{k}\\
        &+2(5k)(5j+1){c}_{i-j-k}{c}_{j}\prescript{\phantom{0}}{3}{{d}_{k}}\\
        &+(5k+1){c}_{i-j-k}{c}_{j}{c}_{k} \bigr].
    \end{aligned}
\end{equation}

We then pick the unique solutions $z_1$, $z_2$, and $z_3$ of the linearized Painlev\'e equations~\eqref{eq:zi2PLT} whose asymptotic expansions are, respectively, given by Eqs.~\eqref{2PLTdiffeq1}, \eqref{2PLTdiffeq2}, and \eqref{2PLTdiffeq3}. Those solutions are universal in the sense that they do not depend upon the primary black hole's spin ${\mathring\chi}$, but the solution to the full differential equation~\eqref{z2PLT} is a ${\mathring\chi}$-dependent linear combination of the $z_i$; refer back to Eq.~\eqref{ziLC}. The behavior of this solution is shown in Fig.~\ref{fig:y0PLT}. 

To obtain the physical forcing function of the orbital frequency, $\check F_{[2]}^{\Delta\check\Omega}(\Delta\check\Omega)$, we rescale the solution $z(u)$ using Eq.~\eqref{F0PLTrenomalization}.

\subsection{Composite forcing functions and mode amplitudes}\label{sec:composite}

By solving the equations in the inspiral and transition regime, we obtain dynamics that are valid in each separate regime. As is standard in the method of matched asymptotic expansions~\cite{KevorkianCole}, one can obtain a uniformly accurate approximation in both regimes by constructing a composite forcing function $\check F_{(0)[2]}^{\check{\Omega}}$ that smoothly interpolates between the 0PA-accurate forcing function $d\check\Omega/d\check{t}=\check\nu\check F_{(0)}^{\check{\Omega}}+\mathcal{O}(\check{\nu}^2)$ and the 2PLT-accurate forcing function $d\check\Omega/d\check{t}=\check\sigma^3\check F_{[0]}^{\Delta\check\Omega}+\check\sigma^5\check F_{[2]}^{\Delta\check\Omega}+\mathcal{O}(\check{\sigma}^6)$. 

The composite forcing function is simply given by the sum of the inspiral and transition forcing functions, minus the common terms that would otherwise be counted twice~\cite{Kuchler:2023jbu,Kuchler:2024esj}:
\begin{align}\label{0PA2PLTcomposite}
    \check F_{(0)[2]}^{\check\Omega}(\check\Omega)&={\check\sigma}^5 \check F_{(0)}^{\check\Omega}(\check\Omega)\nonumber\\
    &\phantom{}+{\check\sigma}^3 \check F_{[0]}^{\Delta\check\Omega}\left(\frac{\check\Omega-\check\Omega_\star}{{\check{\sigma}}^2}\right)+{\check\sigma}^5 \check F_{[2]}^{\Delta\check\Omega}\left(\frac{\check\Omega-\check\Omega_\star}{{\check{\sigma}}^2}\right)\nonumber\\
    &-{\check\sigma}^5\left(\frac{\check F_{(0)}^{(3,-1)}}{\check\Omega-\check\Omega_\star}+\check F_{(0)}^{(5,0)}\right). 
\end{align}
We recall that the coefficients $\check F_{(0)}^{(3,-1)}$ and $\check F_{(0)}^{(5,0)}$ are given explicitly in Eqs.~\eqref{eq:asymptoticOLD}. Such a composite solution is guaranteed, from the theory of matched asymptotic expansions, to be uniformly convergent, with an $o(\check\nu)$ residual  in the entire interval $\check\Omega\in[0,\check\Omega_\star)$, as $\check\nu$ goes to zero. Its basic property is that far from the ISCO, it reduces to the 0PA forcing function plus small, $o(\check\nu)$ residuals; and near the ISCO, it reduces to the 2PLT forcing function plus small, $o(\check\nu)$ residuals. However, as pointed out in Ref.~\cite{Kuchler:2024esj}, the residuals, while subdominant in $\check\nu$, are numerically large and spoil the accuracy of the solution in the inspiral.

First we examine the near-ISCO behavior. Here the first two terms of the near-ISCO expansion of the 0PA forcing function~\eqref{eq:asymptotic0PA} approximately cancel with the last two terms of Eq.~\eqref{0PA2PLTcomposite}, leaving an inspiral residual $\check{F}_{(0)}^{(7,1)}\check\sigma^7\Delta\check\Omega + O(\check\sigma^8)$. The near-ISCO behavior of the composite solution is then
\begin{equation}\label{residuallate}
    \check F_{(0)[2]}^{\check\Omega}(\check\Omega)= {\check\sigma}^3 \check F_{[0]}^{\Delta\check\Omega} (\Delta\check\Omega) + {\check\sigma}^5 \check F_{[2]}^{\Delta\check\Omega}(\Delta\check\Omega) + \mathcal{O}({\check\sigma}^7),
\end{equation}
and the relative error between the composite forcing function and the 2PLT forcing function converges as $\mathcal{O}({\check\sigma}^4)$: 
\begin{equation}\label{0PA2PLTERlate}
    \left|\frac{\check F_{(0)[2]}^{\check\Omega}-\left({\check\sigma}^3 \check F_{[0]}^{\Delta\check\Omega}+{\check\sigma}^5 \check F_{[2]}^{\Delta\check\Omega}\right)}{{\check\sigma}^3 \check F_{[0]}^{\Delta\check\Omega}+{\check\sigma}^5 \check F_{[2]}^{\Delta\check\Omega}}\right|=\mathcal{O}({\check\sigma}^4). 
\end{equation}
In this region, the residual converges to zero at the ISCO, ensuring it is numerically small in the transition region
and does not spoil the transition behavior; see Fig.~\ref{fig:CompositeForcingTerm}.

\begin{figure}
\includegraphics[width=.48\textwidth]{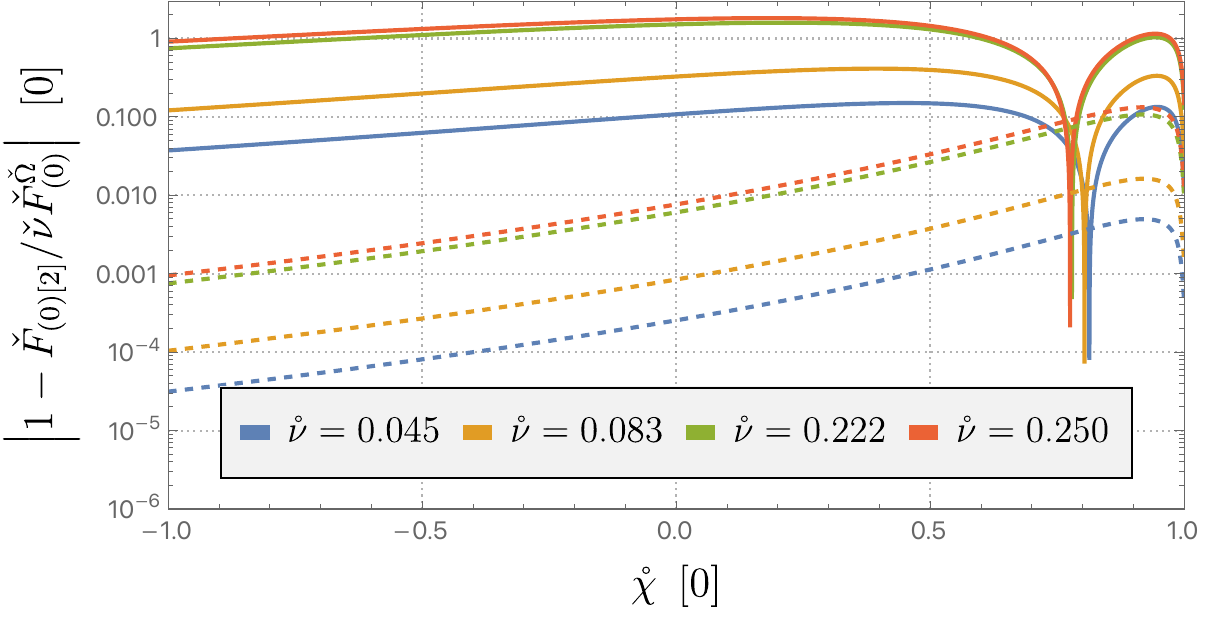}
\caption{Solid curves: relative error between the 0PA-2PLT composite~\eqref{0PA2PLTcomposite} and the 0PA forcing functions~\eqref{FOmega0PA} when the two bodies are separated by twice the ISCO radius, $2r_\star(\mathring\chi)$, for different values of the mass ratio. Dashed curves: the same quantity but using the re-expanded 0PA-2PLT forcing function~\eqref{eq:0PA2PLTcomposite_reexp}.}
\label{fig:2PLTresidualsearly}
\end{figure}

On the other hand, in the early inspiral, when $\Delta\check\Omega \rightarrow -\infty$, the first terms in the asymptotic expansions of the 0PLT~\eqref{eq:asymptotic0PLT} and 2PLT~\eqref{eq:asymptotic2PLT} forcing functions approximately cancel with the last two terms of Eq.~\eqref{0PA2PLTcomposite}. Therefore, the first terms of Eqs.~\eqref{eq:asymptotic0PLT} and \eqref{eq:asymptotic2PLT} that do not cancel are, respectively, $\check{F}_{[0]}^{(3,-6)}\check\sigma^{15}(\check\Omega-\check\Omega_\star)^{-6}$ and $\check{F}_{[2]}^{(5,-5)}\check\sigma^{15}(\check\Omega-\check\Omega_\star)^{-5}$. The early-inspiral behavior of the composite solution is therefore
\begin{equation}\label{residualearly}
    \check F_{(0)[2]}^{\check\Omega}(\check\Omega)= {\check\nu} \check F_{(0)}^{\check\Omega}(\check\Omega)+\mathcal{O}({\check\nu}^3),
\end{equation}
and the relative error between the composite forcing function and the 0PA forcing function converges as $\mathcal{O}({\check\nu}^{2})$:
\begin{equation}\label{0PA2PLTERearly}
    \left|\frac{\check F_{(0)[2]}^{\check\Omega}-{\check\nu} \check F_{(0)}^{\check\Omega}}{{\check\nu} \check F_{(0)}^{\check\Omega}}\right|=\mathcal{O}({\check\nu}^2). 
\end{equation}
This relative error is shown in Fig.~\ref{fig:2PLTresidualsearly} for different values of the mass ratio. We have checked that the relative error converges correctly, as ${\check\nu}^2$, but we observe that the 2PLT forcing function numerically spoils the composite solution for comparable-mass binaries. Indeed, for mass ratios between unity and 1:10 (${\check\nu}= 0.25$ to ${\check\nu}= 0.083$), the transition residual amounts to more than $10\%$, and sometimes well over $100\%$, of the composite forcing function during the early inspiral, where the 0PA forcing function should numerically dominate the entire evolution.

Analogously to our treatment of the forcing functions, we can write the composite mode amplitudes $\check H_{\ell m}^{(1)[7]}$ that smoothly interpolate between 0PA-accurate mode amplitudes $\check H_{\ell m}=\check \nu\check H_{\ell m}^{(1)}+\mathcal{O}(\check{\nu}^2)$ and the 2PLT-accurate mode amplitudes $\check H_{\ell m}=\check \sigma^5\check H_{\ell m}^{[5]}+\check\sigma^7\check H_{\ell m}^{[7]}+\mathcal{O}(\check{\sigma}^8)$:
\begin{align}\label{Hlm composite}
    \check H_{\ell m}^{(1)[7]} =&\, \check\sigma^5 \check H_{\ell m}^{(1)} + \check\sigma^5 \check H_{\ell m}^{[5]} + \check\sigma^7 \check H_{\ell m}^{[7]}\nonumber\\
    &- \check\sigma^5 \left[\check H_{\ell m}^{(1)}(\check\Omega_\star) + (\check\Omega-\check\Omega_\star)\partial_{\check\Omega}\check H_{\ell m}^{(1)}(\check\Omega_\star)\right]\nonumber\\
    =&\, \check\sigma^5 \check H_{\ell m}^{(1)}.
\end{align}
In the last equality we have used the matching conditions~\eqref{PAPLTMatchingHlm}. Crucially, the composite solution simply reduces to the inspiral's first-order mode amplitudes  (a consequence of the fact that at the leading two orders in the transition approximation, the amplitudes are simply Taylor expansions of the inspiral amplitudes around the ISCO frequency). Hence, we do not incur any error in the inspiral in this case.

\section{Curing the composite expansion}\label{sec:reexpansion}

In this section we show how the issue we have described in the previous section, concerning the poor accuracy of the 0PA-2PLT composite solution in the inspiral, can be resolved through a coordinate transformation on phase space in the transition-to-plunge regime. 

In order to motivate the transformation, we start by further explicating the origin of the 0PA-2PLT composite solution's inaccuracy. Using Eqs.~\eqref{eq:asymptotic0PLT} and~\eqref{eq:asymptotic2PLT}, we can write the difference between the 0PA-2PLT composite and the 0PA solution as
\begin{align}\label{earlytime_residual}
    \check F_{(0)[2]}^{\check\Omega} - {\check\sigma}^5\check F_{(0)}^{\check\Omega} =&\, {\check\sigma}^3\left[\frac{\check F^{(3,-6)}_{[0]}}{\Delta\check\Omega^6} + \mathcal{O}(\Delta\check\Omega^{-11})\right]\nonumber\\
    &+ {\check\sigma}^5\left[\frac{\check F^{(5,-5)}_{[2]}}{\Delta\check\Omega^5} + \mathcal{O}(\Delta\check\Omega^{-10})\right].
\end{align}
For the composite expansion to be accurate in the inspiral, we require this to be small. For sufficiently small $\check\nu$, it \emph{is} manifestly small and vanishes uniformly at the rate that is guaranteed by the construction of the composite expansion. However, at any fixed, finite $\check\nu$, and ignoring the magnitude of the numerical coefficients, the residual~\eqref{earlytime_residual} only vanishes in limit $\Delta\check\Omega=(\check\Omega-\check\Omega_\star)/\check\sigma^2\to-\infty$. This limit is formally an early-time limit in the transition because it implies $(\check\Omega_\star-\check\Omega)\gg \check\sigma^2$, meaning the frequency should lie at the low end of or below the transition regime. Such a fact \emph{should} ensure that the residuals in Eq.~\eqref{earlytime_residual} are numerically small once we are sufficiently far outside the transition regime. However, there is a basic problem: $|\Delta\check\Omega|$ is \emph{not} large, and $1/\Delta\check\Omega$ is not small, \emph{anywhere} except in the case of exceedingly small mass ratios. For example, even for $\check\nu=10^{-3}$, $|\Delta\check\Omega|$ only becomes equal to unity around $\check\Omega\approx0$; i.e., at the earliest history of the inspiral. For mass ratios closer to unity, this occurs at unphysical, \emph{negative} frequencies. Worse, for mass ratios above $\check \nu=1/50$, $\Delta\check\Omega$ is numerically small over the entire physical range of frequencies. A more relevant comparison is $\Delta\check\Omega$'s size relative to the inspiral variable $\check\Omega$: for mass ratios above $\check \nu=1/10$, $\Delta\check\Omega$ is comparable to $\check\Omega$ over most of the inspiral.

In short, our variable $\Delta\check\Omega$ is poorly chosen. It is meant to be of order unity in the transition regime and large outside that regime, but instead it is comparable to the unscaled variable $\check\Omega$ over much of the binary evolution, except at extremely small mass ratios. Because of this, the residual~\eqref{earlytime_residual} is numerically comparable to the 0PA forcing function for most moderate mass ratios, spoiling the numerical accuracy of the composite solution, as described below Eq.~\eqref{0PA2PLTERearly} and shown in Fig.~\ref{fig:2PLTresidualsearly}.

Here we address this in the following way: we define a new variable on phase space,
\begin{equation}\label{defDObar}
    \overline{\Delta\Omega} \coloneqq \frac{\Delta\check\Omega}{\check\Omega/\check\Omega_\star} = \frac{\Delta\check\Omega}{1 + \frac{{\check\sigma}^2\Delta\check\Omega}{\check\Omega_\star}}.
\end{equation}
This variable remains of order unity in the transition regime, as required for the transition expansion, and closely approximates $\Delta\check\Omega$ there. Technically, it is of order unity over most of the inspiral (except in the case of very small mass ratios), but again, the more relevant comparison is to the frequency variable of the inspiral, $\check\Omega$. In that comparison, we capture the desired behavior: even at equal mass, $|\overline{\Delta\Omega}|$ is significantly larger than $\check\Omega$ over most of the inspiral. Crucially, by construction, $\overline{\Delta\Omega}$ also never becomes unphysical: the early-time limit $\check\Omega\to0$  of the inspiral corresponds to $\overline{\Delta\Omega}\to-\infty$ for \emph{any} value of $\check\sigma$. This ensures the residual~\eqref{earlytime_residual} vanishes in the infinite past for all $\check\sigma$, not only for $\check\sigma\to0$. 

We now introduce a new set of phase-space coordinates, $\overline{\Delta J^a}=\{\overline{\Delta\Omega}, {\delta \check M}_\text{tot}, {\delta\check \chi}\}$, with $\overline{\Delta\Omega}$ defined in Eq.~\eqref{defDObar}. While the inspiral's post-adiabatic expansion is unaltered by this change of coordinates, the transition dynamics gets re-expanded. The expansion of the transition forcing functions for the orbital frequency in powers of $\check\sigma$ at fixed $\overline{(\Delta) J^a}$ reads
\begin{equation}
    \frac{d\check\Omega}{d{\check t}}(\overline{\Delta J^a},{\check\sigma}) = \check\sigma^3 \sum_{n=0}^\infty \check\sigma^n \bar{F}_{[n]}^{\overline{\Delta \Omega}}(\overline{\Delta J^a}).
\end{equation}
We can relate the forcing functions $\bar{F}^{\overline{\Delta\Omega}}_{[n]}$ and $\check{F}^{\Delta\check\Omega}_{[n]}$ by re-expanding the power series given in Sec.~\ref{sec:PLTexpansion} at fixed $\overline{\Delta J^a}$ after substituting
\begin{align}\label{DODOr}
   \Delta\check\Omega=\frac{\overline{\Delta\Omega}}{1-\frac{{\check\sigma}^2\overline{\Delta\Omega}}{\check\Omega_\star}} 
\end{align}
and expanding for small $\check\sigma$. Up to 2PLT order, we obtain
\begin{subequations}\label{eq:PLTReexpansionBar}
\begin{align}
    \bar{F}^{\overline{\Delta\Omega}}_{[0]}(\overline{\Delta\Omega}) =&\, \check{F}^{\Delta\check\Omega}_{[0]}(\overline{\Delta\Omega}),
    \\
    \bar{F}^{\overline{\Delta\Omega}}_{[1]}(\overline{\Delta\Omega}) =&\, \check{F}^{\Delta\check\Omega}_{[1]}(\overline{\Delta\Omega}) = 0,
    \\
    \bar{F}^{\overline{\Delta\Omega}}_{[2]}(\overline{\Delta\Omega}) =&\, \check{F}^{\Delta\check\Omega}_{[2]}(\overline{\Delta\Omega}) + \frac{\overline{\Delta\Omega}^2}{\check\Omega_\star} \partial_{\overline{\Delta\Omega}}\check{F}^{\Delta\check\Omega}_{[0]} (\overline{\Delta\Omega}).
\end{align}
\end{subequations}
Similarly, for the mode amplitudes $\bar H_{\ell m}^{[n]}$ (i.e., the coefficients of $\check\sigma^n$ in a transition expansion of $H_{\ell m}$ at fixed $\overline{\Delta J^a}$), we have
\begin{subequations}\label{eq:PLTReexpansionHlmbar}
\begin{align}
    \bar{H}_{\ell m}^{[5]}(\overline{\Delta J^a}) &= \check{H}_{\ell m}^{[5]}(\overline{\Delta J ^a}),
    \\
    \bar{H}_{\ell m}^{[7]}(\overline{\Delta\Omega}) &= \check{H}_{\ell m}^{[7]}(\overline{\Delta\Omega}).
\end{align}
\end{subequations}
Lengthier relationships at higher order are given in Appendix~\ref{app:re-expansions}. 

An analogous re-expansion can then also be applied to the composite forcing function~\eqref{0PA2PLTcomposite}, yielding 
\begin{align}\label{eq:0PA2PLTcomposite_reexp}
    \bar F_{(0)[2]}^{\check\Omega}(\check\Omega)&={\check \sigma}^5 \check F_{(0)}^{\check\Omega}(\check\Omega)\nonumber
    \\
    &+{\check\sigma}^3 \bar F_{[0]}^{\overline{\Delta\Omega}}\left(\frac{\check\Omega-\check\Omega_\star}{{\check{\sigma}}^2 \check\Omega/\check\Omega_\star}\right) + {\check\sigma}^5 \bar F_{[2]}^{\overline{\Delta\Omega}}\left(\frac{\check\Omega-\check\Omega_\star}{{\check{\sigma}}^2 \check\Omega/\check\Omega_\star}\right)\nonumber
    \\
    &-{\check\sigma}^5\left(\frac{\bar F_{[0]}^{(3,-1)}}{\check\Omega-\check\Omega_\star}\frac{\check\Omega_\star}{\check\Omega} + \bar F_{[2]}^{(5,0)}\right),
\end{align}
where
\begin{subequations}\label{eq:coefbar}
\begin{align}
    &\bar F_{[0]}^{(3,-1)}=\check F_{[0]}^{(3,-1)},\\
    &\bar F_{[2]}^{(5,0)}=\check F_{[2]}^{(5,0)} - \frac{\check F_{[0]}^{(3,-1)}}{\check\Omega_\star}.
\end{align}
\end{subequations}
The composite solution's residual error in the inspiral is given by
\begin{align}\label{earlytime_residualReexp}
    \bar F_{(0)[2]}^{\check\Omega} -{\check\sigma}^5\check F_{(0)}^{\check\Omega} =&\, {\check\sigma}^3\left[\frac{\bar F^{(3,-6)}_{[0]}}{\overline{\Delta\Omega}^6}+\mathcal{O}(\overline{\Delta\Omega}^{-11})\right]\nonumber\\
    &+ {\check\sigma}^5\left[\frac{\bar F^{(5,-5)}_{[2]}}{\overline{\Delta\Omega}^5} + \mathcal{O}(\overline{\Delta\Omega}^{-10})\right],
\end{align}
with
\begin{subequations}
\begin{align}
    \bar F^{(3,-6)}_{[0]} &= \check F^{(3,-6)}_{[0]} \\
    \bar F^{(5,-5)}_{[2]} &= \check F^{(5,-5)}_{[2]} - \frac{6}{\check\Omega_\star}\check F^{(3,-6)}_{[0]}.
\end{align}
\end{subequations}
This residual now vanishes as $\check\Omega\to0$, independently of the value of $\check\sigma$. The composite mode amplitudes~\eqref{Hlm composite} remain unchanged since they are simply given by the inspiral's first-order mode amplitudes.

\begin{figure*}
\includegraphics[width=.48\textwidth]{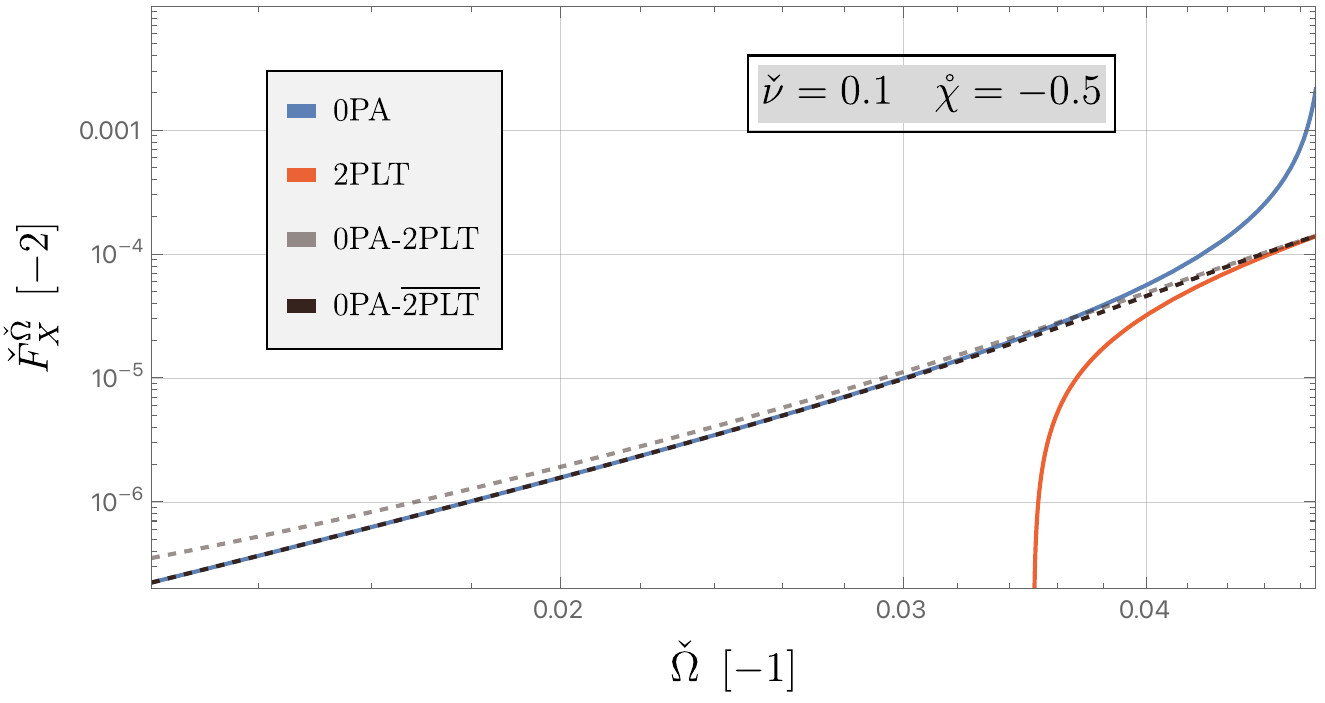}
\includegraphics[width=.48\textwidth]{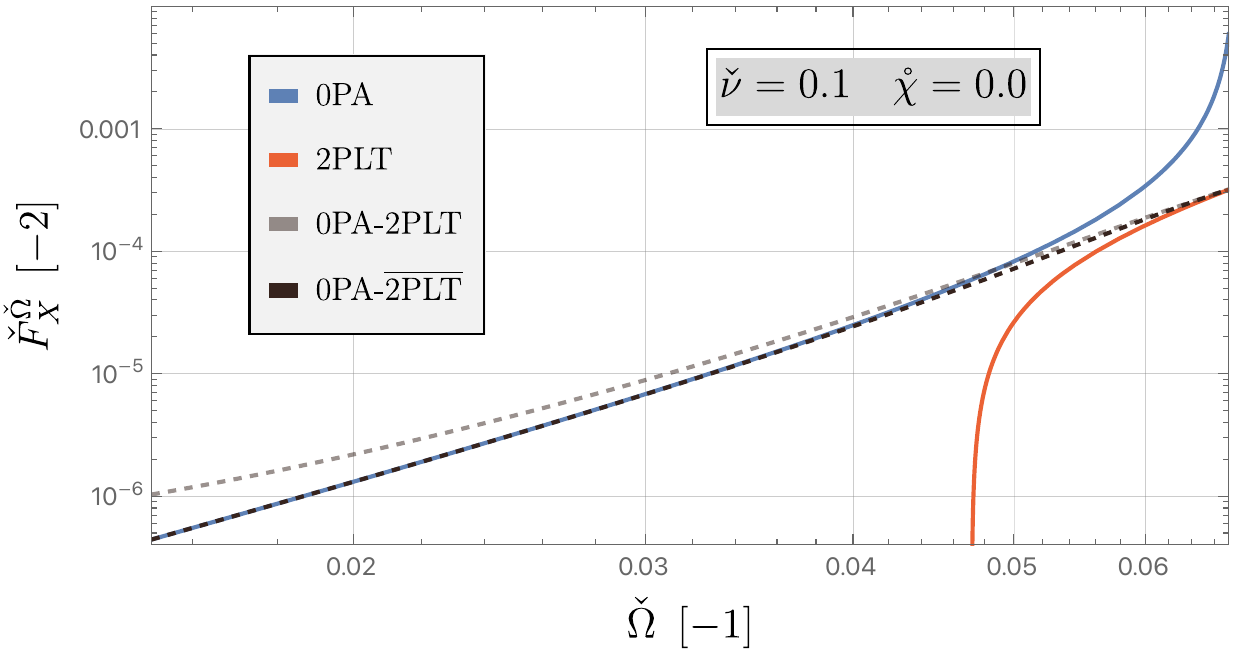}
\includegraphics[width=.48\textwidth]{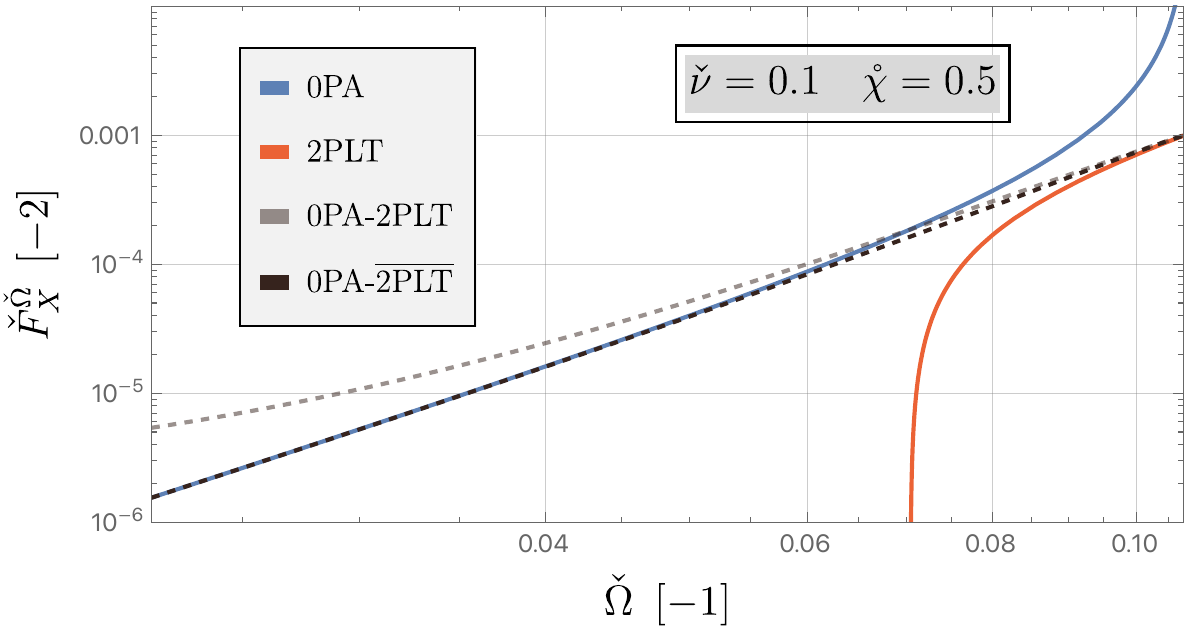}
\includegraphics[width=.48\textwidth]{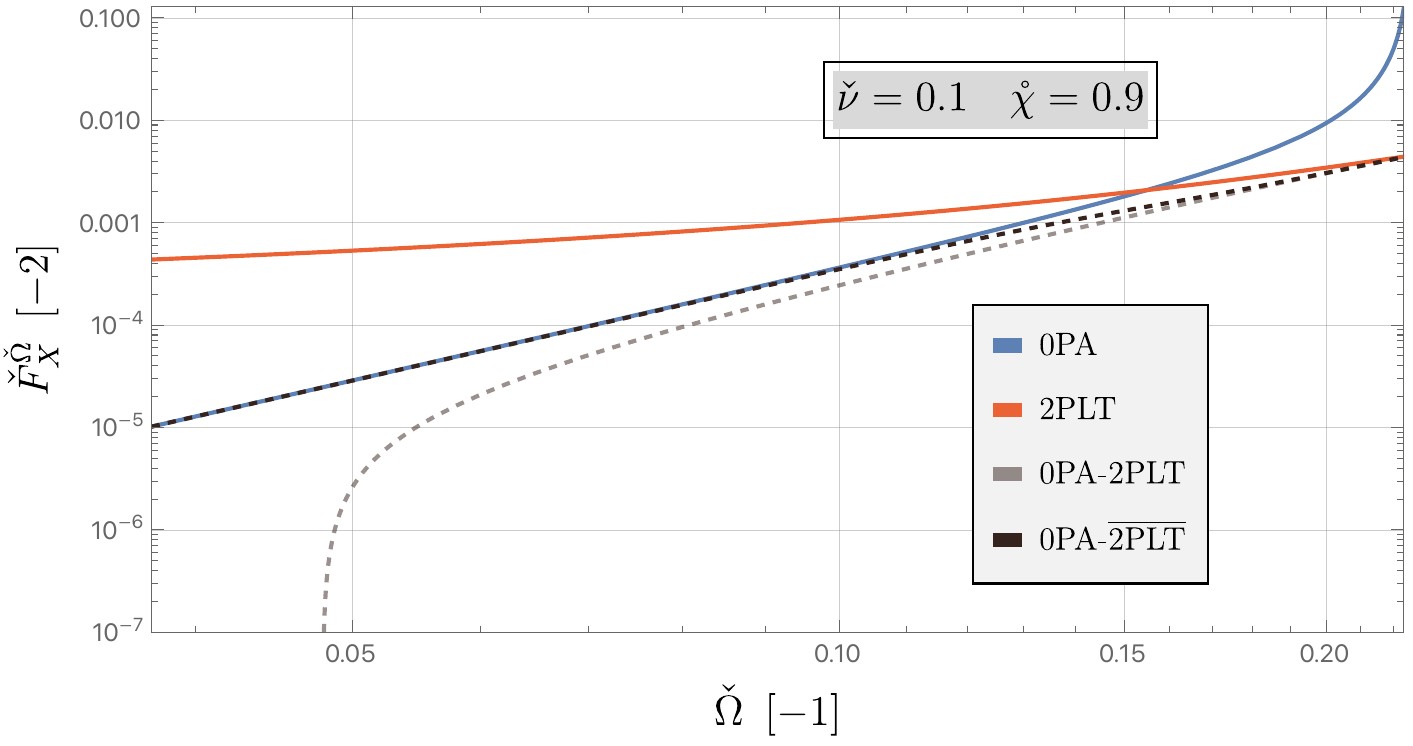}
\caption{Comparisons between the composite 0PA-2PLT forcing function~\eqref{eq:0PA2PLTForcingTerm} (dashed gray lines) and the re-expanded composite 0PA-2PLT forcing function~\eqref{eq:0PA2PLTReexpForcingTerm} (dashed black lines). We also display in blue the 0PA forcing function~\eqref{eq:0PAForcingTerm} and in red the 2PLT forcing function~\eqref{eq:2PLTForcingTerm}. The value of the orbital frequency at the right boundary of each frame corresponds to the ISCO frequency $\check\Omega_\star(\mathring{\chi})$. We have chosen a mass ratio of ${\check\nu}=0.1$ and spin of the primary of $\mathring{\chi}=-0.5$ (top left), $\mathring{\chi}=0$ (top right), $\mathring{\chi}=0.5$ (bottom left), and $\mathring{\chi}=0.9$ (bottom right). For all scenarios, the change of variables to $\overline{\Delta J^a}$ significantly improves the early time agreement of the composite forcing function with the adiabatic one. This behavior is particularly marked in the bottom-right panel, where the 0PA-2PLT composite undergoes a zero crossing at $\check\Omega\approx0.048$; in this form of the composite, the frequency \emph{decreases} with time at low frequencies.}
\label{fig:CompositeForcingTerm}
\end{figure*}

In Figs.~\ref{fig:2PLTresidualsearly} and \ref{fig:CompositeForcingTerm} we show how the change of phase-space coordinates improves the composite solution's behavior by orders of magnitude in the inspiral. In Fig.~\ref{fig:CompositeForcingTerm}, for example, we see that the original composite can differ from the 0PA forcing function by orders of magnitude, while the composite formulated in terms of $\overline{\Delta\Omega}$ hews closely to the 0PA curve over most of the inspiral. 

\section{Numerical implementation}\label{sec:implementation}

We now summarize the waveform-generation scheme in each of our transition and composite models.

\subsection{Models}\label{subsec:models}

The two main models of this paper are the re-expanded 2PLT (which hereafter we will refer to as $\overline{\text{2PLT}}$) and the re-expanded composite 0PA-2PLT (which hereafter we will refer to as 0PA-$\overline{\text{2PLT}}$) models, whose forcing functions have been derived in Sec.~\ref{sec:reexpansion}. In Sec.~\ref{sec:results}, we shall compare them to their non-re-expanded companions, the 2PLT and the 0PA-2PLT composite models, and assess the importance of the change of coordinates introduced in Sec.~\ref{sec:reexpansion}. We shall also assess the importance of including the 2PLT corrections as compared to a leading-order 0PLT model. In this section, we instead describe the content of each of the six waveform models we consider and how we numerically implement them. We postpone the qualitative comparison with NR to Sec.~\ref{sec:results}.

Our use of a multiscale approach on phase space naturally divides the waveform-generation scheme into two steps. First, an offline step which is slow but needs to be performed only once: all ingredients for waveform generation are pre-computed and stored as functions on phase space. Second, a fast online step, which consists of a rapid evolution through phase space once the parameters of the system (in this case, the mass ratio $\mathring\varepsilon$ and the spin of the primary $\mathring\chi$) have been specified. For the orders at which we are working, the primary black hole mass and spin remain constant, and their associated forcing functions vanish. Hence, a model $X$ is fully specified by its orbital frequency forcing function, $\check F^{\check\Omega}_X$, as well as its $(\ell,m)$-mode amplitudes of the GW strain at infinity, $\check H^X_{\ell m}$. Indeed, once the evolution equations through phase space have been solved for $\check\Omega(\check t)$ and $\phi_p(\check t)$, one can generate the $(\ell,m)$-mode contribution to the wavestrain as
\begin{align}\label{waveform}
    \check h^X_{\ell m}(\check t) = \check H^X_{\ell m}(\check\Omega(\check t))e^{-i m \phi_p(\check t)}.
\end{align}
We explicitly state the forcing functions and mode amplitudes used for each model below.

\bigskip

\noindent\textbf{0PA model.} The orbital-frequency forcing function is given by
\begin{equation}\label{eq:0PAForcingTerm}
    \check F^{\check\Omega}_\text{0PA}(\check\Omega) = \check \nu \check F_{(0)}^{\check\Omega}(\check \Omega),
\end{equation}
and the mode amplitudes are given by
\begin{align}\label{amplitudes0PA}
    \check H_{\ell m}^\text{0PA}(\check\Omega) = \check\nu \check H_{\ell m}^{(1)}(\check\Omega).
\end{align}
These are simply the inspiral's adiabatic forcing function and mode amplitudes, respectively; see Eqs.~\eqref{FOmega0PA}, \eqref{eq:F0PAreexp}, \eqref{eq:HlmPA} and \eqref{eq:HlmPAreexp}.

\bigskip

\noindent\textbf{0PLT model.} The orbital frequency is driven by the 0PLT forcing function
\begin{equation}\label{F0PLT}
    \check F_\text{0PLT}^{\check\Omega}(\check\Omega)=\check\nu^{3/5} \check F_{[0]}^{\Delta\check\Omega}\left(\frac{\check\Omega-\check\Omega_\star}{\check\nu^{2/5}}\right);
\end{equation}
see Eqs.~\eqref{eq:F0PLT} and \eqref{eq:PLTReexpansion1}. We use the 2PLT mode amplitudes (as the 0PLT mode amplitudes would simply be a constant),
\begin{equation}
    \check H_{\ell m}^\text{0PLT}(\check\Omega)=\check\nu \left[ \check H^{[5]}_{\ell m}+ \check \nu^{2/5}\check H^{[7]}_{\ell m} \left(\frac{\check\Omega-\check\Omega_\star}{\check\nu^{2/5}}\right)\right];
\end{equation}
see Eqs.~\eqref{eq:PLTReexpansionHlm} and \eqref{PAPLTMatchingHlm}.

\bigskip

\noindent \textbf{2PLT model.} The forcing function now also contains the subleading 2PLT corrections:
\begin{equation}\label{eq:2PLTForcingTerm}
    \check F_\text{2PLT}^{\check\Omega}(\check\Omega) =\check\nu^{3/5} \check F_{[0]}^{\Delta\check\Omega}\left(\frac{\check\Omega-\check\Omega_\star}{\check\nu^{2/5}}\right)+\check\nu \check F_{[2]}^{\Delta\check\Omega}\left(\frac{\check\Omega-\check\Omega_\star}{\check\nu^{2/5}}\right);
\end{equation}
see Eqs.~\eqref{eq:F0PLT}, \eqref{eq:F2PLT}, and \eqref{eq:PLTReexpansion1}. For the mode amplitudes we once again take the 2PLT mode amplitudes
\begin{equation}
    \check H_{\ell m}^\text{2PLT}(\check\Omega)=\check\nu \left[ \check H^{[5]}_{\ell m}+ \check \nu^{2/5}\check H^{[7]}_{\ell m} \left(\frac{\check\Omega-\check\Omega_\star}{\check\nu^{2/5}}\right)\right].
\end{equation}

\bigskip

\noindent \textbf{$\mathbf{\overline{2\text{PLT}}}$ model. } We use the 2PLT-accurate forcing function and mode amplitudes, but now written in terms of the set of mechanical parameters $\overline{\Delta J^a}$ introduced in Sec.~\ref{sec:reexpansion}. Explicitly, the forcing function is given by 
\begin{multline}\label{F2PLT_reexp}
    \check F_{\overline{\text{2PLT}}}^{\check\Omega}(\check\Omega) =\check\nu^{3/5} \bar F_{[0]}^{\overline{\Delta\Omega}}\left(\frac{\check\Omega-\check\Omega_\star}{\check\nu^{2/5}\check\Omega/\check\Omega_\star}\right)\\
    +\check\nu \bar F_{[2]}^{\overline{\Delta\Omega}}\left(\frac{\check\Omega-\check\Omega_\star}{\check\nu^{2/5}\check\Omega/\check\Omega_\star}\right);
\end{multline}
see Eqs.~\eqref{eq:F0PLT}, \eqref{eq:F2PLT}, \eqref{eq:PLTReexpansion1}, and \eqref{eq:PLTReexpansionBar}. The mode amplitudes, on the other hand, are given by
\begin{equation}
    \check H_{\ell m}^{\overline{\text{2PLT}}}(\check\Omega)=\check\nu \left[ \bar H^{[5]}_{\ell m}+ \check \nu^{2/5}\bar H^{[7]}_{\ell m} \left(\frac{\check\Omega-\check\Omega_\star}{\check\nu^{2/5}\check\Omega/\check\Omega_\star}\right)\right];
\end{equation}
see Eqs.~\eqref{eq:PLTReexpansionHlm}, \eqref{PAPLTMatchingHlm}, and \eqref{eq:PLTReexpansionHlmbar}.

\bigskip
\noindent 
\textbf{Composite 0PA-2PLT model.} In this model, we use the 0PA-2PLT composite forcing function that smoothly interpolates between the 0PA and 2PLT solutions:
\begin{equation}\label{eq:0PA2PLTForcingTerm}
    \check F_{\text{0PA-}{\text{2PLT}}}^{\check\Omega}(\check\Omega) = \check F_{(0)[2]}^{\check \Omega}(\check\Omega);
\end{equation}
see Eq.~\eqref{0PA2PLTcomposite}. The composite mode amplitudes are simply given by the 0PA mode amplitudes,
\begin{align}\label{amplitudes0PA2PLT}
    \check H_{\ell m}^\text{0PA-2PLT}(\check\Omega) = \check\nu \check H_{\ell m}^{(1)}(\check\Omega);
\end{align}
see Eq.~\eqref{Hlm composite}.

\bigskip
\noindent 
\textbf{Composite $\mathbf{\text{0PA-}\overline{\text{2PLT}}}$ model.}
In this model, we use the 0PA-2PLT composite forcing function re-expanded using the mechanical parameters $\overline{\Delta J^a}$. It is given by
\begin{equation}\label{eq:0PA2PLTReexpForcingTerm}
    \check F_{\text{0PA-}\overline{\text{2PLT}}}^{\check\Omega}(\check\Omega) = \bar F_{(0)[2]}^{\check \Omega}(\check\Omega);
\end{equation}
see Eq.~\eqref{eq:0PA2PLTcomposite_reexp}. As for the 0PA-2PLT model, the mode amplitudes are given by
\begin{align}\label{amplitudes0PA2PLTbar}
    \check H_{\ell m}^{\text{0PA-}\overline{\text{2PLT}}}(\check\Omega) = \check\nu \check H_{\ell m}^{(1)}(\check\Omega).
\end{align}

\subsection{Offline and online computations}

For the models described in Sec.~\ref{subsec:models}, the offline and online steps of the waveform-generation procedure decompose as follows:

\vspace{6pt}
\noindent \textbf{Offline step}
\begin{itemize}
    \item For a given value of primary spin $\mathring\chi$, tabulate the first-order mode amplitudes $H^{(1)}_{\ell m}(\Omega)$ on a grid (here working in units of $\mathring M=1$). We compute $H^{(1)}_{\ell m}(\Omega)$ through the up solution of the sourced Teukolsky equation~\eqref{eq:HlmTeukMode} as
    \begin{equation}
        H^{(1)}_{\ell m}(\Omega)=2\frac{\prescript{}{-2}{C}^\texttt{up}_{\ell m}(\Omega)}{(m\Omega)^2}
    \end{equation}
    on a grid of $(\Omega,\mathring\chi)$ values and store it as a 2-dimensional interpolating function. We do this using the \texttt{Teukolsky} Mathematica package~\cite{TeukolskyPackage} from the Black Hole Perturbation Toolkit~\cite{BHPToolkit}. Without any additional computational cost, through the matching conditions~\eqref{PAPLTMatchingHlm} we can also tabulate $H_{\ell m}^{[5]}$ and $H_{\ell m}^{[7]}$.
    
    \item Compute the Teukolsky fluxes appearing in the 0PA forcing function~\eqref{FOmega0PA} on a grid of $(\Omega,\mathring\chi)$ values and  store it as a 2-dimensional interpolating function. Again, we employ the Black Hole Perturbation Toolkit's \texttt{Teukolsky} Mathematica package. By combining Eqs.~\eqref{eq:dE0dt1}, \eqref{eq:orthocond}, and \eqref{eq:fluxbalancelaw} we also tabulate the first-order self-force, $f^t_{(1)}$.
    
    \item Solve the normalized 0PLT differential equation~\eqref{eq:y0PLT} for the solution $y(u)$ whose early-time asymptotic expansion is given by Eq.~\eqref{eq:y0PLTansatz} and with the boundary condition~\eqref{0PLTnormalizedmatchingcondition}.
    
    \item Solve the three normalized 2PLT differential equations~\eqref{eq:zi2PLT} for the solutions $z_i(u)$ ($i=1,2,3$) whose early-time asymptotic expansions are given by Eq.~\eqref{eq:ziExpansion} and with boundary conditions~\eqref{2PLTASmatch}.

    \item Re-expand all quantities  $f\left((\Delta) J^a\right)$ to the relevant  parameters of the model in use ($\Delta\check J^a$ or $\overline{\Delta J^a}$).
\end{itemize}

\vspace{6pt}
\noindent 
\textbf{Online step}
\begin{itemize}
    \item Compute the five coefficients $\alpha$, $\beta$ (from Eq.~\eqref{alphabetadef}) and $\gamma$, $\eta$, $\tau$ (from Eq.~\eqref{gammaetataudef}).
    
    \item Compute the inhomogeneous solution $z(u)$ using Eq.~\eqref{ziLC}.
    
    \item Rescale the solutions $y(u)$ and $z(u)$ to $\check F_{[0]}^{\Delta\check\Omega}(\Delta\check \Omega)$ using Eq.~\eqref{F0PLTrenomalization} and $\check F_{[2]}^{\Delta\check\Omega}(\Delta\check\Omega)$ using Eq.~\eqref{F2PLTrenormalization} (and similarly for the re-expanded models, but with the replacement $\check{\cdot}\mapsto\bar{\cdot}$).
    
    \item Solve the differential equations%
    \begin{subequations}%
   \begin{align}
        \frac{d\phi_p}{d\check t} &= \check\Omega,\\
        \frac{d\check\Omega}{d\check t} &= \check F^{\check\Omega}_X(\check\Omega)
    \end{align}
    \end{subequations}
    for a given model $X$.
    
    \item Generate the $(\ell,m)$ mode of the GW strain at infinity from Eq.~\eqref{waveform}.
\end{itemize}

\section{Results}\label{sec:results}

We now benchmark our implementation against NR simulations from the SXS catalog \cite{SXS:catalog} and assess the impact of the change of transition phase-space coordinates to $\overline{\Delta J^a}$ on the composite waveforms.

We have checked that our code exactly reproduces (within numerical precision) the waveforms from the code used in Ref.~\cite{Kuchler:2024esj} when setting the primary black hole spin to zero.

\begin{figure*}[!htb]
\includegraphics[width=.48\textwidth]{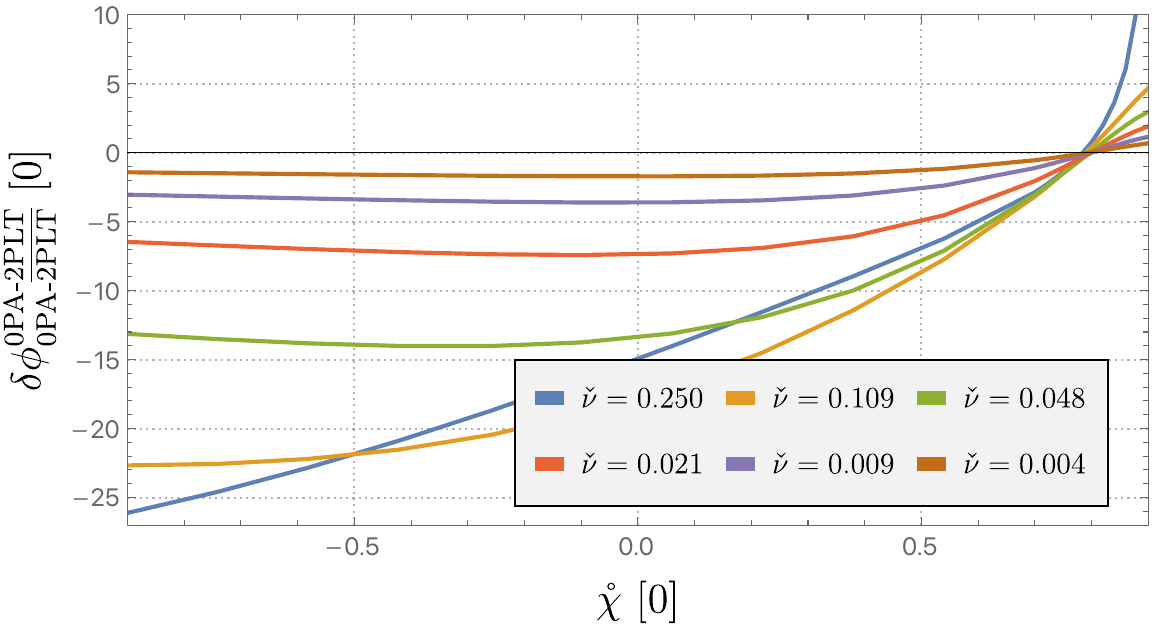}
\hspace{3pt}
\includegraphics[width=.48\textwidth]{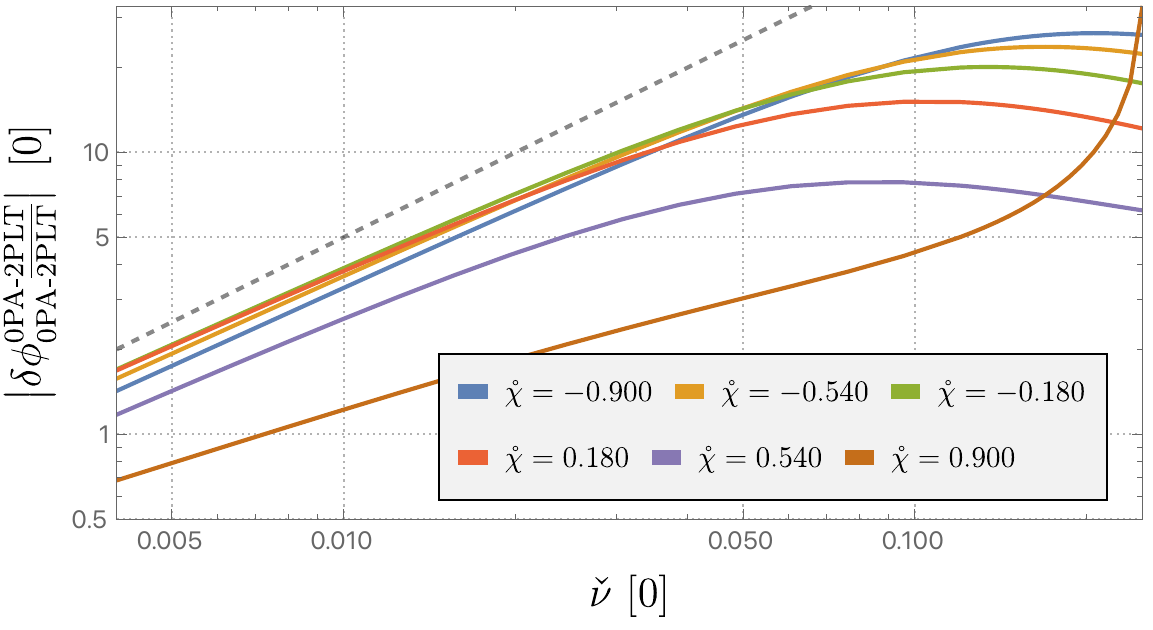}
\caption{Orbital phase dephasing between the 0PA-2PLT model \eqref{eq:0PA2PLTForcingTerm} and the 0PA-$\overline{\text{2PLT}}$ model \eqref{eq:0PA2PLTReexpForcingTerm}; see Eq. \eqref{eq:dephasing}. The frequency range $[\Omega_i,\Omega_f]$ has been fixed such that the secondary black hole starts at twice the ISCO radius and ends at the ISCO radius. That is to say, $\Omega_i=1/\big((2r_\star(\mathring{\chi}))^{3/2}+\mathring{\chi}\big)$ and $\Omega_f=\Omega_\star(\mathring{\chi})$. Left panel: dephasing for fixed values of the dimensionless primary spin $\mathring \chi$ and varying symmetric mass ratio $\check\nu$. Right panel: dephasing for fixed values of the symmetric mass ratio $\check\nu$ and varying primary black hole spin $\mathring\chi$. The reference dashed gray line is a power law $\sim \check\nu^1$.}\label{fig:dephasing}
\end{figure*}

\subsection{Impact of the change of transition-to-plunge coordinates on the composite waveforms}

To measure the impact of the re-expansion to the mechanical parameters $\overline{\Delta J^a}$, we compare the total accumulated orbital phase from both composite models over a frequency range $[\check\Omega_i,\check\Omega_f]$. The model-$X$ accumulated orbital phase reads
\begin{equation}
    \phi_X=\int_{\check\Omega_i}^{\check\Omega_f}\frac{\check{\Omega}}{\check{F}_X^{\check\Omega}}d\check{\Omega},
\end{equation}
and the dephasing between two models $X$ and $Y$ over the same frequency range is then simply
\begin{equation}\label{eq:dephasing}
    \delta\phi^X_Y=\phi_X-\phi_Y=\int_{\check\Omega_i}^{\check\Omega_f}\check{\Omega}\frac{\check{F}_Y^{\check\Omega}-\check{F}_X^{\check\Omega}}{\check{F}_X^{\check\Omega}\check{F}_Y^{\check\Omega}}d\check{\Omega}.
\end{equation}

We show in Fig.~\ref{fig:dephasing} the dephasing \eqref{eq:dephasing} between the 0PA-2PLT and the 0PA-$\overline{\text{2PLT}}$ models as a function of the symmetric mass ratio $\check\nu$ and primary spin $\mathring\chi$ for a secondary black hole orbiting from twice the ISCO radius, $2r_\star(\mathring{\chi})$, to the ISCO radius $r_\star(\mathring{\chi})$. The dephasing converges as $\mathcal{O}(\check\nu)$ with the symmetric mass ratio. This is the expected scaling, as can be seen from the small-mass-ratio expansion of the integrand of Eq.~\eqref{eq:dephasing}, using the early-time asymptotics~\eqref{earlytime_residual} and \eqref{earlytime_residualReexp} as well as the small-mass-ratio expansion of Eq.~\eqref{defDObar}.

Importantly, for intermediate mass ratios $\mathring q \gtrsim 10^{2}$, the correction from the re-expansion to $\overline{\Delta J^a}$ on the total accumulated phase remains relatively high, on the order of one radian over the frequency range $\big[1/\big((2r_\star(\mathring{\chi}))^{3/2}+\mathring{\chi}\big),\Omega_\star(\mathring{\chi})\big]$. This hints towards the fact that the re-expansion of the transition parameters to $\overline{\Delta J^a}$ is a required step for producing a complete IMR
self-force waveform model that meets LISA scientific requirements~\cite{LISA:2024hlh}.

Moreover, the right panel of Fig.~\ref{fig:dephasing} shows that the phase correction due to the re-expansion to $\overline{\Delta J^a}$ is important for any value of the primary spin $\mathring\chi$, except for a narrow window around $\mathring\chi\sim0.78$, where the dephasing undergoes a zero-crossing. This zero-crossing is due to the fact that, around $\mathring\chi\sim0.78$, the early-time difference between the 0PA-2PLT and the 0PA forcing function goes through zero while the 0PA-$\overline{\text{2PLT}}$ forcing function closely matches the 0PA one; see Fig.~\ref{fig:2PLTresidualsearly}. This hints at a second reason why the re-expansion to the $\overline{\Delta J^a}$ variables is a crucial step: the 0PA-2PLT model completely fails to correctly model highly spinning prograde orbits in the early inspiral, as highlighted in the bottom-right panel of Fig.~\ref{fig:CompositeForcingTerm}.

\subsection{Comparison with numerical relativity}\label{subsec:NRComp}

In this section, we compare the different models presented in Sec.~\ref{subsec:models} with NR simulations from the SXS catalog~\cite{SXS:catalog}. Once the waveform of a given model $X$ has been generated through the online step outlined in Sec.~\ref{subsec:models}, we align it with the corresponding NR waveform using the alignment procedure described in Ref.~\cite{Honet2025TBA}. The alignment procedure consists of two steps: first, we find the time shift $\Delta\check t_{\text{min}}$ that minimizes the squared error of half of the $(2,2)$ waveform frequency $\check\omega=-(\partial_t \text{Im}(\check h_{22})\text{Re}(\check h_{22})-\partial_t \text{Re}(\check h_{22})\text{Im}(\check h_{22}))/(2|\check{h}_{22}|^2)$,
\begin{align}\label{eq:SE}
    SE(\Delta {\check t})\coloneqq\int_{\check t_0-\delta {\check t}/2}^{\check t_0+\delta {\check t}/2} dt \left( \check\omega_\text{NR}({\check t}) -\check\omega_X({\check t} + \Delta {\check t})\right)^2,
\end{align}
over a time interval $[\check t_0-\delta {\check t}/2,\check t_0+\delta {\check t}/2]$. Second, we compute the phase shift $\Delta\varphi$ as the average phase difference between the NR and model $X$ waveforms over that same time interval,
\begin{align}
    \Delta\varphi_\text{min} \coloneqq -\frac{1}{\delta {\check t}}\, \int_{\check t_0-\delta {\check t}/2}^{\check t_0+\delta {\check t}/2} dt \,  \text{arg}\left( \frac{\check h_{22}^X({\check t}+\Delta \check t_\text{min})}{\check h_{22}^\text{NR}({\check t})} \right).    
    \end{align}
The model-$X$ aligned waveform is finally given by
\begin{equation}\label{eq:halign}
    \check h_{\ell m}^X(\check t+\Delta \check t_\text{min})e^{-i\Delta\varphi_\text{min}}.
\end{equation}

In our comparisons, we have not observed any notable difference in the performance of our waveform models for different mode numbers. Hence, we only present results for the dominant, $(\ell,m)=(2,2)$ mode.

\begin{figure*}\includegraphics[width=.9\textwidth]{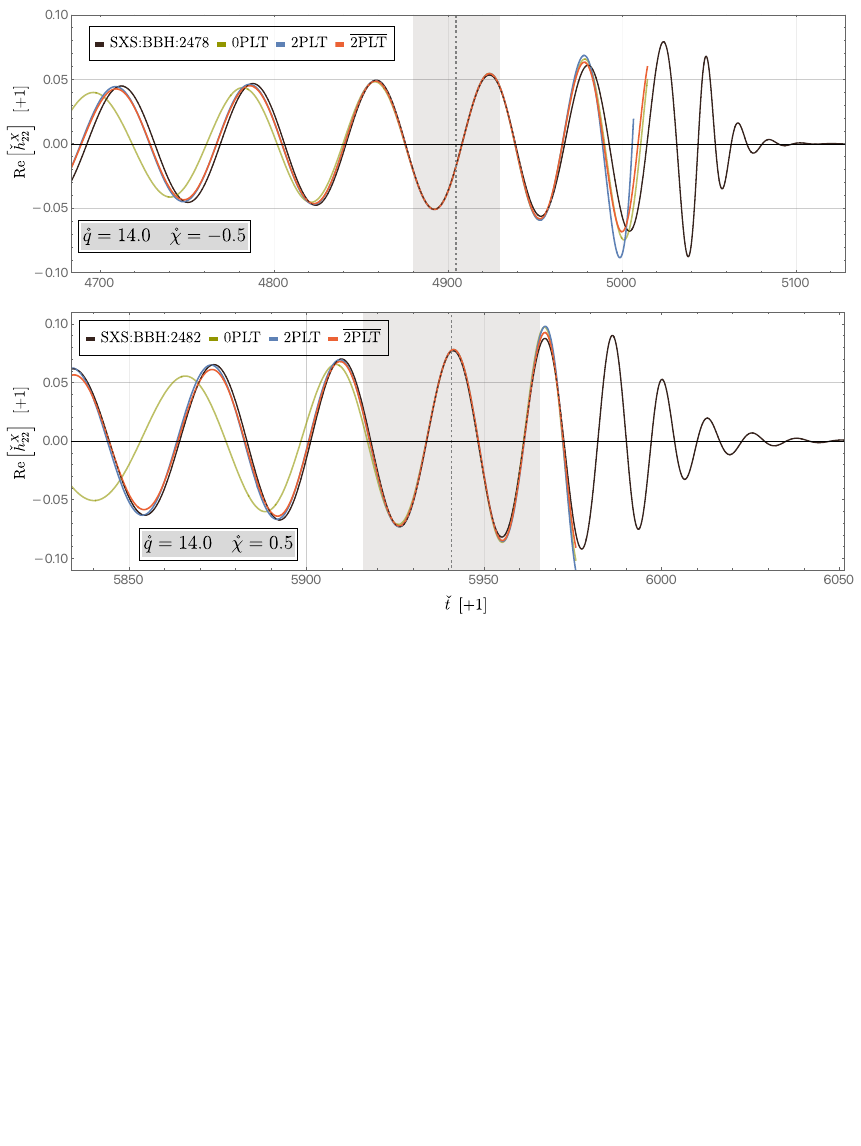}\vspace{-10cm}
\caption{$(\ell,m)=(2,2)$ mode of the gravitational waveform from a binary with large mass ratio $\mathring q=14$ and primary spin $\mathring\chi=-0.5$ (retrograde orbit) or $\mathring\chi=+0.5$ (prograde orbit). The solid black curve shows the NR waveforms SXS:BBH:2478 and SXS:BBH:2482~\cite{sxs_collaboration_2024_13147581,Yoo:2022erv}. The colored curves show the transition-to-plunge waveforms obtained from our three transition models, which are aligned with the NR waveform near the ISCO frequency. The vertical dashed gray line indicates the time at which the NR waveform frequency reaches twice the ISCO orbital frequency, $\omega_\text{NR}=-2\Omega_\star(\mathring{\chi})$, and the gray rectangle shows the time window $[\check t_0-\delta {\check t}/2,\check t_0+\delta {\check t}/2]$ used for the alignment procedure described in Sec.~\ref{subsec:NRComp}.}\label{fig:PLTRetro}
\end{figure*}

We first assess the behavior of our transition models: 0PLT, 2PLT, and $\overline{\text{2PLT}}$. Figure~\ref{fig:PLTRetro} shows these models alongside an NR waveform, where all waveforms are aligned in a window around the ISCO frequency. The large mass ratio is $\mathring q=14$ and the primary's spin is either $\mathring\chi=-0.5$ (retrograde orbit) or $\mathring\chi=+0.5$  (prograde orbit). The 2PLT model dephases from the NR waveform drastically less than the 0PLT model on the left of the ISCO (at early times), while on the right of the ISCO (at late times), the 0PLT model performs slightly better than the 2PLT model. This is consistent with the findings of Ref.~\cite{Kuchler:2024esj} for nonspinning binaries. Moreover, for the same number of cycles around the ISCO, the prograde 2PLT waveform dephases somewhat less than the retrograde 2PLT waveform. Finally, the $\overline{\text{2PLT}}$ model closely matches the 2PLT one. This is to be expected as Fig.~\ref{fig:PLTRetro} shows only a few cycles around the ISCO crossing, where both models are similar. Indeed, close to the ISCO, from Eq.~\eqref{defDObar} we have $\overline{\Delta\Omega}=\Delta\check\Omega + \mathcal{O}(\Delta\check\Omega^2)$.

\begin{figure*}
\includegraphics[width=.9\textwidth]{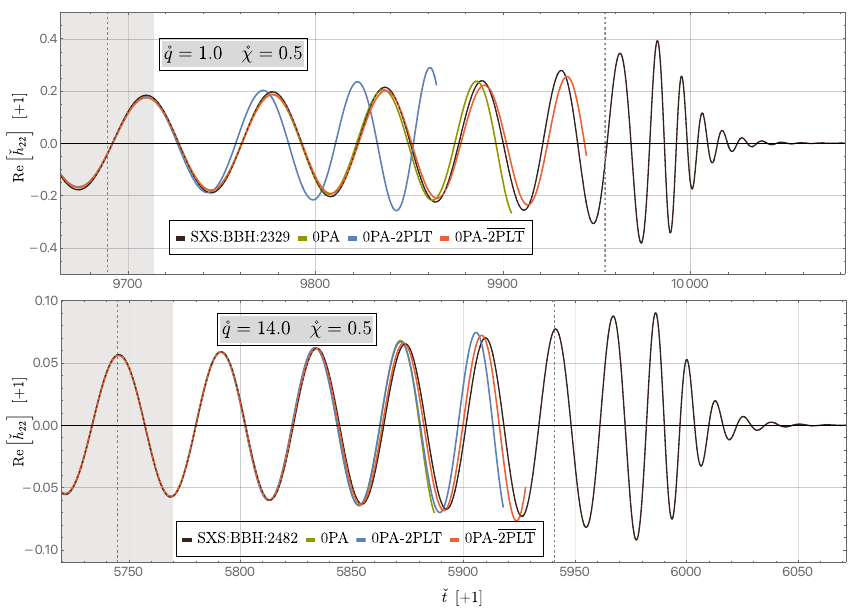}
\caption{$(\ell,m)=(2,2)$ mode of the gravitational waveform from a binary with primary spin $\mathring\chi=+0.5$ (prograde orbit) and large mass ratio $\mathring q=1$ (top) or $\mathring q=14$ (bottom). The NR waveforms (in black with indicated IDs) are taken from the SXS catalog~\cite{SXS:catalog}. The colored curves show the waveforms obtained from the 0PA inspiral model and our composite models 0PA-2PLT and 0PA-$\overline{\text{2PLT}}$, all of which are aligned with the NR waveform in the inspiral regime before the ISCO crossing. The vertical dashed gray line on the right side of the figures indicates the time at which the NR waveform frequency reaches twice the ISCO orbital frequency, $\omega_\text{NR}=-2\Omega_\star(\mathring{\chi})$, and the gray rectangle on the left shows the time window $[\check t_0-\delta {\check t}/2,\check t_0+\delta {\check t}/2]$ used for the alignment procedure of Sec.~\ref{subsec:NRComp}.}\label{fig:CompositeqLow}
\end{figure*}

We next assess the behavior of our composite models. Fig.~\ref{fig:CompositeqLow} shows the performance of 0PA-2PLT and 0PA-$\overline{\text{2PLT}}$ against NR  for a large mass ratio of either $\mathring{q}=1$ or $\mathring{q}=14$, with primary spin $\mathring\chi=+0.5$ in both cases and with the waveforms aligned at a time well before the ISCO. At equal mass, the 0PA-$\overline{\text{2PLT}}$ model is significantly more accurate than the 0PA-2PLT model, but this is solely a feature of it better mimicking the 0PA waveform (in green for comparison); for retrograde orbits (not shown), 0PA performs less well, and 0PA-$\overline{\text{2PLT}}$ hence provides less of an improvement over 0PA-2PLT. This difference between the 0PA-2PLT and 0PA-$\overline{\text{2PLT}}$ models becomes less and less pronounced as the large mass ratio $\mathring q$ increases, as shown in the lower panel. However, as our earlier analyses showed, these models would dephase significantly at earlier times even at $\mathring q=14$, due to the large errors in the 0PA-2PLT model at early times. Hence, the main takeaway from Fig.~\ref{fig:CompositeqLow} is simply this: 0PA-$\overline{\text{2PLT}}$ succeeds in preventing the transition information from spoiling the accuracy of the inspiral model. Since a 0PA model already incurs large errors in the inspiral, the true benefit of our alternative transition variable $\overline{\Delta \Omega}$ will not be realised until higher-order waveforms are constructed. We elucidate this further in the next section.

\section{Outlook}

In this paper, we extended Ref.~\cite{Kuchler:2024esj}'s transition-to-plunge framework to the case of a spinning primary black hole. To do so, we made use of the existence of different timescales: the orbital timescale $\sim1$, the radiation-reaction timescale $\sim 1/\mathring\varepsilon$, and the transition timescale $\sim 1/\mathring\varepsilon^{1/5}$. We performed multiscale expansions of the equations of motion~\eqref{eq:eom} following the phase-space approach of Refs.~\cite{Miller:2020bft,Kuchler:2024esj}. On one hand, for quasicircular inspirals around a Kerr primary, we derived the evolution equations in the inspiral up to 1PA order, extending the results for nonspinning black holes in Appendix~A of Ref.~\cite{Miller:2020bft}. On the other hand, we expanded the evolution equations in the transition up to 2PLT order, generalizing the results for nonspinning black holes in Ref.~\cite{Kuchler:2024esj}.

As a step toward building complete inspiral-merger-ringdown waveforms from first-principles self-force methods, we also constructed a composite waveform model that smoothly evolves from the 0PA inspiral dynamics to the 2PLT transition dynamics. We addressed the issue raised in Ref.~\cite{Kuchler:2024esj} about such composite waveform models: in the inspiral regime, the composite forcing function $\check F^{\check\Omega}_\text{0PA-2PLT}$ in Eq.~\eqref{eq:0PA2PLTForcingTerm} correctly converges toward the 0PA forcing function in the limit $\mathring\varepsilon\to0$ at fixed $\check\Omega$, but with excessively large coefficients in the subleading terms. We found that this behavior can be eliminated with a better choice of rescaled frequency variable, $\overline{\Delta\Omega}$, that is large (compared to $\check\Omega$) throughout the inspiral regime and satisfies
\begin{equation}
     \check\Omega\rightarrow0\iff \overline{\Delta\Omega}\rightarrow-\infty
\end{equation}
independently of the mass ratio.

We numerically implemented our transition model as well as the composite waveform model using both variables,  $\Delta\check\Omega$ and the improved variable $\overline{\Delta\Omega}$. Waveform generation in each model is made effectively immediate by leveraging the offline/online paradigm of the multiscale expansion. In the future we will provide a user-friendly version of the composite 0PA-$\overline{\text{2PLT}}$ waveform model in a further release of the \texttt{WaSABI} Mathematica package~\cite{BHPT_WaSABI}.

Using our models, we estimated the impact that the change to  $\overline{\Delta\Omega}$ has on the dephasing. We showed that, even for intermediate mass ratios $\mathring q\gtrsim 10^2$, the composite model in the original variable (0PA-2PLT) accumulates a few radians of phase error relative to the improved composite  (0PA-$\overline{\text{2PLT}}$) in the frequency interval from twice the ISCO radius to the ISCO radius. This makes the change to the improved variable crucial for constructing an accurate composite model. We also qualitatively compared our $\overline{\text{2PLT}}$ and 0PA-$\overline{\text{2PLT}}$ waveform models against NR simulations from the SXS catalog~\cite{SXS:catalog}. We showed that, as expected, the reformulation in terms of $\overline{\Delta\Omega}$ has little impact in the transition regime but a substantial impact on the quality of the waveform leading up to the transition.

Overall, the composite 0PA-$\overline{\text{2PLT}}$ waveform model does not have high accuracy, which is expected due to the low orders of the inspiral and transition expansions: a 0PA inspiral is insufficient for subradian accuracy~\cite{Hinderer:2008dm,Pound:2021qin,Wardell:2021fyy,Honet2025TBA}, meaning our composite model enters the transition already well out of phase with NR waveforms. Therefore, the next crucial step will be to extend the model to incorporate 1PA effects, which in turn requires carrying the transition equations to 7PLT order for the construction of a composite~\cite{Kuchler:2024esj}. For nonspinning binaries, for which a 1PA model is already built~\cite{Wardell:2021fyy}, the main obstacle is  constructing the 7PLT  model. In contrast, for a spinning primary black hole, the lack of an available 1PA model is a major hurdle. However, the self-force/post-Newtonian hybridization framework developed by some of us \cite{Honet2025TBA,PaperIV} can provide a quite accurate proxy for the complete 1PA dynamics, and it could be used to  approximate the unknown matching coefficients $\check F_{(l)}^{{(n,k)}}$ and $\check F_{[l]}^{{(n,k)}}$ that are needed to proceed to higher PLT orders. This work is currently under active investigation.
    
Finally, the next major step will be to extend our self-force framework for merger-ringdown waveforms~\cite{Kuchler:2025hwx} to the case of a spinning primary black hole. At leading, 0PG order, such a framework should produce the zero-eccentricity case of the merger-ringdown waveforms from Ref.~\cite{faggioli2025peakingabysscharacterizingmerger}. 
With such a  waveform at hand, one can then build a complete inspiral-merger-ringdown waveform model for quasicircular orbits around a Kerr primary, extending the nonspinning model from Ref.~\cite{KuchlerCapra27,KuchlerLISA,KuchlerAEI,KuchlerGR}. Such a model would have limited accuracy if the inspiral is restricted to 0PA order, but until a complete 1PA inspiral model in Kerr is at hand, one could use the post-Newtonian/self-force hybrid model from Refs.~\cite{Honet2025TBA,PaperIV}.

\acknowledgments

We are grateful to Maarten van de Meent for providing first-order self-force data and to Lionel London for helpful discussions. 
L.H. acknowledges the support of the Fonds National pour la Recherche Scientifique through a FRIA doctoral grant.
L.K. and A.P. acknowledge the support of the ERC Consolidator/UKRI Frontier Research Grant GWModels (selected by the ERC and funded by UKRI [grant number EP/Y008251/1]). 
A.P. additionally acknowledges the support of a Royal Society University Research Fellowship.
G.C. is Research Director of the FNRS. 
This work makes use of the Black Hole Perturbation Toolkit.

\appendix

\section{Mass dimensions}\label{app:massdimension}
We provide in Table \ref{tab:massdimension} a short glossary of the different symbols used during the re-expansion procedures, together with their mass dimensions $n$ as defined through Eq.~\eqref{eq:massdimension}.

\begin{table*}[htbp]
\makebox[\textwidth][c]{
    \begin{tabular}{lcc}
\toprule
      Name &    Symbol	&	Mass dimension $n$   \\ \midrule
  Initial primary mass		
  &	$\mathring M$	
  &	$+1$
  \\
  Primary mass		
  &	$M$	
  &	$+1$
  \\
    Secondary mass
  &	$m_p$
  &	$+1$
  \\
    Initial total mass
  &	$\mathring{M}_\text{tot}$
  &	$+1$
  \\
      Total mass
  &	${M}_\text{tot}$
  &	$+1$
  \\
  Dimensionless primary spin
  &	$\mathring{\chi}$
  &	$0$
  \\
    Small mass ratio
  &	$\mathring{\varepsilon}$
  &	$0$
  \\
   Small mass ratio to the power $1/5$
  &	$\mathring{\lambda}$
  &	$0$
  \\
    Symmetric mass ratio
  &	$\check{\nu}$
  &	$0$
  \\
      Symmetric small mass ratio to the power $1/5$
  &	$\check{\sigma}$
  &	$0$
  \\
   Boyer-Lindquist time and radius normalized by the primary mass
  &	$(t,r)$
  &	$+1$
  \\
    Boyer-Lindquist time and radius normalized by the total mass
  &	$(\check t,\check r)$
  &	$+1$
  \\
    Hyperboloidal time normalised by the primary mass  &	$s$
  &	$+1$
  \\
    Hyperboloidal time normalised by the total mass  &	$\check s$
  &	$+1$
  \\
       Orbital phase
  &	$\phi_p$
  &	$0$
  \\
     Orbital frequency normalized by the primary mass
  &	$\Omega$
  &	$-1$
  \\
       Orbital frequency normalized by the total mass
  &	$\check \Omega$
  &	$-1$
  \\
        Transition orbital frequency parameter normalized by the primary mass
  &	$\Delta \Omega$
  &	$-1$
  \\
     Transition orbital frequency parameter normalized by the total mass
  &	$\Delta\check \Omega$
  &	$-1$
  \\
       Reexpanded transition orbital frequency parameter normalized by  the total mass
  &	$\overline{\Delta \Omega}$
  &	$-1$
  \\
    $(\ell,m)$-mode of the wavestrain, normalized by the primary mass
  &	$H_{\ell m}$
  &	$+1$
  \\
      $(\ell,m)$-mode of the wavestrain, normalized by the total mass
  &	$\check H_{\ell m}$
  &	$+1$
  \\
    Orbital frequency forcing function normalized by the primary mass
  &	$F^\Omega=d\Omega/dt$
  &	$-2$
  \\
      Orbital frequency forcing function normalized by the total mass
  &	$\check F^{\check\Omega}=d\check\Omega/d\check t$
  &	$-2$
  \\
    Relative change of the primary mass (fixed $(\mathring{\varepsilon},\mathring{M})$ expansion)    
  &	$\delta M$
  &	$0$
  \\
      Relative change of the total mass (fixed $(\check{\nu},\mathring{M}_\text{tot})$ expansion)    
  &	$\delta \check M_\text{tot}$
  &	$0$
  \\
      Relative change of the primary spin (fixed $(\mathring{\varepsilon},\mathring{M})$ expansion)    
  &	$\delta \chi$
  &	$0$
  \\
      Relative change of the primary spin (fixed $(\check{\nu},\mathring{M}_\text{tot})$ expansion)    
  &	$\delta \check \chi$
  &	$0$
  \\
\bottomrule
\end{tabular}%
}
\caption{\label{tab:massdimension}Glossary of symbols used in the re-expansion procedures. Generally, quantities without overhead ornaments are normalized with respect to the initial primary mass $\mathring M$, while quantities with $\check\cdot$ overhead ornaments are normalized with respect to the initial total mass. The mass dimensions $n$ which enter the re-expansion scheme through Eq. \eqref{eq:massdimension} are also provided.}
\end{table*}%

\allowdisplaybreaks

\section{Transition-to-plunge functions}\label{app:ttp functions}

This appendix contains the expressions for the 3PLT and 4PLT corrections to the orbital radius and the redshift, which are required to derive the 2PLT equation of motion~\eqref{eq:F2PLT}. The solution $F_{[1]}^{\Delta\Omega}=0$ has already been used to simplify the final displayed expressions. At 3PLT order we have
\begin{subequations}\label{eq:3PLTOrb}
\begin{align}
    r_{[3]} &= -\frac{1}{3}\frac{r_\star^2}{U_\star^2\Delta_\star\Omega_\star^2}f_{[5]}^r\label{eq:r3PLT} \\
    U_{[3]} &= 0, \label{eq:U3PLT}
\end{align}
\end{subequations}
while at 4PLT order we obtain
\begin{subequations}\label{eq:4PLTOrb}
\begin{align}
    r_{[4]}=&\frac{1}{6}\left(\frac{d^3r_{(0)}}{d\Omega^3}\right)_\star\Delta\Omega^3\nonumber\\
    &+\frac{r_\star}{18\Omega_\star^2\Delta_\star^2}\left[6\mathcal{E}_\star(dr_{(0)}/d\Omega)_\star^2+\Delta_\star r_\star^{3/2}U_\star^2(dD/d\Omega)_\star\right.\nonumber\\
    &\phantom{000000000\frac{}{}}+6\Delta_\star r_\star\left(U_\star^{-1}(dr_{(0)}/d\Omega)_\star(dU_{(0)}/d\Omega)_\star\right.\nonumber\\
    &\left.\phantom{\frac{000000000}{}}\left.+(d^2r_{(0)}/d\Omega^2)_\star\right)\right]\left(F_{[0]}^{\Delta\Omega}\right)^2\nonumber\\
    &+\frac{r_\star^{2}(dr_{(0)}/d\Omega)_\star}{3\Omega_\star^2\Delta_\star}\left(\partial_{\Delta\Omega}F_{[0]}^{\Delta\Omega}F_{[2]}^{\Delta\Omega}+F_{[0]}^{\Delta\Omega}\partial_{\Delta\Omega}F_{[2]}^{\Delta\Omega}\right)\nonumber\\
    &+\frac{r_\star U_\star}{12\Omega_\star^4}\left(-4 r_\star^{3/2}\Omega_\star^2(dU_{(0)}/d\Omega)_\star+4U_\star\right.\nonumber\\
    &\left.\phantom{\frac{}{}}+3r_\star^2\Omega_\star^4U_\star(d^2r_{(0)}/d\Omega^2)_\star\right)\Delta\Omega F_{[0]}^{\Delta\Omega}\partial_{\Delta\Omega}F_{[0]}^{\Delta\Omega},\label{eq:r4PLT}\\
    U_{[4]}=&\frac{1}{6}\left(\frac{d^3U_{(0)}}{d\Omega^3}\right)_\star\Delta\Omega^3 + \frac{1}{2}\frac{r_\star^2U_\star^3 (dr_{(0)}/d\Omega)^2}{\Delta_\star}\left(F_{[0]}^{\Delta\Omega}\right)^2.\label{eq:U4PLT}
\end{align}
\end{subequations}

\section{Re-expansions of transition-to-plunge functions}\label{app:re-expansions}

We present the relations up to 7PLT order between the forcing functions $\check F_{[n]}^{\Delta\check\Omega}$ and $F_{[n]}^{\Delta\Omega}$, extending the discussion from Sec.~\ref{subsec:PLTreexp}:
\begin{subequations}\label{eq:PLTReexpansion2}
\begin{align}
    \check F_{[n]}^{\Delta\check\Omega}(\Delta\check J^a) &= F_{[n]}^{\Delta\Omega}(\Delta\check J^a),\phantom{0000}n=3,4,
    \\
    \check F_{[5]}^{\Delta\check\Omega}(\Delta\check J^a) =&\, F_{[5]}^{\Delta\Omega}(\Delta\check J^a)+\frac{16}{5}F_{[0]}^{\Delta\Omega}(\Delta\check\Omega)\nonumber\\
    &-\frac{9}{5}\Delta\check\Omega\,\partial_{\Delta\check\Omega}
    F_{[0]}^{\Delta\Omega}(\Delta\check\Omega),\\
    \check F_{[6]}^{\Delta\check\Omega}(\Delta\check J^a) =&\, F_{[6]}^{\Delta\Omega}(\Delta\check J^a),\\
    \check F_{[7]}^{\Delta\check\Omega}(\Delta\check J^a) =&\, F_{[7]}^{\Delta\Omega}(\Delta\check J^a)+4F_{[2]}^{\Delta\Omega}(\Delta\check\Omega)\nonumber\\
    &-\frac{9}{5}\Delta\check\Omega\,\partial_{\Delta\check\Omega}
    F_{[2]}^{\Delta\Omega}(\Delta\check\Omega).
\end{align}
\end{subequations}
Similarly, the relationship between the mode amplitudes $\check H_{\ell m}^{[n]}$ and $H_{\ell m}^{[n]}$ is
\begin{subequations}\label{eq:PLTReexpansionHlm}
\begin{align}
    \check H_{\ell m}^{[n]}(\Delta\check J^a) =&\, H_{\ell m}^{[n]}(\Delta\check J^a),\phantom{0000} n=8,9,\\
    \check H_{\ell m}^{[10]}(\Delta\check J^a) =&\, H_{\ell m}^{[10]}(\Delta\check J^a) + H_{\ell m}^{[5]}(\Delta\check J^a)\nonumber\\
    &-\delta\check M_\text{tot}\partial_{\delta\check M_\text{tot}}
    H_{\ell m}^{[5]}(\Delta\check J^a)\nonumber\\
    &-2\delta\check\chi\partial_{\delta\check\chi}
    H_{\ell m}^{[5]}(\Delta\check J^a),\\
    \check H_{\ell m}^{[11]}(\Delta\check J^a) =&\, H_{\ell m}^{[11]}(\Delta\check J^a),\\
    \check H_{\ell m}^{[12]}(\Delta\check J^a) =&\, H_{\ell m}^{[12]}(\Delta\check J^a)+\frac{9}{5}H_{\ell m}^{[7]}(\Delta\check\Omega)\nonumber\\
    &-\frac{9}{5}\Delta\check\Omega\partial_{\Delta\check\Omega}
    H_{\ell m}^{[7]}(\Delta\check\Omega).
\end{align}
\end{subequations}

We also provide the relations up to 4PLT order between the forcing functions $\bar{F}^{\overline{\Delta\Omega}}_{[n]}$ and $\check F^{\Delta\check\Omega}_{[n]}$, extending the discussion in Sec.~\ref{sec:reexpansion}:
\begin{align}\label{eq:PLTReexpansionBar appendix}
    \bar{F}^{\overline{\Delta\Omega}}_{[3]}(\overline{\Delta J^a}) =&\, \check{F}^{\Delta\check\Omega}_{[3]}(\overline{\Delta J^a}),
    \\
    \bar{F}^{\overline{\Delta\Omega}}_{[4]}(\overline{\Delta J^a}) =&\, \check{F}^{\Delta\check\Omega}_{[4]}(\overline{\Delta J^a}) + \frac{\overline{\Delta\Omega}^2}{\check\Omega_\star} \partial_{\overline{\Delta\Omega}}\check{F}^{\Delta\check\Omega}_{[2]}(\overline{\Delta\Omega})\nonumber\\
    &+ \frac{\overline{\Delta\Omega}^3}{\check\Omega_\star^2} \partial_{\overline{\Delta\Omega}}\check{F}^{\Delta\check\Omega}_{[0]}(\overline{\Delta\Omega})\nonumber\\
    &+ \frac{\overline{\Delta\Omega}^4}{2\check\Omega_\star^2} \partial^2_{\overline{\Delta\Omega}}\check{F}^{\Delta\check\Omega}_{[0]}(\overline{\Delta\Omega}),
\end{align}
Similarly, the relationship between the mode amplitudes $\bar{H}_{\ell m}^{[n]}$ and $\check{H}_{\ell m}^{[n]}$ is
\begin{align}\label{eq:PLTReexpansionHlmbar appendix}
    \bar{H}_{\ell m}^{[8]}(\overline{\Delta\Omega}) &= \check{H}_{\ell m}^{[8]}(\overline{\Delta \Omega}),
    \\
    \bar{H}_{\ell m}^{[9]}(\overline{\Delta\Omega}) &= \check{H}_{\ell m}^{[9]}(\overline{\Delta\Omega}) + \frac{\overline{\Delta\Omega}^2}{\check\Omega_\star} \partial_{\overline{\Delta\Omega}}\check{H}_{\ell m}^{[7]} (\overline{\Delta\Omega}).
\end{align}

\clearpage
\bibliographystyle{utphys}
\bibliography{ThisBib.bib}

\providecommand{\href}[2]{#2}\begingroup\raggedright\begin{thebibliography}{10}

\bibitem{LIGOScientific:2016aoc}
{\bf LIGO Scientific, Virgo} Collaboration, B.~P. Abbott {\em et al.}, ``{Observation of Gravitational Waves from a Binary Black Hole Merger},'' {\em Phys. Rev. Lett.} {\bf 116} (2016), no.~6, 061102, \href{http://www.arXiv.org/abs/1602.03837}{{\tt 1602.03837}}.

\bibitem{Abbott_2023}
{\bf KAGRA, LIGO Scientific, Virgo} Collaboration, B.~P. Abbott {\em et al.}, ``Gwtc-3: Compact binary coalescences observed by ligo and virgo during the second part of the third observing run,'' {\em Physical Review X} {\bf 13} (Dec., 2023).

\bibitem{GraceDB}
{\bf LIGO Scientific} Collaboration, ``Gravitational-wave candidate event database.'' \url{https://gracedb.ligo.org}.
\newblock Accessed: 2025-02-04.

\bibitem{toubiana2024indistinguishabilitycriterionestimatingpresence}
A.~Toubiana and J.~R. Gair, ``Indistinguishability criterion and estimating the presence of biases,'' 2024.

\bibitem{PhysRevD.57.4566}
E.~E. Flanagan and S.~A. Hughes, ``Measuring gravitational waves from binary black hole coalescences. ii. the waves' information and its extraction, with and without templates,'' {\em Phys. Rev. D} {\bf 57} (Apr, 1998) 4566--4587.

\bibitem{PhysRevD.78.124020}
L.~Lindblom, B.~J. Owen, and D.~A. Brown, ``Model waveform accuracy standards for gravitational wave data analysis,'' {\em Phys. Rev. D} {\bf 78} (Dec, 2008) 124020.

\bibitem{PhysRevD.82.024014}
S.~T. McWilliams, B.~J. Kelly, and J.~G. Baker, ``Observing mergers of nonspinning black-hole binaries,'' {\em Phys. Rev. D} {\bf 82} (Jul, 2010) 024014.

\bibitem{PhysRevD.95.104004}
K.~Chatziioannou, A.~Klein, N.~Yunes, and N.~Cornish, ``Constructing gravitational waves from generic spin-precessing compact binary inspirals,'' {\em Phys. Rev. D} {\bf 95} (May, 2017) 104004.

\bibitem{PhysRevResearch.2.023151}
M.~P\"urrer and C.-J. Haster, ``Gravitational waveform accuracy requirements for future ground-based detectors,'' {\em Phys. Rev. Res.} {\bf 2} (May, 2020) 023151.

\bibitem{Wardell:2021fyy}
B.~Wardell, A.~Pound, N.~Warburton, J.~Miller, L.~Durkan, and A.~Le~Tiec, ``{Gravitational Waveforms for Compact Binaries from Second-Order Self-Force Theory},'' {\em Phys. Rev. Lett.} {\bf 130} (2023), no.~24, 241402, \href{http://www.arXiv.org/abs/2112.12265}{{\tt 2112.12265}}.

\bibitem{LVKWhitePaper}
{\bf LVK} Collaboration, ``{The LSC-Virgo-KAGRA Observational Science White Paper (2024 Edition)},'' \href{http://www.arXiv.org/abs/LIGO-T2300406-v1}{{\tt LIGO-T2300406-v1}}.

\bibitem{LISA:2024hlh}
{\bf LISA} Collaboration, M.~Colpi {\em et al.}, ``{LISA Definition Study Report},'' \href{http://www.arXiv.org/abs/2402.07571}{{\tt 2402.07571}}.

\bibitem{ET:2019dnz}
{\bf ET} Collaboration, M.~Maggiore {\em et al.}, ``{Science Case for the Einstein Telescope},'' {\em JCAP} {\bf 03} (2020) 050, \href{http://www.arXiv.org/abs/1912.02622}{{\tt 1912.02622}}.

\bibitem{PaperII}
J.~Mathews, B.~Wardell, A.~Pound, and N.~Warburton, ``Post-adiabatic self-force waveforms: slowly spinning primary and precessing secondary,'' {\em to appear}.

\bibitem{Chapman-Bird:2025xtd}
C.~E.~A. Chapman-Bird {\em et al.}, ``{The Fast and the Frame-Dragging: Efficient waveforms for asymmetric-mass eccentric equatorial inspirals into rapidly-spinning black holes},'' \href{http://www.arXiv.org/abs/2506.09470}{{\tt 2506.09470}}.

\bibitem{Hinderer:2008dm}
T.~Hinderer and E.~E. Flanagan, ``{Two timescale analysis of extreme mass ratio inspirals in Kerr. I. Orbital Motion},'' {\em Phys. Rev.} {\bf D78} (2008) 064028,
\href{http://www.arXiv.org/abs/0805.3337}{{\tt 0805.3337}}.

\bibitem{Miller:2020bft}
J.~Miller and A.~Pound, ``{Two-timescale evolution of extreme-mass-ratio inspirals: waveform generation scheme for quasicircular orbits in Schwarzschild spacetime},'' {\em Phys. Rev. D} {\bf 103} (2021), no.~6, 064048, \href{http://www.arXiv.org/abs/2006.11263}{{\tt 2006.11263}}.

\bibitem{Pound:2021qin}
A.~{Pound} and B.~{Wardell}, ``{Black Hole Perturbation Theory and Gravitational Self-Force},'' in {\em Handbook of Gravitational Wave Astronomy}, p.~38.
\newblock 2022.
\newblock \href{http://www.arXiv.org/abs/2101.04592}{{\tt 2101.04592}}.

\bibitem{Mathews:2025nyb}
J.~Mathews and A.~Pound, ``{Post-adiabatic waveform-generation framework for asymmetric precessing binaries},'' \href{http://www.arXiv.org/abs/2501.01413}{{\tt 2501.01413}}.

\bibitem{Katz:2021yft}
M.~L. Katz, A.~J.~K. Chua, L.~Speri, N.~Warburton, and S.~A. Hughes, ``{Fast extreme-mass-ratio-inspiral waveforms: New tools for millihertz gravitational-wave data analysis},'' {\em Phys. Rev. D} {\bf 104} (2021), no.~6, 064047, \href{http://www.arXiv.org/abs/2104.04582}{{\tt 2104.04582}}.

\bibitem{Warburton:2021kwk}
N.~Warburton, A.~Pound, B.~Wardell, J.~Miller, and L.~Durkan, ``{Gravitational-Wave Energy Flux for Compact Binaries through Second Order in the Mass Ratio},'' {\em Phys. Rev. Lett.} {\bf 127} (2021), no.~15, 151102, \href{http://www.arXiv.org/abs/2107.01298}{{\tt 2107.01298}}.

\bibitem{Albertini:2022rfe}
A.~Albertini, A.~Nagar, A.~Pound, N.~Warburton, B.~Wardell, L.~Durkan, and J.~Miller, ``{Comparing second-order gravitational self-force, numerical relativity, and effective one body waveforms from inspiralling, quasicircular, and nonspinning black hole binaries},'' {\em Phys. Rev. D} {\bf 106} (2022), no.~8, 084061, \href{http://www.arXiv.org/abs/2208.01049}{{\tt 2208.01049}}.

\bibitem{Kuchler:2024esj}
L.~K\"uchler, G.~Comp\`ere, L.~Durkan, and A.~Pound, ``{Self-force framework for transition-to-plunge waveforms},'' {\em SciPost Phys.} {\bf 17} (4, 2024) 056, \href{http://www.arXiv.org/abs/2405.00170}{{\tt 2405.00170}}.

\bibitem{Compere:2021iwh}
G.~Comp\`ere and L.~K\"uchler, ``{Self-consistent adiabatic inspiral and transition motion},'' {\em Phys. Rev. Lett.} {\bf 126} (2021), no.~24, 241106, \href{http://www.arXiv.org/abs/2102.12747}{{\tt 2102.12747}}.

\bibitem{PhysRevLett.128.029901}
G.~Comp\`ere and L.~K\"uchler, ``Erratum: Self-consistent adiabatic inspiral and transition motion [phys. rev. lett. 126, 241106 (2021)],'' {\em Phys. Rev. Lett.} {\bf 128} (Jan, 2022) 029901.

\bibitem{Compere:2021zfj}
G.~Comp\`ere and L.~K\"uchler, ``{Asymptotically matched quasi-circular inspiral and transition-to-plunge in the small mass ratio expansion},'' {\em SciPost Phys.} {\bf 13} (2022), no.~2, 043, \href{http://www.arXiv.org/abs/2112.02114}{{\tt 2112.02114}}.

\bibitem{Lewis:2025ydo}
J.~Lewis, T.~Kakehi, A.~Pound, and T.~Tanaka, ``{Post-adiabatic dynamics and waveform generation in self-force theory: an invariant pseudo-Hamiltonian framework},'' \href{http://www.arXiv.org/abs/2507.08081}{{\tt 2507.08081}}.

\bibitem{Ori:2000zn}
A.~Ori and K.~S. Thorne, ``{The Transition from inspiral to plunge for a compact body in a circular equatorial orbit around a massive, spinning black hole},'' {\em Phys. Rev.} {\bf D62} (2000) 124022,
\href{http://www.arXiv.org/abs/gr-qc/0003032}{{\tt gr-qc/0003032}}.

\bibitem{Apte:2019txp}
A.~Apte and S.~A. Hughes, ``{Exciting black hole modes via misaligned coalescences: I. Inspiral, transition, and plunge trajectories using a generalized Ori-Thorne procedure},'' {\em Phys. Rev. D} {\bf 100} (2019), no.~8, 084031, \href{http://www.arXiv.org/abs/1901.05901}{{\tt 1901.05901}}.

\bibitem{Kuchler:2025hwx}
L.~K\"uchler, G.~Comp\`ere, and A.~Pound, ``{Self-force framework for merger-ringdown waveforms},'' \href{http://www.arXiv.org/abs/2506.02189}{{\tt 2506.02189}}.

\bibitem{Hadar:2009ip}
S.~Hadar and B.~Kol, ``{Post-ISCO Ringdown Amplitudes in Extreme Mass Ratio Inspiral},'' {\em Phys. Rev. D} {\bf 84} (2011) 044019, \href{http://www.arXiv.org/abs/0911.3899}{{\tt 0911.3899}}.

\bibitem{Folacci:2018cic}
A.~Folacci and M.~Ould El~Hadj, ``{Multipolar gravitational waveforms and ringdowns generated during the plunge from the innermost stable circular orbit into a Schwarzschild black hole},'' {\em Phys. Rev. D} {\bf 98} (2018), no.~8, 084008, \href{http://www.arXiv.org/abs/1806.01577}{{\tt 1806.01577}}.

\bibitem{Roy:2025kra}
A.~Roy, L.~K{\"u}chler, A.~Pound, and R.~Panosso~Macedo, ``{Black hole mergers beyond general relativity: a self-force approach},'' \href{http://www.arXiv.org/abs/2510.11793}{{\tt 2510.11793}}.

\bibitem{KuchlerCapra27}
L.~K\"uchler, ``Progress towards merger-ringdown waveforms from self-force theory.'' Talk given at 27th Capra Meeting on Radiation Reaction in General Relativity (National University of Singapore), https://www.caprameeting.org/capra-meetings/capra-27/abstracts, 2024.

\bibitem{KuchlerLISA}
L.~K\"uchler, ``Self-force framework for transition-to-plunge waveforms.'' Talk given at 15th International LISA Symposium (University College Dublin), https://www.lisasymposium2024.ie/programme/, 2024.

\bibitem{KuchlerAEI}
L.~K\"uchler, ``Self-force framework for transition-to-plunge waveforms.'' Talk given at Fundamental Physics Meets Waveforms with LISA Workshop (Max Planck Institute for Gravitational Physics, Potsdam), https://workshops.aei.mpg.de/fpmeetswavelisa/program/, 2024.

\bibitem{KuchlerGR}
L.~K\"uchler, ``Inspiral-merger-ringdown waveforms from gravitational self-force theory.'' Talk given at the 24th International Conference on General Relativity and Gravitation and the 16th Edoardo Amaldi Conference on Gravitational Waves (Glasgow, UK), https://iop.eventsair.com/gr24-amaldi16/conference-agenda, 2025.

\bibitem{Rifat:2019ltp}
N.~E.~M. Rifat, S.~E. Field, G.~Khanna, and V.~Varma, ``{Surrogate model for gravitational wave signals from comparable and large-mass-ratio black hole binaries},'' {\em Phys. Rev. D} {\bf 101} (2020), no.~8, 081502, \href{http://www.arXiv.org/abs/1910.10473}{{\tt 1910.10473}}.

\bibitem{Islam:2022laz}
T.~Islam, S.~E. Field, S.~A. Hughes, G.~Khanna, V.~Varma, M.~Giesler, M.~A. Scheel, L.~E. Kidder, and H.~P. Pfeiffer, ``{Surrogate model for gravitational wave signals from nonspinning, comparable-to large-mass-ratio black hole binaries built on black hole perturbation theory waveforms calibrated to numerical relativity},'' {\em Phys. Rev. D} {\bf 106} (2022), no.~10, 104025, \href{http://www.arXiv.org/abs/2204.01972}{{\tt 2204.01972}}.

\bibitem{Rink:2024swg}
K.~Rink, R.~Bachhar, T.~Islam, N.~E.~M. Rifat, K.~Gonzalez-Quesada, S.~E. Field, G.~Khanna, S.~A. Hughes, and V.~Varma, ``{Gravitational wave surrogate model for spinning, intermediate mass ratio binaries based on perturbation theory and numerical relativity},'' {\em Phys. Rev. D} {\bf 110} (2024), no.~12, 124069, \href{http://www.arXiv.org/abs/2407.18319}{{\tt 2407.18319}}.

\bibitem{Sundararajan:2007jg}
P.~A. Sundararajan, G.~Khanna, and S.~A. Hughes, ``{Towards adiabatic waveforms for inspiral into Kerr black holes. I. A New model of the source for the time domain perturbation equation},'' {\em Phys. Rev. D} {\bf 76} (2007) 104005, \href{http://www.arXiv.org/abs/gr-qc/0703028}{{\tt gr-qc/0703028}}.

\bibitem{Detweiler:2011tt}
S.~Detweiler, ``{Gravitational radiation reaction and second order perturbation theory},'' {\em Phys. Rev. D} {\bf 85} (2012) 044048, \href{http://www.arXiv.org/abs/1107.2098}{{\tt 1107.2098}}.

\bibitem{Upton:2021oxf}
S.~D. Upton and A.~Pound, ``{Second-order gravitational self-force in a highly regular gauge},'' {\em Phys. Rev. D} {\bf 103} (2021), no.~12, 124016, \href{http://www.arXiv.org/abs/2101.11409}{{\tt 2101.11409}}.

\bibitem{Pound:2014xva}
A.~Pound and J.~Miller, ``{Practical, covariant puncture for second-order self-force calculations},'' {\em Phys. Rev. D} {\bf 89} (2014), no.~10, 104020, \href{http://www.arXiv.org/abs/1403.1843}{{\tt 1403.1843}}.

\bibitem{Detweiler:2000gt}
S.~L. Detweiler, ``{Radiation reaction and the selfforce for a point mass in general relativity},'' {\em Phys. Rev. Lett.} {\bf 86} (2001) 1931--1934, \href{http://www.arXiv.org/abs/gr-qc/0011039}{{\tt gr-qc/0011039}}.

\bibitem{Pound_2012}
A.~Pound, ``Second-order gravitational self-force,'' {\em Physical Review Letters} {\bf 109} (July, 2012).

\bibitem{Pound:2017psq}
A.~Pound, ``{Nonlinear gravitational self-force: second-order equation of motion},'' {\em Phys. Rev. D} {\bf 95} (2017), no.~10, 104056, \href{http://www.arXiv.org/abs/1703.02836}{{\tt 1703.02836}}.

\bibitem{Pound:2012nt}
A.~Pound, ``{Second-order gravitational self-force},'' {\em Phys. Rev. Lett.} {\bf 109} (2012) 051101, \href{http://www.arXiv.org/abs/1201.5089}{{\tt 1201.5089}}.

\bibitem{Pound_2015}
A.~Pound, ``Gauge and motion in perturbation theory,'' {\em Physical Review D} {\bf 92} (Aug., 2015).

\bibitem{Miller:2023ers}
J.~Miller, B.~Leather, A.~Pound, and N.~Warburton, ``{Worldtube puncture scheme for first- and second-order self-force calculations in the Fourier domain},'' {\em Phys. Rev. D} {\bf 109} (2024), no.~10, 104010, \href{http://www.arXiv.org/abs/2401.00455}{{\tt 2401.00455}}.

\bibitem{PhysRevLett.129.161101}
A.~Mummery and S.~Balbus, ``Inspirals from the innermost stable circular orbit of kerr black holes: Exact solutions and universal radial flow,'' {\em Phys. Rev. Lett.} {\bf 129} (Oct, 2022) 161101.

\bibitem{Pound:2009sm}
A.~Pound, ``{Self-consistent gravitational self-force},'' {\em Phys. Rev. D} {\bf 81} (2010) 024023, \href{http://www.arXiv.org/abs/0907.5197}{{\tt 0907.5197}}.

\bibitem{Kuchler:2023jbu}
{K\"uchler, Lorenzo}, {\em Inspiral, transition and plunge: a framework for complete waveforms in the small-mass-ratio expansion}.
\newblock PhD thesis, U. Brussels, 2023.

\bibitem{Teukolsky:1974yv}
S.~Teukolsky and W.~Press, ``{Perturbations of a rotating black hole. III - Interaction of the hole with gravitational and electromagnet ic radiation},'' {\em Astrophys. J.} {\bf 193} (1974) 443--461.

\bibitem{TeukolskyPackage}
B.~Wardell, N.~Warburton, K.~Cunningham, L.~Durkan, B.~Leather, Z.~Nasipak, C.~Kavanagh, T.~Torres, A.~Ottewill, and M.~Casals, ``Teukolsky (1.0.4),'' 2023.

\bibitem{Tiec:2014lba}
A.~Le~Tiec, ``{The Overlap of Numerical Relativity, Perturbation Theory and Post-Newtonian Theory in the Binary Black Hole Problem},'' {\em Int. J. Mod. Phys.} {\bf D23} (2014), no.~10, 1430022,
\href{http://www.arXiv.org/abs/1408.5505}{{\tt 1408.5505}}.

\bibitem{LeTiec:2011bk}
A.~Le~Tiec, A.~H. Mroue, L.~Barack, A.~Buonanno, H.~P. Pfeiffer, N.~Sago, and A.~Taracchini, ``{Periastron Advance in Black Hole Binaries},'' {\em Phys. Rev. Lett.} {\bf 107} (2011) 141101, \href{http://www.arXiv.org/abs/1106.3278}{{\tt 1106.3278}}.

\bibitem{LeTiec:2011dp}
A.~Le~Tiec, E.~Barausse, and A.~Buonanno, ``{Gravitational Self-Force Correction to the Binding Energy of Compact Binary Systems},'' {\em Phys. Rev. Lett.} {\bf 108} (2012) 131103, \href{http://www.arXiv.org/abs/1111.5609}{{\tt 1111.5609}}.

\bibitem{LeTiec:2013uey}
A.~Le~Tiec {\em et al.}, ``{Periastron Advance in Spinning Black Hole Binaries: Gravitational Self-Force from Numerical Relativity},'' {\em Phys. Rev. D} {\bf 88} (2013), no.~12, 124027, \href{http://www.arXiv.org/abs/1309.0541}{{\tt 1309.0541}}.

\bibitem{Nagar:2013sga}
A.~Nagar, ``{Gravitational recoil in nonspinning black hole binaries: the span of test-mass results},'' {\em Phys. Rev. D} {\bf 88} (2013), no.~12, 121501, \href{http://www.arXiv.org/abs/1306.6299}{{\tt 1306.6299}}.

\bibitem{LeTiec:2017ebm}
A.~Le~Tiec and P.~Grandcl\'ement, ``{Horizon Surface Gravity in Corotating Black Hole Binaries},'' {\em Class. Quant. Grav.} {\bf 35} (2018), no.~14, 144002, \href{http://www.arXiv.org/abs/1710.03673}{{\tt 1710.03673}}.

\bibitem{vandeMeent:2020xgc}
M.~van~de Meent and H.~P. Pfeiffer, ``{Intermediate mass-ratio black hole binaries: Applicability of small mass-ratio perturbation theory},'' {\em Phys. Rev. Lett.} {\bf 125} (2020), no.~18, 181101, \href{http://www.arXiv.org/abs/2006.12036}{{\tt 2006.12036}}.

\bibitem{Buonanno:2000ef}
A.~Buonanno and T.~Damour, ``{Transition from inspiral to plunge in binary black hole coalescences},'' {\em Phys. Rev.} {\bf D62} (2000) 064015,
\href{http://www.arXiv.org/abs/gr-qc/0001013}{{\tt gr-qc/0001013}}.

\bibitem{Compere:2019cqe}
G.~Comp\`ere, K.~Fransen, and C.~Jonas, ``{Transition from inspiral to plunge into a highly spinning black hole},'' {\em Class. Quant. Grav.} {\bf 37} (2020), no.~9, 095013, \href{http://www.arXiv.org/abs/1909.12848}{{\tt 1909.12848}}.

\bibitem{Gralla:2016qfw}
S.~E. Gralla, S.~A. Hughes, and N.~Warburton, ``{Inspiral into Gargantua},'' {\em Class. Quant. Grav.} {\bf 33} (2016), no.~15, 155002,
\href{http://www.arXiv.org/abs/1603.01221}{{\tt 1603.01221}}.

\bibitem{KevorkianCole}
J.~Kevorkian and J.~D. Cole, {\em Multiple Scale and Singular Perturbation Methods}.
\newblock Springer, 1996.

\bibitem{BHPToolkit}
``{Black Hole Perturbation Toolkit}.'' (\href{http://bhptoolkit.org/}{bhptoolkit.org}).

\bibitem{SXS:catalog}
``Sxs catalog.'' "\url{data.black-holes.org}".

\bibitem{Honet2025TBA}
L.~Honet, G.~Compère, and A.~Pound, ``{Hybrid waveform model for asymmetric spinning binaries},'' {\em to appear}.

\bibitem{sxs_collaboration_2024_13147581}
{SXS Collaboration}, ``Binary black-hole simulation sxs:bbh:2482,'' Aug., 2024.
\newblock {10.5281/zenodo.13147581}.

\bibitem{Yoo:2022erv}
J.~Yoo, V.~Varma, M.~Giesler, M.~A. Scheel, C.-J. Haster, H.~P. Pfeiffer, L.~E. Kidder, and M.~Boyle, ``{Targeted large mass ratio numerical relativity surrogate waveform model for GW190814},'' {\em Phys. Rev. D} {\bf 106} (2022), no.~4, 044001, \href{http://www.arXiv.org/abs/2203.10109}{{\tt 2203.10109}}.

\bibitem{BHPT_WaSABI}
B.~Wardell, J.~Mathews, and L.~Honet, ``\href{https://doi.org/10.5281/zenodo.16692237}{WaSABI},'' Aug., 2025.

\bibitem{PaperIV}
L.~Honet, J.~Mathews, G.~Comp\`ere, A.~Pound, B.~Wardell, G.~Piovano, M.~van~de Meent, and N.~Warburton, ``Spin-aligned inspiral waveforms from self-force and post-newtonian theory,'' {\em to appear}.

\bibitem{faggioli2025peakingabysscharacterizingmerger}
G.~Faggioli, M.~van~de Meent, A.~Buonanno, and G.~Khanna, ``Peaking into the abyss: Characterizing the merger of equatorial-eccentric-geodesic plunges in rotating black holes,'' 2025.

\end{thebibliography}\endgroup

\end{document}